\documentclass[a4paper,11pt]{article}

\usepackage{jheppub} 
\usepackage{tikz}
\usepackage{dsfont}
\usepackage{extarrows}
\usepackage{hyperref}
\usepackage{cleveref}
\usepackage{supertabular}
\usepackage{todonotes}
\usepackage{mathrsfs}
\usepackage{array}
\usepackage{lipsum}
\usepackage{breqn}
\usepackage{physics}
\usepackage{tikz-cd}
\usepackage{slashed}
\usepackage{tabularx}
\usepackage{multirow}
\usetikzlibrary{decorations.markings}

\usepackage{helvet} 
\usepackage{amsmath}
\usepackage{amssymb} 

\newcommand{\newreptheorem}[2]{%
\newenvironment{rep#1}[1]{%
 \def\rep@title{#2 \ref{##1}}%
 \begin{rep@theorem}}%
 {\end{rep@theorem}}}
\makeatother

\newreptheorem{lemma}{Lemma}

\newreptheorem{conj}{Conjecture}

\usepackage{lscape} 
\usepackage{braket}

\usepackage[T1]{fontenc} 

\newcommand{\Hom}{{\text{Hom}}}

\newcommand{\Rep}{\textbf{Rep}}

\newcommand{\be}{\begin{equation}}
\newcommand{\ee}{\end{equation}}
\newcommand{\ba}{\begin{aligned}}
\newcommand{\ea}{\end{aligned}}
\newcommand{\bea}{\begin{eqnarray}}
\newcommand{\eea}{\end{eqnarray}}

\def\Tr{\mathop{\mathrm{Tr}}\nolimits}
\def\tr{\mathop{\mathrm{tr}}\nolimits}

\def\bp{\begin{pmatrix}}
\def\ep{\end{pmatrix}}

\def\ker{\mathrm{ker}}

\def\bbC{\mathbb{C}}

\def\Hom{\textrm{Hom}}

\def\hilb{{\rm Hilb}}
\newcommand{\Ad}{\mathrm{Ad}}

\title{\boldmath Hilbert Space and Defect Hilbert Spaces \\ Associated with Categorical Symmetries
}

\author[b]{Qiang Jia}
\author[a]{Jiahua Tian}
\affiliation[a]{School of Physics, East China Normal University, \\
Shanghai, China, 200241}
\affiliation[b]{Department of Physics, Korea Advanced Institute of Science Technology, \\
Daejeon 34141, Korea}
\emailAdd{jtian1905@gmail.com}

\abstract{We present a quantum mechanical approach to understanding the Hilbert space and the defect Hilbert spaces associated with line operators of BF theory combined with level-$k$ Chern-Simons theory. The defect Hilbert spaces are closely related to the category of $*$-representations of the $C^*$-algebra of the compactly supported sections of the Fell line bundle over the conjugation action groupoid $G//_\Ad G$, and the structure of this category and the groupoid action on the objects of this category is interpreted quantum mechanically. We show that the action of the line operators on the Hilbert space of the $BF+kCS$ TQFT is given concretely by a convolution between the kernels that represent the line operators, and that the codimension-$2$ twist and the codimension-$1$ prequantum line bundle arise as two transgressions of the same universal level $k\in H^4(BG,\mathbb{Z})$. For finite gauge group, the resulting convolution-eigenvalue formula is identified with the Verlinde formula for the (twisted) Drinfeld double $D^\omega(G)$ via an explicit phase-by-phase match with the known finite modular data. For compact Lie group, the convolution-kernel eigenvalues coincide in the regular sector with the semiclassical Hopf-link $S$-kernel, identifying two complementary derivations of the same modular data.}

\begin{document}
\maketitle
\flushbottom

\section{Introduction}\label{sec:Intro}

The modern notion of symmetry in quantum field theory has expanded well beyond ordinary group actions on local operators. Higher-form symmetries, higher-group symmetries, and non-invertible symmetries are all most naturally described in terms of topological defects and their fusion, rather than solely by invertible operators acting on pointlike fields~\cite{Gaiotto:2014kfa,Cordova:2018cvg,Freed:2022qnc,Schafer-Nameki:2023jdn,Shao:2023gho}. This viewpoint connects physical constructions with the tensor-categorical language of topological phases, where topological orders, gapped boundaries, anyon condensation, and anomalous symmetry actions are organized by fusion categories, Drinfeld centers, and related higher-categorical structures~\cite{Chen:2011pg,Chen:2012ctz,Gu:2012ib,Kong:2015flk,Kong:2020cie,Ji:2019jhk}. In this sense, categorical symmetry is not merely a refinement of group symmetry, but a common language for symmetry, anomaly, and extended topological degrees of freedom.

The SymTFT perspective encodes symmetry data into a topological quantum field theory (TQFT) in one higher dimension. In this picture, the symmetry operators and their action on the physical theory are realized through topological boundary conditions of the TQFT, while the bulk defects encode the data of generalized charges, anomalies, gauging, and condensation~\cite{Apruzzi:2021nmk,Kaidi:2022cpf,Kaidi:2023maf,Bhardwaj:2023ayw,Apruzzi:2023uma}. Recent work has developed this framework for non-invertible defects, generalized charges, boundary and interface sectors, tube algebras, and continuous versions of non-invertible symmetry~\cite{Bhardwaj:2022yxj,Bhardwaj:2024igy,Choi:2024tri,Delmastro:2025ksn,Bonetti:2025dvm}. In the continuous-symmetry setting relevant here, the $BF$ and $BF+kCS$ SymTFTs have been developed through non-Abelian $BF$ constructions, flavor-symmetry SymTFTs, categorical continuous symmetry, defect networks, operator-algebraic formulations, and candidate gaugings~\cite{Bonetti:2024cjk,Jia:2025jmn,Jia:2025vrj,Jia:2025uun,Jia:2026vcr,Jia:2026jmt}. Much of this progress is structural and categorical. The purpose of the present paper is complementary: we work directly with the Hilbert spaces and line-operator kernels of these SymTFTs, making the action of categorical symmetry concrete at the level of convolution operators, centralizer representation theory, and their twisted projective analogues.

For a finite symmetry group $G$ of a 2-dimensional theory, the line operators in the associated 3D SymTFT studied in this paper are the familiar simple objects of the representation category of the untwisted Drinfeld double $D(G)$ or the twisted Drinfeld double $D^{\omega}(G)$. Equivalently, the corresponding categories are the Drinfeld centers $\mathcal{Z}(\text{Vec}_G)$ and $\mathcal{Z}(\text{Vec}^{\omega}_G)$. These categories also admit a groupoid description in terms of the conjugation groupoid $G//_{\text{Ad}}G$ equipped with the corresponding Fell line bundle twist. Their classification by conjugacy classes together with irreducible centralizer representations or irreducible projective centralizer representations, as well as the associated character and modular data, is standard and goes back to the work of Dijkgraaf-Witten, Dijkgraaf-Pasquier-Roche, Freed-Quinn, and subsequent developments~\cite{Dijkgraaf:1989pz,Roche:1990hs,FreedQuinn:1993,Coste:2000tq,Willerton:2008gyk}. More recently, SymTFT formulations have clarified how such codimension-$2$ line operators act on codimension-$1$ Hilbert spaces in a more abstract categorical language~\cite{Kaidi:2022cpf,Jia:2026vcr}.

The emphasis of this paper is operational and quantum mechanical. We formulate the action of line operators on the Hilbert spaces of $BF$ and $BF+kCS$ theory directly in terms of explicit convolution kernels on the relevant centralizer groups, so that their diagonalization by characters of the centralizer group, and in the twisted case by projective characters, becomes transparent. For finite groups, this gives a groupoid-and-kernel approach of the familiar character-theoretic data of the twisted double. For compact Lie groups, it leads to concrete formulas that can be worked out beyond the purely finite setting. The same groupoid formalism also makes explicit that the codimension-$2$ projective transport and the codimension-$1$ projective Hilbert-space action are governed by the same transgressed cocycle $\sigma_k$, and that the mixed $BF+kCS$ theory is naturally described by a magnetic cotangent-bundle deformation of the usual geometric quantization picture for Chern-Simons theory.

The work is organized as follows. In section~\ref{sec:Conj_groupoid_basis} we introduce the conjugation action groupoid, the untwisted Fell line bundle, and the simple basis for the codimension-$2$ defect Hilbert spaces, together with the corresponding arrow action and braiding data. In section~\ref{sec:HilbGG_on_HBF} we explain how these untwisted defect data act on the Hilbert space of the pure $BF$ theory, first through gauge-invariant line operators and then through their stacking or convolution product, and we illustrate the formalism with explicit examples. In section~\ref{sec:Twisted_conj_groupoid_basis} we turn on the level-$k$ Chern-Simons twist and develop the twisted Fell line bundle, the projective simple basis, and the resulting twisted braiding. In section~\ref{sec:Twisted_HilbGG_on_HBF_section} we study the Hilbert space of the $BF+kCS$ theory: we first discuss its quantization on a spatial surface and then describe the action of twisted line operators on the resulting Hilbert space. In section~\ref{sec:Transgression_of_CS_Level} we explain how the same universal level $k\in H^4(BG,\mathbb{Z})$ produces both the codimension-$2$ twisting datum and the codimension-$1$ prequantum line bundle by transgression along $S^1$ and along a closed oriented surface $\Sigma$. In section~\ref{sec:comparison_known_cases} we compare the resulting convolution-eigenvalue formalism with known finite-group modular data and with the regular-sector compact Lie-group Hopf-link kernel. Finally, section~\ref{sec:Conclusion} summarizes the main results and comments on possible extensions.

\section{The Conjugation Groupoid, the Untwisted Fell Line Bundle, and the Simple Basis}\label{sec:Conj_groupoid_basis}

Before discussing the action of line operators on the Hilbert space of the pure $BF$ theory, it is useful to write explicitly the simple objects of the untwisted groupoid algebra $C^*(G//_\Ad G)$ of the conjugation action groupoid $G//_\Ad G$. Throughout this section, we keep the theory untwisted, namely we do \emph{not} turn on the additional level-$k$ Chern-Simons term. In particular, the relevant Fell line bundle is the trivial one. For notational simplicity, we first phrase the explicit basis formulas for finite $G$, so that all vector spaces are finite-dimensional and all convolutions are finite sums. For compact Lie $G$, finite orbit sums are replaced by $L^2$-sections, equivalently by direct integrals over the corresponding conjugacy classes and the precise replacement of the finite sum~\eqref{eq:H_g_rho_directsum} will be stated below. In that setting, the same ket notation is used only as distributional shorthand.

It is important, however, to keep track of what kind of state space is being described here. For a finite group, the relevant symmetry category is $\mathrm{Vec}_G$, while for a compact Lie group we use its analogue $\hilb(G)$, the category of measurable fields of Hilbert spaces over $G$ \cite{Jia:2026vcr}. The vectors introduced in this section, such as $\ket{g_i,\mu}$ below, are basis vectors in a simple object $H_{[g],\rho_g}\in \mathcal{Z}(\mathrm{Vec}_G)$, or in the compact Lie group setting $H_{[g],\rho_g}\in \mathcal{Z}(\hilb(G))$. They should be interpreted as the internal state space attached to the codimension-$2$ topological line defect labeled by $([g],\rho_g)$, not as basis vectors of the physical Hilbert space obtained by quantizing the $BF$ theory on a spatial Riemann surface to be discussed later. Throughout the paper we mostly use the notation $\hilb(G)$ and $\mathcal{Z}(\hilb(G))$, which is adapted to the compact Lie group setting; in the finite group case, these specialize to $\mathrm{Vec}_G$ and $\mathcal{Z}(\mathrm{Vec}_G)$, respectively.

We consider the conjugation action groupoid
\begin{equation}
    \mathcal{G}:= G//_{\mathrm{Ad}}G\,.
\end{equation}
Its objects are the elements $x\in G$, and an arrow from $x$ to $uxu^{-1}$ is denoted by
\begin{equation}
    (u,x):x\longrightarrow uxu^{-1}\,.
\end{equation}
The source, target, and composition maps are
\begin{equation}
    s(u,x)=x\,,\qquad t(u,x)=uxu^{-1}\,,
\end{equation}
\begin{equation}\label{eq:groupoid_comp_conj}
    (v,uxu^{-1})\circ (u,x)=(vu,x)\,.
\end{equation}

Following \cite{Jia:2026vcr}, we use the Fell line bundle $\Sigma_k$ over the conjugation groupoid $\mathcal{G}$ as a geometric incarnation for the cocycle phases appearing in the twisted groupoid convolution algebra. In the untwisted case considered in this section, the Fell line bundle over $\mathcal{G}$ is simply the trivial complex line bundle
\begin{equation}
    \Sigma_0:= (G\times G)\times \mathbb{C}\longrightarrow G\times G\,,
\end{equation}
whose fiber over an arrow $(u,x)$ is a one-dimensional complex vector space
\begin{equation}
    (\Sigma_0)_{(u,x)}\cong \mathbb{C}\,.
\end{equation}
Therefore a compactly supported section of $\Sigma_0$ is simply a complex-valued function on the set of arrows of $\mathcal{G}$. If $\delta_{(u,x)}$ denotes the delta-section supported on the single arrow $(u,x)$, then the convolution product in the untwisted groupoid algebra is
\begin{equation}\label{eq:delta_conv_conj_groupoid}
    \delta_{(v,y)} * \delta_{(u,x)} = \delta_{y,uxu^{-1}}\,\delta_{(vu,x)}\,,
\end{equation}
which is just the algebraic incarnation of the groupoid composition~(\ref{eq:groupoid_comp_conj}). Namely we must require $t(u,x) = s(v,y)$ and the result is supported only on the arrow $(vu, x)$.

A representation of the trivial Fell line bundle $\Sigma_0$ is therefore the same as a collection of vector spaces $\{E_x\}_{x\in G}$, one for each object of the groupoid, together with linear isomorphisms (see~\cite{Kumjian:1998umx})
\begin{equation}\label{eq:T_u_x_untwisted}
    T_u(x):E_x\longrightarrow E_{uxu^{-1}}
\end{equation}
for every arrow $(u,x)$, obeying the untwisted composition rule
\begin{equation}\label{eq:T_u_x_comp_untwisted}
    T_v(uxu^{-1})\,T_u(x)=T_{vu}(x)\,.
\end{equation}
Equivalently, the operator corresponding to the basis element $\delta_{(u,x)}$ of the groupoid algebra is precisely the transport map $T_u(x)$ from the fiber over $x$ to the fiber over $uxu^{-1}$. In the twisted case, these transport maps would compose only projectively, with the discrepancy controlled by the $U(1)$-valued $2$-cocycle of the Fell line bundle~\cite{Willerton:2008gyk}. In the present untwisted $BF$ theory, however, there is no additional phase in~\eqref{eq:T_u_x_comp_untwisted}.

\subsection{The Simple Basis from a Conjugacy Class}\label{sec:simple_basis_conj_groupoid}

We now construct the simple object labeled by a conjugacy class $[g]$ and an irreducible representation $\rho_g$ of the centralizer $C_G(g)$. The conjugacy class of $g$ is the orbit of $g$ under conjugation,
\begin{equation}
    [g]=\{ugu^{-1}:u\in G\}\cong G/C_G(g)\,.
\end{equation}
Choose a set of representatives $\{r_i\}$ for the left cosets in $G/C_G(g)$, and define
\begin{equation}\label{eq:g_i_from_cosets}
    g_i:=r_i g r_i^{-1}\,.
\end{equation}
Then each $g_i$ is an element of the conjugacy class $[g]$, and every element of $[g]$ appears exactly once in this way.

Let $V_{\rho_g}$ be the representation space of $\rho_g$. The simple object $H_{[g],\rho_g}$ is a $G$-equivariant Hilbert-space field over $G$ supported on the conjugacy class $[g]$. Its fiber over the object of $\mathcal{G}$ is
\begin{equation}\label{eq:fiber_H_g_rho}
    (H_{[g],\rho_g})_x=
    \begin{cases}
        V_{\rho_g}\,, & x\in [g]\,,\\
        0\,, & x\notin [g]\,.
    \end{cases}
\end{equation}
After choosing the coset representatives $r_i$, we may identify each nonzero fiber $(H_{[g],\rho_g})_{g_i}$ with a copy of $V_{\rho_g}$ and write
\begin{equation}\label{eq:H_g_rho_directsum}
    H_{[g],\rho_g}\cong \bigoplus_i V_{\rho_g}^{(i)}\,.
\end{equation}
For compact Lie $G$, the above equation is replaced by the induced Hilbert field over the conjugacy class. After choosing a measurable trivialization, such a Hilbert field may be written schematically as the direct integral
\begin{equation}
    H_{[g],\rho_g} \simeq \int_{G/C_G(g)}^{\oplus} V_{\rho_g}\,d\bar r
\end{equation}
where $d\bar r$ is the quotient measure on $G/C_G(g) \cong [g]$. Thus the finite label $i$ in~\eqref{eq:H_g_rho_directsum} is replaced for compact Lie $G$ by a point $rC_G(g)\in G/C_G(g)$. The symbols $|rgr^{-1},\mu\rangle$ should then be understood as distributional Dirac sections localized at $rgr^{-1}\in[g]$, and not as normalizable Hilbert-space vectors. Operator identities involving these kets are to be interpreted after smearing against $L^2$-sections. In the finite group notation, if $\{e_\mu\}$ is a basis of $V_{\rho_g}$, we denote by:
\begin{equation}\label{eq:def_basis_g_i_mu}
    \ket{g_i,\mu}
\end{equation}
the basis vector $e_\mu$ sitting in the $i$-th copy $V_{\rho_g}^{(i)}$, or equivalently in the fiber over the object $g_i$. Thus the first label directly records \emph{which representatives} $g_i$ in the orbit $[g]$ we are sitting over, while the second label $\mu$ records a basis vector in the representation space of the centralizer irrep $\rho_g$.

The reason for introducing the coset representatives is therefore purely bookkeeping: the simple object $H_{[g],\rho_g}$ is not supported only at the single object $g$, but on the entire orbit $[g]$. The representatives $r_i$ provide a concrete way to identify the fiber over each representative $g_i$ with the fixed vector space $V_{\rho_g}$, but the basis ket itself is more naturally labeled by $g_i$.

The same data also determine a representation of the untwisted groupoid algebra on the underlying vector space of the simple object:
\begin{equation}\label{eq:def_pi_g_rhog}
    \pi_{[g],\rho_g}: C^*(G//_{\mathrm{Ad}}G)\longrightarrow \mathrm{End}\big(H_{[g],\rho_g}\big)\,.
\end{equation}
The notation emphasizes that this representation depends on the label $([g],\rho_g)$ of the simple object. Different conjugacy classes $[g]$ have different supports in the object set of the groupoid, and even for fixed $[g]$, different irreducible representations $\rho_g$ of the centralizer $C_G(g)$ give different actions on the internal fiber $V_{\rho_g}$. Thus $([g],\rho_g)$ is precisely the datum that classifies which representation of the groupoid algebra we are talking about.

\subsection{How Groupoid Arrows Act on the Basis}\label{sec:arrow_action_basis}

We now describe explicitly how an arrow of the conjugation groupoid acts on the basis~(\ref{eq:def_basis_g_i_mu}). Since $(u,x)$ has source $x$, it can only act nontrivially on the fiber over $x$. Hence, if $x\neq g_i$ for all $i$, then $(u,x)$ annihilates $H_{[g],\rho_g}$. If $x=g_i$, then the arrow
\begin{equation}
    (u,g_i):g_i\longrightarrow ug_i u^{-1}
\end{equation}
must map the $i$-th fiber of~\eqref{eq:fiber_H_g_rho} to the fiber over the conjugated element $ug_i u^{-1}$.

To make this map explicit, note that $ur_i$ belongs to a unique left coset of $C_G(g)$, so there exist a unique index $j$ and a unique element $c\in C_G(g)$ such that
\begin{equation}\label{eq:uri_equals_rjc}
    ur_i=r_j c\,.
\end{equation}
Because $c$ centralizes $g$, we have
\begin{equation}\label{eq:ugi_to_gj}
    ug_i u^{-1}=ur_i g r_i^{-1}u^{-1}=r_j c g c^{-1} r_j^{-1}=r_j g r_j^{-1}=g_j\,.
\end{equation}
Therefore the arrow $(u,g_i)$ lands in the $j$-th fiber. Its action is precisely the action of $c$ through the representation $\rho_g$:
\begin{equation}\label{eq:arrow_action_on_basis}
    \pi_{[g],\rho_g}\big(\delta_{(u,g_i)}\big)\ket{g_i,\mu}
    =
    \sum_{\nu} \big[\rho_g(c)\big]_{\nu\mu}\ket{g_j,\nu}\,.
\end{equation}
Equivalently, if we want a formula for a general arrow $(u,x)$, then
\begin{equation}\label{eq:general_arrow_action_on_basis}
    \pi_{[g],\rho_g}\big(\delta_{(u,x)}\big)\ket{g_i,\mu}
    =
    \delta_{x,g_i}
    \sum_{\nu}\big[\rho_g(r_j^{-1}ur_i)\big]_{\nu\mu}\ket{g_j,\nu}\,,
\end{equation}
where $j$ is determined by the condition $ug_i u^{-1}=g_j$, so that $r_j^{-1}ur_i\in C_G(g)$.

This formula gives the precise relation between the arrow notation and the basis notation. The first entry of the arrow transports $g_i$ to its conjugate $g_j$, while the residual factor in the centralizer acts on the internal charge label $\mu$ through $\rho_g$.

As a consistency check, one can verify directly that~\eqref{eq:arrow_action_on_basis} reproduces the groupoid multiplication. Suppose that
\begin{equation}
    ur_i=r_j c\,,\qquad vr_j=r_k d\,,
    \qquad c,d\in C_G(g)\,.
\end{equation}
Then
\begin{equation}
    vur_i = vr_j c = r_k dc\,,
\end{equation}
and hence
\begin{equation}
    \begin{split}
        \pi_{[g],\rho_g}\big(\delta_{(v,g_j)}\big)\pi_{[g],\rho_g}\big(\delta_{(u,g_i)}\big)\ket{g_i,\mu}
        &= \sum_{\lambda,\nu}\big[\rho_g(d)\big]_{\nu\lambda}\big[\rho_g(c)\big]_{\lambda\mu}\ket{g_k,\nu}\\
        &= \sum_{\nu}\big[\rho_g(dc)\big]_{\nu\mu}\ket{g_k,\nu}\\
        &= \pi_{[g],\rho_g}\big(\delta_{(vu,g_i)}\big)\ket{g_i,\mu}\,,
    \end{split}
\end{equation}
which is exactly the representation-theoretic version of the convolution law~(\ref{eq:delta_conv_conj_groupoid}). This is the untwisted formula relevant for the pure $BF$ theory. If one later turns on the level-$k$ Chern-Simons twist, the same transport law is modified by the $U(1)$-phase of the corresponding Fell line bundle, and the centralizer representation $\rho_g$ is replaced by the appropriate projective representation.

\subsection{Half-Braiding and Braiding in the Basis}\label{sec:half_braiding_basis}

We now explain how the braiding in $\mathcal{Z}(\hilb(G))$ is written in the notation introduced above. In the untwisted case relevant for the pure $BF$ theory, the category $\mathcal{Z}(\hilb(G))$ can be described in terms of $G$-equivariant Hilbert-space fields $\hilb_G(G)$ over $G$. The braiding is obtained by using the class representative of the first factor to transport the second factor by conjugation. In terms of the equivariant vector bundle description used above, this transport is exactly the arrow action encoded by the maps $T_u(x)$ in~(\ref{eq:T_u_x_untwisted}).

Let $X$ and $Y$ be two objects of $\mathcal{Z}(\hilb(G))$. Their tensor product is the convolution tensor product
\begin{equation}\label{eq:tensor_product_equivariant_bundle}
    (X\otimes Y)_z = \bigoplus_{xy=z} X_x\otimes Y_y\,.
\end{equation}
Thus if $v_x\in X_x$ and $w_y\in Y_y$, then the elementary tensor $v_x\otimes w_y$ lies in the component $(X\otimes Y)_{xy}$.

To distinguish clearly the half-braiding from the full braiding, let us first introduce the one-dimensional degree-$x$ object $\delta_x$ of the \emph{underlying} category $\hilb(G)$, namely
\begin{equation}
    (\delta_x)_{x}=\bbC\,,\qquad (\delta_x)_{z}=0\quad \text{for}\ z\neq x\,.
\end{equation}
Here $\delta_x$ is \emph{not} the delta-section $\delta_{(e,x)}$ of the groupoid algebra supported on the identity arrow at $x$, nor for generic $x$ is it an object of $\mathcal{Z}(\hilb(G))$ itself: an honest object of $\mathcal{Z}(\hilb(G))$ supported on $x$ must in general be supported on the full conjugacy class $[x]$. Rather, because $Y\in \mathcal{Z}(\hilb(G))$ is a centered object, it comes equipped with a half-braiding against every object of $\hilb(G)$, in particular against the degree-$x$ line $\delta_x$. If $1_x$ denotes its basis vector and $w_y\in Y_y$ is a homogeneous vector of degree $y$, then the half-braiding of the centered object $Y$ with the degree-$x$ line $\delta_x\in \hilb(G)$ is
\begin{equation}\label{eq:elementary_half_braiding}
    c^Y_x:\delta_x\otimes Y \longrightarrow Y\otimes \delta_x\,,
    \qquad
    c^Y_x(1_x\otimes w_y):= T^Y_x(y)(w_y)\otimes 1_x\,.
\end{equation}
Here $T^Y_x(y):Y_y\rightarrow Y_{xyx^{-1}}$ is the transport map assigned by $Y$ to the arrow $(x,y):y\to xyx^{-1}$. In other words, this is not a new kind of map: it is precisely the earlier groupoid transport of~(\ref{eq:T_u_x_untwisted}) specialized to the representation $E=Y$ and the arrow label $u=x$, so one may read
\begin{equation}
    T^Y_x(y)=T_u(y)\big|_{u=x,\;E=Y}\,.
\end{equation}
The target must be $Y_{xyx^{-1}}$ because $c^Y_x$ is a morphism in $\hilb(G)$: the source $1_x\otimes w_y$ has total degree $xy$, while an element of $Y_z\otimes \delta_x$ has total degree $zx$, so degree preservation forces $zx=xy$, hence $z=xyx^{-1}$.

The relevant \emph{center coherence} is the tensor-compatibility axiom for the half-braiding of the centered object $Y$:
\begin{equation}\label{eq:center_coherence_vecg}
    c^Y_{X\otimes Z}
    =
    \big(c^Y_X\otimes \mathrm{id}_Z\big)\circ \big(\mathrm{id}_X\otimes c^Y_Z\big),
    \qquad
    c^Y_{\mathbf{1}}=\mathrm{id}_Y\,.
\end{equation}
Specializing to $X=\delta_u$ and $Z=\delta_v$, and evaluating on $1_u\otimes 1_v\otimes w_y$, this gives
\begin{equation}\label{eq:center_coherence_transport}
    T^Y_u(vyv^{-1})\,T^Y_v(y)=T^Y_{uv}(y)\,,
\end{equation}
which is exactly the same composition law as the groupoid-representation rule~(\ref{eq:T_u_x_comp_untwisted}), now read in the center language.

So the logic is the following. The \emph{half}-braiding belongs to the Drinfeld-center structure of an object $Y\in \mathcal{Z}(\hilb(G))$, and is evaluated against arbitrary objects of the underlying category $\hilb(G)$. The \emph{full} braiding, by contrast, is the braiding internal to $\mathcal{Z}(\hilb(G))$ itself, obtained by applying this half-braiding to the underlying $\hilb(G)$-object of the first factor.

The full braiding with a general object $X$ is obtained by decomposing $X$ into its homogeneous sectors
\begin{equation}
    X=\bigoplus_{x\in G} X_x
\end{equation}
and applying the half-braiding of $Y$ with $x$ on each $X_x$-sector. Concretely, for $v_x\in X_x$ one treats $v_x$ as carrying $x$ label, so the braiding is
\begin{equation}\label{eq:braiding_from_half_braiding}
    c_{X,Y}(v_x\otimes w_y):= T^Y_x(y)(w_y)\otimes v_x\,.
\end{equation}
Thus the full braiding in $\mathcal{Z}(\hilb(G))$ is assembled from the half-braidings with the elementary degree-$x$ lines in the underlying category $\hilb(G)$. Only when $x\in Z(G)$, where $Z(G)\in G$ is the center subgroup so that $[x] = \{x\}$, can such a line also be viewed as an honest object of $\mathcal{Z}(\hilb(G))$ supported at the single group element $x$.

The right-hand side of~(\ref{eq:braiding_from_half_braiding}) indeed lies in the correct homogeneous component of $Y\otimes X$, because
\begin{equation}
    T^Y_x(y)(w_y)\otimes v_x\in Y_{xyx^{-1}}\otimes X_x \subset (Y\otimes X)_{(xyx^{-1})x}=(Y\otimes X)_{xy}\,.
\end{equation}
Hence the braiding preserves the total grading while moving the second factor by conjugation with $x$ carried by the first factor.

We now specialize to two simple objects
\begin{equation}
    X=H_{[g],\rho_g}\,,\qquad Y=H_{[h],\rho_h}\,.
\end{equation}
For the first object we keep the notation of the previous subsection
\begin{equation}
    g_i=r_i g r_i^{-1}\,,\qquad \ket{g_i,\mu}\in (H_{[g],\rho_g})_{g_i}\,.
\end{equation}
For the second object we choose representatives $\{q_k\}$ of the left cosets in $G/C_G(h)$ and define
\begin{equation}
    h_k:= q_k h q_k^{-1}\,,\qquad \ket{h_k,\nu}\in (H_{[h],\rho_h})_{h_k}\,.
\end{equation}
Then the tensor product basis vector
\begin{equation}
    \ket{g_i,\mu}\otimes \ket{h_k,\nu}
\end{equation}
lies in the homogeneous component of $H_{[g],\rho_g}\otimes H_{[h],\rho_h}$ labeled by $g_i h_k$.

To compute its braiding, we must evaluate the action of the arrow $(g_i,h_k)$ on the second factor. By the same argument as in~(\ref{eq:uri_equals_rjc})--(\ref{eq:arrow_action_on_basis}), there exist a unique index $\ell$ and a unique element $c_{ik}\in C_G(h)$ such that
\begin{equation}\label{eq:g_i_q_k_equals_q_l_cik}
    g_i q_k = q_\ell c_{ik}\,,\qquad c_{ik}\in C_G(h)\,.
\end{equation}
Equivalently,
\begin{equation}\label{eq:g_i_h_k_to_h_l}
    g_i h_k g_i^{-1}= h_\ell\,.
\end{equation}
Therefore the transport of the second factor by the arrow $(g_i,h_k)$ is
\begin{equation}\label{eq:T_gi_on_H_h}
    T^{H_{[h],\rho_h}}_{g_i}(h_k)\ket{h_k,\nu}
    =
    \sum_{\lambda}\big[\rho_h(c_{ik})\big]_{\lambda\nu}\ket{h_\ell,\lambda}\,.
\end{equation}
Thus the elementary half-braiding of $H_{[h],\rho_h}$ with the object $\delta_{g_i}$ is
\begin{equation}\label{eq:half_braiding_simple_flux}
    c^{H_{[h],\rho_h}}_{g_i}\big(1_{g_i}\otimes \ket{h_k,\nu}\big)
    =
    \sum_{\lambda}\big[\rho_h(c_{ik})\big]_{\lambda\nu}\ket{h_\ell,\lambda}\otimes 1_{g_i}\,.
\end{equation}
The full braiding between the two simples is obtained by tensoring this half-braiding with the internal vector in the $g_i$-sector. More explicitly, under the decomposition
\begin{equation}
    H_{[g],\rho_g}=\bigoplus_i \delta_{g_i}\otimes V_{\rho_g}^{(i)}\,,
\end{equation}
one may regard
\begin{equation}
    \ket{g_i,\mu}=1_{g_i}\otimes e_\mu\,.
\end{equation}
Substituting~(\ref{eq:half_braiding_simple_flux}) into the general formula~(\ref{eq:braiding_from_half_braiding}), we obtain the braiding between the two simples:
\begin{equation}\label{eq:braiding_basis_formula}
    c_{H_{[g],\rho_g},H_{[h],\rho_h}}\big(\ket{g_i,\mu}\otimes \ket{h_k,\nu}\big)
    =
    \sum_{\lambda}\big[\rho_h(c_{ik})\big]_{\lambda\nu}\ket{h_\ell,\lambda}\otimes \ket{g_i,\mu}\,,
\end{equation}
where $\ell$ and $c_{ik}$ are determined by~(\ref{eq:g_i_q_k_equals_q_l_cik}).

This formula should be compared directly with the action of groupoid arrows discussed in the previous subsection. The representative $g_i$ carried by the first factor acts on the second factor exactly as the arrow $(g_i,h_k)$:
\begin{equation}
    (g_i,h_k): h_k \longrightarrow g_i h_k g_i^{-1}=h_\ell\,.
\end{equation}
After this transport is performed, one flips the two tensor factors. Therefore the braiding is not, in general, simply multiplication by $\rho_h(g_i)$: the element $g_i$ does not usually belong to the centralizer $C_G(h)$. Instead one must first rewrite its action on the $h_k$-fiber in the form~(\ref{eq:g_i_q_k_equals_q_l_cik}), and only the residual centralizer element $c_{ik}$ acts through the representation $\rho_h$.

It is useful to isolate the commuting case. Suppose that $g_i$ commutes with $h_k$. Then $g_i h_k g_i^{-1}=h_k$, so $\ell=k$, and~(\ref{eq:g_i_q_k_equals_q_l_cik}) becomes
\begin{equation}
    g_i q_k = q_k c_{ik}\,,\qquad c_{ik}=q_k^{-1}g_i q_k\in C_G(h)\,.
\end{equation}
In this case the braiding reduces to
\begin{equation}\label{eq:braiding_commuting_case}
    c_{H_{[g],\rho_g},H_{[h],\rho_h}}\big(\ket{g_i,\mu}\otimes \ket{h_k,\nu}\big)
    =
    \sum_{\lambda}\big[\rho_h(q_k^{-1}g_i q_k)\big]_{\lambda\nu}\ket{h_k,\lambda}\otimes \ket{g_i,\mu}\,.
\end{equation}
In particular, if we choose the base representative $q_1=e$ so that $h_1=h$, and if $g_i\in C_G(h)$, then
\begin{equation}\label{eq:braiding_base_fiber}
    c_{H_{[g],\rho_g},H_{[h],\rho_h}}\big(\ket{g_i,\mu}\otimes \ket{h,\nu}\big)
    =
    \sum_{\lambda}\big[\rho_h(g_i)\big]_{\lambda\nu}\ket{h,\lambda}\otimes \ket{g_i,\mu}\,.
\end{equation}
This is the precise sense in which, in a commuting sector, the braiding looks like ``the second charge transforms by $\rho_h(g_i)$ and then the two factors are exchanged''.

Since we are still in the pure $BF$ theory, there is no additional phase in~(\ref{eq:braiding_basis_formula}). If one later replaces the trivial Fell line bundle by a nontrivial Fell line bundle $\Sigma_k$, then the transport law~(\ref{eq:T_gi_on_H_h}) is multiplied by the corresponding transgression phase, so~(\ref{eq:braiding_basis_formula}) acquires an extra $\Sigma_k$-dependent factor while keeping the same conjugation pattern of the basis labels.

Throughout the discussion above, the braiding was written on homogeneous basis
vectors $|g_i,\mu\rangle$ and $|h_k,\nu\rangle$ of the two simple objects $H_{[g],\rho_g}$ and $H_{[h],\rho_h}$.
Before closing this section, let us relate the basis-level braiding formula
\eqref{eq:braiding_basis_formula} to the usual fusion-channel language in the finite group case with the Drinfeld center $\mathcal{Z}(\mathrm{Vec}_G)$. After decomposing the tensor product $H_{[g],\rho_g}\otimes H_{[h],\rho_h}$ into
simple objects $H_{[r],\rho_r}$, the same braiding gives the corresponding $R$-matrices $R_{H_{[g],\rho_g},H_{[h],\rho_h}}^{H_{[r],\rho_r}}$ on the
fusion multiplicity spaces, usually depicted as the following diagram
\begin{equation}
    \begin{gathered}
    \begin{tikzpicture}
        \draw[thick] (-0.75,-0.75)--(0,0);
        \draw[thick] (-0.75,-0.75)--(0.75,-2.25);
        \draw[thick] (0.75,-0.75)--(0,0);
        \draw[thick] (0.75,-0.75)--(0.1,-1.4);
        \draw[thick] (-0.1,-1.6)--(-0.75,-2.25);
        \draw[thick] (0,0)--(0,1.25);
        \filldraw[black] (0,0) circle (1.5pt); 
        \node at (-1.5,-2) {$H_{[h],\rho_h}$};
        \node at (1.5,-2) {$H_{[g],\rho_g}$};    
        \node at (0,1.75) {$H_{[r],\rho_r}$};
        \node at (0.5,0.2) {$\alpha$};
    \end{tikzpicture} 
    \end{gathered} \quad = \quad \sum_{\beta} \left(R_{H_{[g],\rho_g},H_{[h],\rho_h}}^{H_{[r],\rho_r}}\right)_{\alpha \beta} 
    \begin{gathered}
    \begin{tikzpicture}
        \draw[thick] (-1.5,-1.5)--(0,0);
        \draw[thick] (1.5,-1.5)--(0,0);
        \draw[thick] (0,0)--(0,1.25);
        \filldraw[black] (0,0) circle (1.5pt); 
        \node at (-1.5,-2) {$H_{[h],\rho_h}$};
        \node at (1.5,-2) {$H_{[g],\rho_g}$};    
        \node at (0,1.75) {$H_{[r],\rho_r}$};
        \node at (0.5,0.2) {$\beta$};
    \end{tikzpicture} 
    \end{gathered}    
\end{equation}

Consider the fusion rule
    \begin{equation}
        H_{[g],\rho_g} \otimes H_{[h],\rho_h} = \bigoplus_{([k],\rho_k)} N_{H_{[g],\rho_g},H_{[h],\rho_h}}^{H_{[r],\rho_r}} H_{[r],\rho_r}\,,
    \end{equation}
where the fusion coefficient is
\begin{equation}
    N_{H_{[g],\rho_g},H_{[h],\rho_h}}^{H_{[r],\rho_r}}
    =
    \dim
    \Hom_{\mathcal{Z}(\mathrm{Vec}_G)}
    \big(
        H_{[r],\rho_r},
        H_{[g],\rho_g}\otimes H_{[h],\rho_h}
    \big)\,,
\end{equation}
which measures the number of independent ways to embed $H_{[r],\rho_r}$ into $H_{[g],\rho_g} \otimes H_{[h],\rho_h}$. Choose the channel embedding maps
\begin{equation}
    i_{\alpha}:H_{[r],\rho_r}
    \longrightarrow
    H_{[g],\rho_g}\otimes H_{[h],\rho_h}\,,
    \qquad
    j_{\beta}:H_{[r],\rho_r}
    \longrightarrow
    H_{[h],\rho_h}\otimes H_{[g],\rho_g}\,,
\end{equation}
where $\alpha,\beta=1,\cdots,N_{H_{[g],\rho_g},H_{[h],\rho_h}}^{H_{[r],\rho_r}}$ label the different copies of the same simple object
$H_{[r],\rho_r}$ in the two tensor products. Since
$c_{H_{[g],\rho_g},H_{[h],\rho_h}}$ is a morphism in $\mathcal{Z}(\mathrm{Vec}_G)$, the composite
\begin{equation}
    c_{H_{[g],\rho_g},H_{[h],\rho_h}}\circ i_{\alpha}
    :
    H_{[r],\rho_r}
    \longrightarrow
    H_{[h],\rho_h}\otimes H_{[g],\rho_g}
\end{equation}
can be expanded using $j_{\beta}$ as
\begin{equation}
    c_{H_{[g],\rho_g},H_{[h],\rho_h}}\circ i_{\alpha}
    =
    \sum_{\beta}
    \left(R_{H_{[g],\rho_g},H_{[h],\rho_h}}^{H_{[k],\rho_k}}\right)_{\alpha\beta} \, j_{\beta}\,,
\end{equation}
and the matrix
$R_{H_{[g],\rho_g},H_{[h],\rho_h}}^{H_{[k],\rho_k}}$
is the braiding matrix in the $H_{[k],\rho_k}$ fusion channel. Using the braiding in basis \eqref{eq:braiding_basis_formula}, we can also write 
\begin{equation}
    \left(R_{H_{[g],\rho_g},H_{[h],\rho_h}}^{H_{[r],\rho_r}}\right)_{\alpha \beta} = p^{\beta} \left( \sum_{\substack{i,j,\mu,\nu}}C^{\alpha}_{i\mu,j\nu}  \sum_{} [\rho_h(c_{ik})]_{\lambda \nu} \, |h_{\ell},\lambda \rangle \otimes |g_i,\mu\rangle \right)\,,
\end{equation}
where $p^{\beta}$ is the projection operator in $H_{[h],\rho_h}\otimes H_{[g],\rho_g}$ satisfying $p^{\beta} \circ j_{\beta'} = \delta^{\beta}_{\beta'}$, and the coefficients $C^{\alpha}_{i\mu,j\nu}$ are the basis coefficients of the embedding $i_{\alpha}$
    \begin{equation}
        i_{\alpha} ( |r_m,\delta \rangle) =  \sum_{\substack{i,j,\mu,\nu\\g_ih_j=r_m}}C^{\alpha}_{i\mu,j\nu,m\delta} \,|g_i,\mu\rangle\otimes |h_j,\nu\rangle\,,
    \end{equation}
and previous formula should be applied to each $|r_m,\delta\rangle$.

For compact Lie groups, the analogous statement requires replacing finite fusion sums by direct integral decompositions, and finite-dimensional multiplicity spaces by the appropriate multiplicity Hilbert spaces. We will not consider this analytic refinement in this paper.

\section{The Actions of Extended Operators on the Hilbert Space of the SymTFT}\label{sec:HilbGG_on_HBF}

We now turn to the genuine Hilbert space of the untwisted $BF$ theory on a spatial torus. This is a different state space from the simple objects discussed in section~\ref{sec:Conj_groupoid_basis}. There the kets $\ket{g_i,\mu}$ live in $H_{[g],\rho_g}\in \mathcal{Z}(\hilb(G))$ and encodes codimension-$2$ line-defect data in the Drinfeld center. Here, by contrast, the kets $\ket{x(a),x(b)}_{BF}$ form a position basis for the physical Hilbert space assigned by quantization to the codimension-$1$ spatial manifold which we choose to be $T^2$. The relation between the two structures is that an object of $\mathcal{Z}(\hilb(G))$ labels a topological line operator, and this line operator acts on the physical Hilbert space $H_{BF}(T^2)$.

For simplicity, we will fix the 3D spacetime to be a solid torus $D^2\times S^1$ with the ``time-direction'' being the radial direction of $D^2$. In this case, the spatial manifold to which one assigns the Hilbert space of the theory is topologically a torus. We will denote by $a$ and $b$ the $a$- and $b$-cycles of the spatial manifold $T^2$. The configuration space of the theory is given by a flat connection $x$ on $T^2$ specified by the holonomies $x(a),x(b)\in G$ satisfying $x(a)x(b) = x(b)x(a)$, modulo gauge transformations that act on $x(a)$ and $x(b)$ as simultaneous conjugations. I.e., the configuration space of the theory is:
\begin{equation}\label{eq:Config_Space_ST2}
    \widetilde{\mathcal{A}}_0 = \{ (x(a), x(b)) \in G\times G | x(a)x(b) = x(b)x(a) \}/\sim
\end{equation}
with $(x(a), x(b))\sim (gx(a)g^{-1}, gx(b)g^{-1})$, $\forall g\in G$. We will denote by $\mathcal{A}_0$ the configuration space before modding out gauge transformations and by $\bra{x(a),x(b)}_{BF}$ (and its dual $\ket{x(a),x(b)}_{BF}$) with $(x(a),x(b))\in\mathcal{A}_0$ a representative of a gauge-equivalent class of configuration of the theory. The Hilbert space of the theory is the space of functions on $\widetilde{\mathcal{A}}_0$ and is represented as $\psi(x) := \braket{x(a),x(b)|\psi}_{BF} \in H_{BF}$. When no confusion can arise, we will suppress the subscript $BF$ on these physical basis states. This function must satisfy $\psi(x) = \psi(gxg^{-1})$ in order to respect the structure of $\widetilde{\mathcal{A}}_0$ in~(\ref{eq:Config_Space_ST2}).

To match the notation of section~\ref{sec:Conj_groupoid_basis}, we continue to label untwisted line operators by the same pair $([g],\rho_g)$, where $g$ is a chosen representative of the conjugacy class and $\rho_g$ is an irreducible representation of $C_G(g)$. In section~\ref{sec:Gauge_Inv_LineOp}, we will define the representation of gauge invariant line operators, which live in $\mathcal{Z}(\hilb(G))$, on $H_{BF}$ by looking at their action on $\widetilde{\mathcal{A}}_0$. In section~\ref{sec:Stacking_LineOp}, we will show that the representation we define indeed respects the structure of $\mathcal{Z}(\hilb(G))$. We show in particular that the representation of gauge invariant line operators on $H_{BF}$ is simply a convolution in the space of functions on $\widetilde{\mathcal{A}}_0$. In section~\ref{sec:Examples} we will provide concrete examples.

\subsection{The Gauge Invariant Line Operator}\label{sec:Gauge_Inv_LineOp}

We start with the line operator $v_{g,\rho_g}$ whose action on a representative $\bra{x}$ of a physical state is as follows
\begin{equation}
	\bra{x(a),x(b)} \rho(v_{g,\rho_g}) = \mathbf{1}_{C_G(x(b))}(g) \ \chi_{\rho_g}(x(b)) \bra{gx(a),x(b)}\,,
\end{equation}
where $\mathbf{1}_{H}(x)$ is a $\mathbf{1}$-function supported on $H\subset G$. Here we always assume that the line operator $v_{g,\rho_g}$ is placed along the $b$-cycle, and we have implicitly assumed that $\rho$ acts as a representation mapping the line operators into an endomorphism of the Hilbert space. To understand this action, notice that the line operator $v_{g,\rho_g}$ is written as
    \begin{equation}
        v_{g,\rho_g} = \textrm{tr}_{\rho_g} \mathcal{P}\exp\left(i \oint_{b} A \right) \exp\left( i\oint_{b} (\alpha, B) \right)\,,
    \end{equation}
in the $BF$ theory, where $g=e^{i\alpha}$ and $\textrm{tr}_{\rho_g} \mathcal{P}\exp\left(i \oint_{b} A \right)$ is the Wilson loop of representation $\rho_g$ under the $C_G(g)$ subgroup. The insertion of $\exp\left( i\oint_{b} (\alpha, B) \right)$ shifts the holonomy $x(a)$ along the $a$-cycle to $g x(a)$. Since $gx(a)$ should commute with $x(b)$, which implies the action gives zero unless $g \in C_G(x(b))$. On the other hand, the Wilson loop $\textrm{tr}_{\rho_g} \mathcal{P}\exp\left(i \oint_{b} A \right)$ depends on the holonomy along $b$-cycle, and is evaluated to be the character $\chi_{\rho_g}(x(b))$.

Unfortunately, $\rho(v_{g,\rho_g})$ is not a genuine physical operation on $\bra{x(a),x(b)}$, since in general
\begin{equation}
    \bra{x(a),x(b)}\rho(v_{g,\rho_g}) \neq \bra{kx(a)k^{-1},kx(b)k^{-1}}\rho(v_{g,\rho_g})\,,
\end{equation}
due to $\mathbf{1}_{C_G(kx(b)k^{-1})}(g) = \mathbf{1}_{C_G(x(b))}(k^{-1}gk)$ being not necessarily equal to $\mathbf{1}_{C_G(x(b))}(g)$ for arbitrary $k\in G$. In other words, $v_{g,\rho_g}$ acts differently on different representatives of the same physical state. To resolve this issue, we define:
\begin{equation}\label{eq_line_operator}
	v_{[g],\rho_g} := \int_{[g]} d\mu(g)\ v_{g,\rho_{g}}\,, \quad \textrm{with} \quad\mu(kgk^{-1})=\mu(g),\quad \forall k \in G\,,
\end{equation}
where $\mu(g)$ is a measure on the conjugacy class $[g]$, which is chosen to be invariant under the conjugation of $k\in G$.

One way to construct a conjugate-invariant measure $\mu(g)$ is via the measure of the coset $[G/C_G(g)]$\cite{tornier2020haar}. Consider the coset decomposition $k=r h$ with $k\in G, r\in[G/C_G(g)],h\in C_G(g)$, and decompose the measure $dk=d\lambda dh$. Begin with the operator
\begin{equation}
 	v_{[g],\rho_g} := \int_G dk\ v_{kgk^{-1},\rho_{kgk^{-1}}} = \int_{[G/C_G(g)]} dr \int_{C_G(g)} dh \, v_{rgr^{-1},\rho_{rgr^{-1}}}\,,
 \end{equation}
where $\rho_{kgk^{-1}}(x) := \rho_g(k^{-1}xk)$. By construction, $v_{[g],\rho_g}$ depends only on the conjugacy class of $g$. We assume $dk$ and $dh$ are left-invariant measure on $G$ and $C_G(g)$, respectively, and we normalize $\int_{C_G(g)}dh=1$. The integral is independent on the choice of the coset representative $r$, since if we shift $r\rightarrow r h', h'\in C_G(g)$, we have $dk=d(r h'h) = d r d(h'h) = dr dh$. Moreover, $dr$ is also left-invariant under the action of $G$ as $dk$ is left-invariant. Since the elements in $[G/C_G(g)]$ have a one-to-one correspondence to the elements in the conjugacy class $[g]$ via the map $r \rightarrow r g r^{-1}$, therefore $d r$ also defines a measure $d \mu(g)$ on $[g]$ which is invariant under conjugation. We will write down the measure explicitly for the conjugacy class of $SU(2)$ in the example section.

The action of $v_{[g],\rho_g}$ on a representative $\bra{x(a),x(b)}$ of a physical state is:
\begin{equation}\label{eq:def_rho_[g]}
	\bra{x(a),x(b)} \rho(v_{[g],\rho_g}) = \int_{[g]\cap C_G(x(b))} d\mu(g) \ \chi_{\rho_{g}}(x(b)) \bra{gx(a),x(b)}\,,
\end{equation}
where $d\mu(g)$ here is understood as the induced measure on the submanifold $[g]\cap C_G(x(b))$ from $[g]$. For finite $G$, the integral in (3.7) is simply the finite sum over $[g]\cap C_G(x(b))$. For compact Lie $G$, the same notation denotes the distributional localization of the orbital integral over $[g]$ to the commuting locus with $x(b)$, with appropriate normalization~\footnote{More explicitly, for compact Lie $G$ one may write this localization schematically as $\int_{[g]\cap C_G(v)} d\mu_{[g],v}(\gamma)\,F(\gamma) := \int_{[g]}d\mu_{[g]}(\gamma)\, \delta_G(\gamma v\gamma^{-1}v^{-1})\,F(\gamma)$, where the delta function is understood as a distributional constraint imposing $\gamma\in C_G(v)$. In regular transverse sectors this reduces to a Weyl-orbit sum after fixing Haar normalizations. For example, for regular $SU(2)$ holonomies this is the convention under which $[g]\cap C_G(v)=\{g,g^{-1}\}$ contributes the two terms appearing in later in~\eqref{eq:su(2)_kernel_center}. Singular or excess-intersection sectors require the corresponding distributional measure and are not analyzed in full here.} One can check that:
\begin{equation}
	\begin{split}
		&\bra{hx(a)h^{-1},hx(b)h^{-1}} \rho(v_{[g],\rho_g}) \\
		=& \int_{[g]\cap C_G(hx(b)h^{-1})} d\mu(g) \ \chi_{\rho_{g}}(hx(b)h^{-1}) \bra{ghx(a)h^{-1},hx(b)h^{-1}}\\
		=& \int_{[hgh^{-1}]\cap C_G(hx(b)h^{-1})} d\mu(hgh^{-1}) \ \chi_{\rho_{hgh^{-1}}}(hx(b)h^{-1}) \bra{hgx(a)h^{-1},hx(b)h^{-1}} \\
		=& \int_{[g]\cap C_G(x(b))} d\mu(g) \ \chi_{\rho_{g}}(x(b)) \bra{gx(a),x(b)} = \bra{x(a),x(b)} \rho(v_{[g],\rho_g})\,,
	\end{split}
\end{equation}
where in the third line we shifted $g \rightarrow h g h^{-1}$ and the fourth line we used the fact that $d\mu(g)$ is invariant under conjugation. Thus $v_{[g],\rho_g}$ is a gauge invariant line operator. We will denote by $\mathcal{L}$ the space of such line operators formed by tensoring $v_{[g],\rho_g}$'s with different conjugacy classes. We are yet to show that $\rho$ respects the tensor product in $\mathcal{L}$ for it to be a genuine representation of $\mathcal{L}$ in $\text{End}(H_{BF})$.

\subsection{Stacking the Line Operators}\label{sec:Stacking_LineOp}

We now consider stacking two line operators on the $b$-cycle to form $v_{[g],\rho_g}\otimes v_{[h],\rho_h} \in \mathcal{L}$ acting on a representative $\bra{x}$ from the right. Since $Z(\mathrm{Hilb}(G))$ is braided, the two stacked line operators are canonically related by the braiding $c_{v_{[g],\rho_g},v_{[h],\rho_h}}:  v_{[g],\rho_g}\otimes v_{[h],\rho_h}\xrightarrow{\sim} v_{[h],\rho_h}\otimes v_{[g],\rho_g}$. Thus, in the Hilbert-space representation considered below, braided commutativity should appear as the operator relation:
\begin{equation}\label{eq:Commutativity}
	\rho(v_{[g],\rho_g}\otimes v_{[h],\rho_h}) \cong \rho(v_{[h],\rho_h}\otimes v_{[g],\rho_g})\,.
\end{equation}
Namely, after applying the Hilbert-space action and identifying the two stackings through this braiding, the corresponding endomorphisms of
$H_{\mathrm{BF}}$ should match.

If $\rho:\mathcal{L}\rightarrow\text{End}(H_{BF})$ were a genuine representation of $\mathcal{L}$, it must preserve the tensor product of $\mathcal{L}$, i.e. the following must hold
\begin{equation}\label{eq:True_Rep}
	\rho(v_{[g],\rho_g}\otimes v_{[h],\rho_h}) = \rho(v_{[g],\rho_g})\ \rho(v_{[h],\rho_h})\,,
\end{equation}
where the RHS is set to be a multiplication from the right. As a sanity check, we must require that~(\ref{eq:True_Rep}) respects~(\ref{eq:Commutativity}). To see this, by~(\ref{eq:True_Rep}) we have
\begin{equation}\label{eq:Check_Comm}
	\begin{split}
		\bra{x} \rho(v_{[g],\rho_g}\otimes v_{[h],\rho_h}) &= \bra{x} \rho(v_{[g],\rho_g})\ \rho(v_{[h],\rho_h}) \\
		&= \int_{[g]\cap C_G(x(b))} d\mu(g)\ \chi_{\rho_{g}}(x(b)) \bra{gx(a),x(b)} \rho(v_{[h],\rho_h}) \\
		&= \int_{[g]\cap C_G(x(b))} d\mu(g)\int_{[h]\cap C_G(x(b))} d\mu(h) \\ &\qquad \chi_{\rho_{g}}(x(b)) \chi_{\rho_{h}}(x(b))\bra{hgx(a),x(b)}\\
        &= \int_{[g]\cap C_G(x(b))} d\mu(g)\int_{[h]\cap C_G(x(b))} d\mu(ghg^{-1}) \\ &\qquad \chi_{\rho_{g}}(x(b)) \chi_{\rho_{ghg^{-1}}}(x(b))\bra{ghx(a),x(b)}\\
        &= \bra{x} \rho(v_{[h],\rho_h})\ \rho(v_{[g],\rho_g}) = \bra{x} \rho(v_{[h],\rho_h}\otimes v_{[g],\rho_g})
	\end{split}
\end{equation}
where in the fourth line we shift $h$ to $ghg^{-1}$ and in the fifth line we used the fact that $g$ is restricted to be in $C_G(x(b))$ by the first integral $\int_{[g]\cap C_G(x(b))} d\mu(g)$, so that $\chi_{\rho_{ghg^{-1}}}(x(b)) = \chi_{\rho_h}(g^{-1}x(b)g)=\chi_{\rho_h}(x(b))$. Therefore, as long as $\rho:\mathcal{L}\rightarrow\text{End}(H_{BF})$ defined in~(\ref{eq:def_rho_[g]}) is a genuine representation of $\mathcal{L}$ that preserves the tensor product in $\mathcal{L}$, the requirement of commutativity~(\ref{eq:Commutativity}) is trivially satisfied.

Unfortunately, one cannot simply require~(\ref{eq:True_Rep}) be true, since there are extra consistency conditions to be checked. Schematically, for $v_{[g],\rho_g}, v_{[h],\rho_h}\in\mathcal{L}$, we have
\begin{equation}\label{eq:v1_tensor_v2}
	v_{[g],\rho_g}\otimes v_{[h],\rho_h} = \bigoplus_{[k],\rho_k} N_{([g],\rho_g), ([h],\rho_h)}^{[k],\rho_k}\ v_{[k],\rho_k}\,,
\end{equation}
where $[k]$ and $\rho_k$ depend implicitly on $([g],\rho_g)$ and $([h],\rho_h)$. It is natural to expect that $[k]$ runs over all conjugacy classes that appear in $[h]\cdot [g]$, the set of which coincides with those obtained from $[g]\cdot [h]$. Hence~(\ref{eq:v1_tensor_v2}) is automatically invariant under $[g]\leftrightarrow [h]$. Given~(\ref{eq:v1_tensor_v2}), for $\rho$ to be a genuine representation of $\mathcal{L}$, we must require
\begin{equation}\label{eq:Require_1}
		\rho(v_{[g],\rho_g})\ \rho(v_{[h],\rho_h}) = \rho(v_{[g],\rho_g}\otimes v_{[h],\rho_h}) = \sum_{[k]\in [h]\cdot[g],\rho_k} N_{([g],\rho_g), ([h],\rho_h)}^{[k],\rho_k}\ \rho(v_{[k],\rho_k})\,.
\end{equation}

One must also ask what centralizer representations $\rho_k$ could have appeared in the direct sum. A natural construction would be to first construct the induced $G$-representations $\text{Ind}_{C_G(g)}^G \rho_g$ and $\text{Ind}_{C_G(h)}^G \rho_h$, and consider the tensor product decomposition of $\text{Ind}_{C_G(g)}^G \rho_g\otimes \text{Ind}_{C_G(h)}^G \rho_h$, for which we have:
\begin{equation}\label{eq:tensor_decomp_G}
	\text{Ind}_{C_G(g)}^G \rho_g\otimes \text{Ind}_{C_G(h)}^G \rho_h = \bigoplus_{[k]\in [h]\cdot [g]} \bigoplus_{\rho_k} {N'}^{([k],\rho_k)}_{([g],\rho_g),([h],\rho_h)} \text{Ind}_{C_G(k)}^G \rho_k\,,
\end{equation}
where the sum runs over all conjugacy classes obtained from $[h]\cdot[g]$, and all representations of each class $[k]$ that come from the tensor product decomposition of $\text{Ind}_{C_G(g)}^G \rho_{g}\otimes \text{Ind}_{C_G(h)}^G \rho_{h}$ with representative $k = hg$. It is natural to expect that $N'$'s in~(\ref{eq:tensor_decomp_G}) are proportional to the corresponding $N$'s in~(\ref{eq:Require_1}), which, if verified, will ensure that $\rho:\mathcal{L}\rightarrow \text{End}(H_{BF})$ defined in~(\ref{eq:def_rho_[g]}) respects the tensor product in $\mathcal{L}$.

Let us revisit the action~(\ref{eq:def_rho_[g]}). For a fixed $x(b)$, we define the wave function:
\begin{equation}
	\psi_{v}(u) = \braket{u,v|\psi}\,,
\end{equation}
for $v := x(b)$, $u := x(a) \in C_G(v)$. One can think of $a$-cycle as the temporal direction so that $v$ label the twisted Hilbert space. Moreover, since
\begin{equation}
	\psi_v(kuk^{-1}) = \braket{kuk^{-1},v|\psi} = \braket{kuk^{-1},kvk^{-1}|\psi} = \braket{u,v|\psi} = \psi_v(u),\ \forall k\in C_G(v)\,,
\end{equation}
$\psi_v$ is a class function of $C_G(v)$, i.e. $\psi_v\in Cl(C_G(v))$. One can now write the action of $\rho(v_{[g],\rho_g})$ on $Cl(C_G(v))$ as
\begin{equation}\label{eq:rho_as_conv}
	\begin{split}
		(\rho(v_{[g],\rho_g})\psi_v)(u) &:= \bra{u,v} \rho(v_{[g],\rho_g}) \ket{\psi} \\
		&= \int_{[g]\cap C_G(v)}\ d\mu(g) \ \chi_{\rho_{g}}(v)\ \psi_v(gu) \\
		&= \int_{[g]\cap C_G(v)}\ d\mu(g) \ \chi_{\rho_{g}}(v)\ \left(\int_{C_G(v)} dh\ \delta^{C_G(v)}_{h,g} \right) \ \psi_v(gu) \\
		&= \int_{C_G(v)} dh \left(\int_{[g]\cap C_G(v)}\ d\mu(g) \ \chi_{\rho_{g}}(v)\  \delta^{C_G(v)}_{h,g} \right) \ \psi_v(gu) \\
		&:= \int_{C_G(v)} dh\ \mathcal{K}^{[g],\rho_g}_v(h^{-1}) \  \psi_v(hu)\,,
	\end{split}
\end{equation}
with kernel $\mathcal{K}_v$ is defined as
\begin{equation}\label{eq:Define_Kernel}
	\mathcal{K}^{[g],\rho_g}_v(h) = \int_{[g]\cap C_G(v)}\ d\mu(g) \ \chi_{\rho_{g}}(v)\  \delta^{C_G(v)}_{h^{-1},g}\,,
\end{equation}
where $\int_{C_G(v)} dx\ \delta^{C_G(v)}_{x,h}f(x) = f(h)\,, \forall h\in C_G(v)$. For any $\ell\in C_G(v)$ we have
\begin{equation}
	\begin{split}	\mathcal{K}^{[g],\rho_g}_v(\ell h \ell^{-1}) &= \int_{[g]\cap C_G(v)}\ d\mu(g) \ \chi_{\rho_{g}}(v)\  \delta^{C_G(v)}_{\ell h^{-1} \ell^{-1},g} \\
    &= \int_{[g]\cap C_G(v)}\ d\mu(\ell g \ell^{-1}) \ \chi_{\rho_{\ell g \ell^{-1}}}(v)\  \delta^{C_G(v)}_{\ell h^{-1} \ell^{-1},\ell g \ell^{-1}} 
    \\
	&= \mathcal{K}^{[g],\rho_g}_v(h)\,,
	\end{split}
\end{equation}
where in the second line we change $g \rightarrow \ell g \ell^{-1}$, and in the last step we use $\delta^{C_G(v)}_{\ell h^{-1} \ell^{-1},\ell g \ell^{-1}} = \delta^{C_G(v)}_{h^{-1},g}$ and $\chi_{\rho_{\ell g \ell^{-1}}}(v) = \chi _{\rho_g}(\ell^{-1}v \ell) = \chi_{\rho_g}(v)$. Hence $\mathcal{K}^{[g],\rho_g}_v(h)$ is a class function on $C_G(v)$ and~(\ref{eq:rho_as_conv}) is a convolution by $\mathcal{K}^{[g],\rho_g}_v$ in the $v$-sector of $H_{BF}\cong L^2(\widetilde{\mathcal{A}}_0)$, i.e. when $x(b) = v$. Thus, the action of $v_{[g],\rho_g}$ on $\psi_v \in Cl(C_G(v))$ can be written as
\begin{equation}\label{eq:rho_as_convolution}
	(\rho(v_{[g],\rho_g})\psi_v)(u) = (\mathcal{K}_v^{[g],\rho_g} * \psi_v) (u)\,,
\end{equation}
which is a convolution on $Cl(C_G(v))$.

We now look for eigenfunctions and eigenvalues of~(\ref{eq:rho_as_convolution}). Since $\mathcal{K}_v^{[g],\rho_g}$ is itself a class function of $C_G(v)$, it can be expanded as
\begin{equation}
	\mathcal{K}_v^{[g],\rho_g}(x) = \sum_{R_v} K^{[g],\rho_g}_{R_v} \chi_{R_v}(x) \,,
\end{equation}
where $\chi_{R_v}(x)$ is the character of $R_v \in \text{Irrep}(C_G(v))$, the set of the isomorphism classes of the irreducible representations of $C_G(v)$. Thus the convolution~(\ref{eq:rho_as_convolution}) becomes
\begin{equation}\label{eq:convolution_expansion}
	(\mathcal{K}_v^{[g],\rho_g} * \psi_v) (u) = \sum_{R_v} K^{[g],\rho_g}_{R_v} (\chi_{R_v} * \psi_v)(u)\,.
\end{equation}
Recall that we have\footnote{It follows from the first orthogonal relation of characters
\begin{equation}
    \frac{1}{|G|}\int d g \,\chi_{\rho}(g)\chi_{\nu}(g^{-1}h)  = \delta_{\rho,\nu} \frac{\chi_{\rho}(h)}{\chi_{\rho}(e)}\,,
\end{equation} where $\rho,\nu$ labels the irreducible representation of $G$.}
\begin{equation}
	(\chi_{R_v} * \chi_{R'_v})(u) = \delta_{R_v,R'_v} \frac{\text{Vol}(C_G(v))}{\dim R_v} \chi_{R_v}(u)\,.
\end{equation}
Therefore, when $\psi_v = \chi_{R_v}$ in~(\ref{eq:convolution_expansion}), we have
\begin{equation}\label{eq:Convolution_with_ChiR}
	(\mathcal{K}_v^{[g],\rho_g} * \chi_{R_v}) (u) = \frac{\text{Vol}(C_G(v)) K^{[g],\rho_g}_{R_v}}{\dim R_v} \chi_{R_v}(u) := \lambda^{[g],\rho_g}_{R_v} \chi_{R_v}(u)\,.
\end{equation}
Then the eigenfunctions of~(\ref{eq:rho_as_convolution}) are the characters $\chi_{R_v}$ of $R_v \in \text{Irrep}(C_G(v))$, and the corresponding eigenvalues are $\lambda^{[g],\rho_g}_{R_v} := \text{Vol}(C_G(v)) K^{[g],\rho_g}_{R_v}/\dim R_v$.

Actually, for any class function $\mathcal{K}$ of any group $G$, the convolution with the character is
\begin{equation}
	\begin{split}
		(\mathcal{K} * \chi_R)(x) &= \int_G dh\ \mathcal{K}(h^{-1}) \tr(\rho_R(h)\rho_R(x)) \\
		&= \tr \left( \int_G dh\ \mathcal{K}(h^{-1}) \rho_R(h) \right) \rho_R(x) := \tr \mathcal{A}_{\mathcal{K}} \rho_R(x)\,.
	\end{split} 
\end{equation}
For the operator $\mathcal{A}_{\mathcal{K}}$ in the $R$ representation of $G$, we have
\begin{equation}
	\begin{split}
		\rho_R(x)^{-1} \mathcal{A}_{\mathcal{K}} \rho_R(x) &= \int_G dh\ \mathcal{K}(h^{-1}) \rho_R(x^{-1}hx) \stackrel{h\rightarrow xhx^{-1}}{=} \int_G dh\ \mathcal{K}(h^{-1}) \rho_R(h) = \mathcal{A}_{\mathcal{K}}\,,
	\end{split}
\end{equation}
for any $x\in G$. Therefore, $\mathcal{A}_{\mathcal{K}}$ commutes with all $\rho_R(x)$ hence by Schur's lemma it is proportional to the identity, i.e. $\mathcal{A}_{\mathcal{K}} = \lambda_{\mathcal{K}}\mathbf{1}_{d_R}$. Taking trace of both sides, we have
\begin{equation}\label{eq:general_conv_with_chi}
	\lambda_{\mathcal{K}} = \frac{1}{d_R} \tr \int_G dh\ \mathcal{K}(h^{-1}) \rho_R(h) = \frac{1}{d_R} \int_G dh\ \mathcal{K}(h^{-1}) \chi_R(h)\,.
\end{equation}
The above general form of convolution with the character matches our result~(\ref{eq:Convolution_with_ChiR}).

It is now obvious that
\begin{equation}
	(\rho(v_{[h],\rho_h})\rho(v_{[g],\rho_g})\psi_v)(u) = (\rho(v_{[g],\rho_g})\rho(v_{[h],\rho_h})\psi_v)(u)\,,
\end{equation}
because
\begin{equation}
	\begin{split}
		(\mathcal{K}_v^{[h],\rho_h} * (\mathcal{K}_v^{[g],\rho_g} * \psi_v)) (u) &= ((\mathcal{K}_v^{[h],\rho_h} * (\mathcal{K}_v^{[g],\rho_g}) * \psi_v) (u) \\
		&= ((\mathcal{K}_v^{[g],\rho_g}*\mathcal{K}_v^{[h],\rho_h}) * \psi_v) (u) = (\mathcal{K}_v^{[g],\rho_g} * (\mathcal{K}_v^{[h],\rho_h} * \psi_v)) (u)\,,
	\end{split}
\end{equation}
using the associativity of the convolution and the commutativity of convolution by class functions. Therefore we have proved that $\rho:\mathcal{L}\rightarrow \text{End}(H_{BF})$ preserves the commutativity of $\mathcal{L}$.

The tensor product \eqref{eq:v1_tensor_v2} can be rewritten as
    \begin{equation}\label{eq:v1_tensor_v2_kernel}
        \mathcal{K}_v^{[g],\rho_g} * \mathcal{K}_v^{[h],\rho_h} = \sum_{[k]\in [h]\cdot[g],\rho_k} N_{([g],\rho_g), ([h],\rho_h)}^{([k],\rho_k)} \mathcal{K}_v^{[k],\rho_k}\,,
    \end{equation}
which should hold for all $v$. Acting both sides on an eigenfunction $\chi_{R_v}$, the LHS is
\begin{equation}
	(\mathcal{K}_v^{[g],\rho_g} * \mathcal{K}_v^{[h],\rho_h} * \chi_{R_v}) (u) = \lambda^{[g],\rho_g}_{R_v} \lambda^{[h],\rho_h}_{R_v} \chi_{R_v}(u)\,,
\end{equation} 
while the RHS leads to
\begin{equation}
		\sum_{[k]\in [h]\cdot[g],\rho_k} N_{([g],\rho_g), ([h],\rho_h)}^{([k],\rho_k)}\ (\mathcal{K}_v^{[g],\rho_g} * \chi_{R_v})(u) = \sum_{[k]\in [h]\cdot[g],\rho_k} N_{([g],\rho_g), ([h],\rho_h)}^{[k],\rho_k}\ \lambda^{[k],\rho_k}_{R_v} \chi_{R_v}(u)\,.
\end{equation}
Therefore, proving the preservation of tensor product \eqref{eq:v1_tensor_v2} or \eqref{eq:v1_tensor_v2_kernel} amounts to proving\begin{equation}\label{eq:To_Prove}
		\lambda^{[g],\rho_g}_{R_v} \lambda^{[h],\rho_h}_{R_v} = \sum_{[k]\in [h]\cdot[g],\rho_k} N_{([g],\rho_g), ([h],\rho_h)}^{[k],\rho_k}\ \lambda^{[k],\rho_k}_{R_v}
\end{equation}
for any $R_v \in \text{Irrep}(C_G(v))$.

\subsection{Untwisted Examples}\label{sec:Examples}

We now give concrete untwisted examples of the kernels defined in~(\ref{eq:Define_Kernel}).

\paragraph{Discrete Group}
For a discrete group $G$, the kernel in \eqref{eq:Define_Kernel} is 
    \begin{equation}
        \mathcal{K}_{v}^{([g],\rho_g)} (h) = \sum_{g'\in [g]\cap C_G(v)}\delta^{C_G(v)}_{h^{-1},g'} \chi_{\rho_{g'}}(v) \,,
    \end{equation}
and the convolution between kernels is
    \begin{equation}
        \mathcal{K}_{v}^{([g],\rho_g)} *\mathcal{K}_{v}^{([h],\rho_h)} (k) = \sum_{k'\in C_G(v)} \mathcal{K}_{v}^{([g],\rho_g)} (k') \mathcal{K}_{v}^{([h],\rho_h)} (k k'{}^{-1}).
    \end{equation}
Let us consider the order-6 symmetric group $S_3=\{e,a,a^2,b,ab,a^2b\}$ as an example. Here $a$ and $b$ are the generators of $\mathbb{Z}_3$ and $\mathbb{Z}_2$ satisfying $a^3=b^2=1$ and the constraint $ba = a^2 b$. There are three conjugacy classes represented by $[e],[a],[b]$
    \begin{equation}
        [e] = \{ e \},\quad [a] = \{ a , a^2 \},\quad [b] = \{ b , ab, a^2 b\}
    \end{equation}
and the centralizer group for $e,a,b$ are respectively:
    \begin{equation}
        C_{S_3}(e) = S_3,\quad C_{S_3}(a) = \mathbb{Z}_3, \quad C_{S_3}(b) = \mathbb{Z}_2\,,
    \end{equation}
where $\mathbb{Z}_3$ and $\mathbb{Z}_2$ are also generated by $a$ and $b$. There are in total 8 line operators labeled by
    \begin{equation}
        ([e],\rho_+),([e],\rho_-),([e],\rho_E),([b],\rho_0),([b],\rho_1),([a],\rho_0),([a],\rho_1),([a],\rho_2)\,.
    \end{equation}
Setting $v=e,b,a$, respectively, we have the following tables of the kernel.
\begin{table}[!h]
\centering
\begin{tabularx}{\textwidth}{|c|c|c|c|c|c|c|c|c|c|}
\cline{1-10}
$v$&$h$&$([e],\rho_+)$ & $([e],\rho_-)$ & $([e],\rho_E)$ & $([b],\rho_0)$ & $([b],\rho_1)$ & $([a],\rho_0)$ & $([a],\rho_1)$ & $([a],\rho_2)$\\
\cline{1-10}
\multirow{3}{*}{$e$}&$e$&1&1&2&0&0&0&0&0\\
\cline{2-10}&$b$&0&0&0&1&1&0&0&0\\
\cline{2-10}&$a$&0&0&0&0&0&1&1&1\\
\cline{1-10}
\multirow{2}{*}{$b$}&$e$&1&-1&0&0&0&0&0&0\\
\cline{2-10}&$b$&0&0&0&1&-1&0&0&0\\
\cline{1-10}
\multirow{3}{*}{$a$}&$e$&1&1&-1&0&0&0&0&0\\
\cline{2-10}&$a$&0&0&0&0&0&1&$\omega^2$&$\omega$\\
\cline{2-10}&$a^2$&0&0&0&0&0&1&$\omega$&$\omega^2$\\
\cline{1-10}
\end{tabularx}
\caption{The kernel $\mathcal{K}^{([g],\rho)}_{v}(h)$ for $S_3$ group.}
\end{table}

One can check the kernels satisfy the fusion rule \eqref{eq:v1_tensor_v2_kernel}. In fact, using the explicit values of $\mathcal{K}^{([g],\rho)}_{v}(h)$ given in the table, one can solve the fusion coefficient $N_{([g],\rho_g), ([h],\rho_h)}^{([k],\rho_k)}$ from the relation \eqref{eq:v1_tensor_v2_kernel}, which matches the fusion rule of the $\mathcal{Z}(\text{Vec}_{S_3})$ obtained from the Verlinde formula.

\paragraph{$U(1)$ Group}
For $U(1)$ group parametrized by $g=e^{i\theta}$ with $\theta\in [0,2\pi)$, each conjugacy class $[e^{i\theta}]$ has only a single element $e^{i\theta}$, and the centralizer group $C_{U(1)}(e^{i \theta})=U(1)$. The kernel in \eqref{eq:Define_Kernel} is
\begin{equation}
    K_v^{(\theta,n)}(\theta') = e^{i n \theta_v} \delta_{2\pi}(\theta'+\theta)\,,\quad (v=e^{i \theta_v})
\end{equation}
where $\delta_{2\pi}(\theta) = \sum_{n\in \mathbb{Z}} \delta(\theta -2\pi n)$ and $n\in \mathbb{Z}$ labels the irreducible representation of $U(1)$. The convolution between two kernels give
    \begin{equation}
    \begin{split}
    &\mathcal{K}_v^{(\theta,n)} * K_v^{(\varphi,m)}(\theta')\\ =& \frac{1}{2\pi} \int d\theta''\, \mathcal{K}_v^{(\theta,n)}(-\theta'') \mathcal{K}_v^{(\varphi,m)}(\theta'+\theta'')\\
    =& e^{in \theta_v+im\theta_v} \frac{1}{2\pi} \int d\theta'' \delta_{2\pi} (\theta-\theta'') \delta_{2\pi}(\varphi+\theta'+\theta'')\\
    =& e^{i(n+m)\theta_v} \delta_{2\pi} (\theta'+\theta+\varphi) = \mathcal{K}_v^{\theta+\varphi,n+m}(\theta')\,,
    \end{split}
    \end{equation}
from which one can read the fusion rule
    \begin{equation}
        v_{\theta,n} \otimes v_{\varphi,m}=v_{\theta+\varphi,n+m}\,.
    \end{equation}

\paragraph{$SU(2)$ Group}

Then let us move to less trivial example. It is convenient to use the Euler-angle parametrization, where each $g\in SU(2)$ can be written as
    \begin{equation}
        g= g_{\phi} a_{\theta} g_{\psi}\,,
    \end{equation}
with
    \begin{equation}
        g_{\phi} = \left(\begin{array}{cc}
            e^{\frac{i\phi}{2}} & 0 \\
            0 & e^{-\frac{i\phi}{2}} 
        \end{array} \right) \,,\quad a_{\theta}=\left(\begin{array}{cc}
            \cos \frac{\theta}{2} & -\sin \frac{\theta}{2} \\
             \sin \frac{\theta}{2} & \cos \frac{\theta}{2}
        \end{array} \right)\,,
    \end{equation}
and
    \begin{equation}
        0\leq \theta \leq \pi\,,\quad 0 \leq \phi \leq 2\pi\,, \quad -2\pi \leq \psi \leq 2\pi\,.
    \end{equation}
The metric $g_{ij}$ defined by $g_{ij} = -2\textrm{tr} (\partial_i g \partial_j g)$ is
    \begin{equation}
        ds^2 = d\theta^2 + \sin^2 \theta d\phi^2 + (d\psi + \cos \theta d \phi)^2\,.
    \end{equation}
Given the conjugacy class $[g]$, we will choose a representative $g$ in $T/W$, where $T=U(1)$ is the Cartan torus and $W$ is the $\mathbb{Z}_2$ Weyl group. For a generic $g\neq \pm 1$, the centralizer group is also $T$, which can be represented by $g_{\psi}$ in the Euler-angle parametrization. On the other hand, when $g=\pm 1$, the centralizer group is the whole $SU(2)$. Let us consider the coset $SU(2)/U(1)=S^2$, which is a 2-sphere whose measure is obtained by integrating $\psi$
    \begin{equation}
        d\mu(\theta,\phi) = \sin \theta d\theta d\phi \,,
    \end{equation}
which can be defined as the measure on the conjugacy class $[g]$ parametrized by $g(\theta,\phi) = r(\theta,\phi) g r(\theta,\phi)^{-1}$. Here the elements of the coset are $r(\theta,\phi)= g_{\phi} a_{\theta}$, and it is related to the spherical coordinates of $S^2$ via
    \begin{equation}
        r(\theta,-\phi) \sigma_3 r(\theta,-\phi)^{-1} = x \sigma_1+y \sigma_2+z\sigma_3\,,
    \end{equation}
with
    \begin{equation}
        x=\sin\theta \cos\phi\,,\quad y=\sin\theta \sin\phi\,, \quad z=\cos \theta\,,
    \end{equation}
and $\sigma_1,\sigma_2,\sigma_3$ are the Pauli matrices. The conjugation action of $SU(2)$ on $g(\theta,\phi)$ is thus the isometry group of $S^2$ generated by the rotation along $x,y,z$-axes.

For the state $\langle u,v|$ with $uv=vu$, we will choose both of them lying along the Cartan torus $T$. The kernel $\mathcal{K}_v^{[g],\rho_g}$ in \eqref{eq:Define_Kernel} is evaluated as following. When $g=\pm 1$, the conjugacy class $[\pm1]=\{ \pm 1\}$ contains only a single element, and therefore $[\pm 1]\cap C_G(v) = \{\pm 1\}$ is a point regardless of $v$. We then have
    \begin{equation}\label{eq:su(2)_kernel_center}
        \mathcal{K}_{v}^{[\pm 1],\rho_{\pm 1}}(h) = \delta^{C_G(v)}_{h^{-1},\pm 1} \chi_{\rho_{\pm 1}}(v)\,, \quad \rho_{\pm 1}\in \textbf{Rep}(SU(2))\,.
    \end{equation}
When both $g\neq \pm 1$ and $v \neq \pm 1$ are generic, the intersection $[g] \cap C_G(v) = \{g,g^{-1} \}$ contains two points and we have
\begin{equation}\label{eq:su(2)_kernel_generic}
    \mathcal{K}_{v}^{[g],\rho_g}(h) = \delta^{C_G(v)}_{h^{-1},g} \chi_{\rho_g}(v) + \delta^{C_G(v)}_{h^{-1},g^{-1}} \chi_{\rho_{g^{-1}}}(v)\,,\quad \rho_g \in \textbf{Rep}(U(1))\,.
\end{equation}
When $g\neq \pm 1$ is generic but $v=\pm 1$, the intersection $[g] \cap C_G(v) = [g]$ is the conjugacy class and we have
    \begin{equation}
        \mathcal{K}_{\pm 1}^{[g],\rho_g}(h) = \int_{[g]} d\mu(\theta,\phi) \chi_{\rho_g}(v) \delta^G_{h^{-1},g}\,, \quad \rho_g \in \textbf{Rep}(U(1))\,,
    \end{equation}
where $d\mu(g(\theta,\phi))$ is the measure introduced above.

We will work out the fusion rule of $SU(2)$ in the following. To do this, let us first consider the convolution of $\mathcal{K}_{\pm1}^{[g],\rho_g}(h)$ given by
    \begin{equation}
    \begin{split}
        &\mathcal{K}_{\pm 1}^{[g],\rho_g}*\mathcal{K}_{\pm 1}^{[h],\rho_h}(k)\\
        =& \int_G dk' \int_{[g]} d\mu(g) \int_{[h]} d\mu(h) \chi_{\rho_g}(\pm 1) \chi_{\rho_h}(\pm 1) \delta^G_{{k'}^{-1},g} \delta^G_{k'k^{-1},h}\\
        =&\int_{[g]} d\mu(\theta,\phi) \int_{[h]} d\mu(\theta',\phi') \chi_{\rho_g}(\pm 1) \chi_{\rho_h}(\pm 1)  \delta^G_{k^{-1},gh}\,,
    \end{split}
    \end{equation}
where we assume both $g,h$ are regular element whose centralizer groups are $U(1)$ parametrized from $[0,4\pi)$, thus $\rho_g,\rho_h \in \frac{1}{2}\mathbb{Z}$ label the irreducible representations of $U(1)$. To proceed, we need to reduce the double integral $\int_{[g]} d\mu(g) \int_{[h]} d\mu(h)$. In the coset decomposition, we can write $g,h$ as
    \begin{equation}
        g= r(\theta,\phi)g_{\psi_g}r^{-1}(\theta,\phi)\,, \quad h=r(\theta',\phi') g_{\psi_h} r^{-1}(\theta',\phi')\,,\quad \psi_g,\psi_h\in (0,2\pi)\,,
    \end{equation}
with $r(\theta,\phi) = g_{\phi} a_{\theta}$. Then $gh$ is
    \begin{equation}
        gh = r(\theta,\phi)g_{\psi_g}r^{-1}(\theta,\phi)r(\theta',\phi') g_{\psi_h} r^{-1}(\theta',\phi')\,.
    \end{equation}
Denote $gh=\ell$ whose conjugacy class is represented by $g_{\psi_{\ell}}$, we can rewrite the condition as
\begin{equation}
    r(\theta,\phi)^{-1} \ell r(\theta,\phi) = g_{\psi_g} r^{-1}(\theta,\phi)r(\theta',\phi') g_{\psi_h} r^{-1}(\theta',\phi')r(\theta,\phi)\,.
\end{equation}
For a fixed $(\theta,\phi)$, the element $r^{-1}(\theta,\phi)r(\theta',\phi')$ is a left translation acting on $r(\theta',\phi')$. Let us introduce new coordinates $(\theta'',\phi'')$ defined by $r^{-1}(\theta,\phi)r(\theta',\phi')=r(\theta'',\phi'')U(1)$, and rewrite the condition as
\begin{equation}
     \ell  = r(\theta,\phi)g_{\psi_g} r(\theta'',\phi'') g_{\psi_h} r^{-1}(\theta'',\phi'')r(\theta,\phi)^{-1}\,,
\end{equation}
which can be further written as
\begin{equation}\label{eq:su(2)_conjugacy_fuse_1}
        \ell = g(\phi,\theta,\phi'')g_{\psi_{g}} a_{\theta''} g_{\psi_h} a_{\theta''}^{-1}g(\phi,\theta,\phi'')^{-1}\,,
    \end{equation}
where $g(\phi,\theta,\phi'')$ is the group element defined by $g(\phi,\theta,\phi'')=g_{\phi}a_{\theta}g_{\phi''}$. Taking trace of both sides, we get a relation between $\psi_{\ell}$ with $\psi_{g},\psi_{h}$
\begin{equation}\label{eq:su(2)_conjugacy_fuse_2}
    \cos \frac{\psi_{\ell}}{2} = \cos \frac{\psi_{g}}{2} \cos \frac{\psi_{h}}{2} - \cos \theta'' \sin \frac{\psi_g}{2} \sin \frac{\psi_h}{2}\,,
\end{equation}
which implies the range of $\psi_{\ell}$
    \begin{equation}
        \psi_{\ell} \in \mathcal{I}_{g,h} \equiv  [\, |\psi_g-\psi_h| \, , \, \min(\psi_g+\psi_h,4\pi - \psi_g -\psi_h)\,]\,.
    \end{equation}
Taking the derivative on both side, we have
    \begin{equation}\label{eq:psi_and_theta}
        -\frac{1}{2}\sin \frac{\psi_{\ell}}{2} d\psi_{\ell} = -\sin \theta'' \sin \frac{\psi_g}{2} \sin \frac{\psi_h}{2}d\theta''\,.
    \end{equation}
    
We then decompose the integral $\int_{[g]} d\mu(\theta,\phi) \int_{[h]} d\mu(\theta',\phi')$ as
    \begin{equation}
        \int_{[g]} d\mu(\theta,\phi) \int_{[h]} d\mu(\theta',\phi') = \int \sin\theta  d\theta d\phi d\phi''  \int \sin \theta'' d\theta''\,.
    \end{equation}
Here $\theta''$ determines the conjugacy class $g_{\psi_{\ell}}$ via \eqref{eq:su(2)_conjugacy_fuse_2}, and $g(\phi,\theta,\phi'')$ will act as a conjugation and map $g_{\psi_{g}} a_{\theta''} g_{\psi_h} a_{\theta''}^{-1}$ to the whole conjugacy class in $[g_{\psi_{\ell}}]$. Notice that since $\phi''\in [0,2\pi)$, $g(\phi,\theta,\phi'')$ only parametrize half of $SU(2)$, namely $SO(3)=SU(2)/\mathbb{Z}_2$. For a given $\theta''$, we can decompose $g(\phi,\theta,\phi'')$ via the coset of the Cartan torus $T$ associated to $g_{\psi_{g}} a_{\theta''} g_{\psi_h} a_{\theta''}^{-1}$. For simplicity, we will still label the elements in $[\ell]$ using $(\theta,\phi)$, and the centralizer $C_G(\ell)$ direction using $\phi''$. Then the integral can be written as
    \begin{equation}
        \int_{[g]} d\mu(\theta,\phi) \int_{[h]} d\mu(\theta',\phi') = \int d\phi'' \int_{[\ell]} d\mu (\theta,\phi) \int \frac{\sin \frac{\psi_{\ell}}{2}}{2 \sin \frac{\psi_g}{2} \sin \frac{\psi_h}{2}} d\psi_{\ell}\,.
    \end{equation}
Here $\phi''$ only parametrize half of the centralizer $C_G(\ell)$, namely $U(1)/\mathbb{Z}_2 \subset C_G(\ell)$. Instead of integrating $\phi''$ out, one should understand $\int \phi''$ as a direct integral $\int^{\oplus} d\phi''$, which is an infinite dimensional vector space furnishing the regular representation of $U(1)/\mathbb{Z}_2$, which is decomposed as $\int^{\oplus} d \phi''\simeq \oplus_{\rho \in \mathbb{Z}} \,\rho$. Taken into account the representation $\rho_g,\rho_h\in \frac{1}{2}\mathbb{Z}$, we have
\begin{equation}
    \left(\int^{\oplus} d\phi''\right)\otimes \rho_g \otimes \rho_h = \bigoplus_{\rho_{\ell}\in \mathbb{Z}+\rho_g+\rho_h}\rho_{\ell}\,,
\end{equation}
as the representation of $C_G(\ell)$. Therefore, one candidate for the fusion rule is\cite{Koornwinder:1998xg,Bais:1998yn}
    \begin{equation}\label{eq:su(2)-fusion-kernel-fake}
        \mathcal{K}_{v}^{[g],\rho_g}*\mathcal{K}_{v}^{[h],\rho_h} = \bigoplus_{\rho_{\ell}=(\rho_g+\rho_h)\ \text{mod}\ \mathbb{Z}} \int_{\mathcal{I}_{g,h}} \frac{\sin \frac{\psi_{\ell}}{2}}{2 \sin \frac{\psi_g}{2} \sin \frac{\psi_h}{2}} d\psi_{\ell} \,\mathcal{K}^{[\ell],\rho_{\ell}}_{v}\,.
    \end{equation}


However, the formula above should be interpreted as the regular, interior part of the
fusion. There is an important subtlety at the endpoints of $\mathcal{I}_{g,h}$, which happens when $\theta''=0,\pi$ in \eqref{eq:su(2)_conjugacy_fuse_2} and $\psi_{\ell} = |\psi_g -\psi_h|$ or $\text{min}(\psi_g+\psi_h,4\pi - \psi_g -\psi_h)$. At that point, the residual circle fiber parametrized by $\phi''$ no longer appears, and $g,h$ are aligned or anti-aligned in the same Cartan
torus, thus $g,h,\ell$ share the same centralizer group $C_G(g)=C_G(h)=C_G(\ell)=U(1)$\footnote{We assume $\psi_{\ell} \neq 0,2\pi$ generically.}. Therefore we should have exactly $\rho_{\ell}=\rho_g+\rho_h$ when $\psi = \text{min}(\psi_g+\psi_h,4\pi - \psi_g -\psi_h)$ and $\rho_{\ell}=\rho_g-\rho_h$ or $\rho_h - \rho_g$ when $\psi_{\ell} = |\psi_g -\psi_h|$, depending on the sign of $\psi_g-\psi_h$, because they should transform in the same way under the common centralizer group $U(1)$. 

This endpoint contribution is essential when the convolution kernels \eqref{eq:su(2)-fusion-kernel-fake} are
tested in a regular $v$-sector.  For generic $v\neq\pm1$, the
$\rho_{\ell}$ tower contributes through the distribution
\begin{equation}
    \sum_{m\in\mathbb Z} \chi_{\rho_g+\rho_h+m}(v) = \sum_{m\in \mathbb{Z}} e^{i (\rho_g+\rho_h)\varphi_v} e^{i m \varphi_{v}} = 2\pi e^{i (\rho_g+\rho_h)\varphi_v}  \delta_{2\pi}(\varphi_v),
\end{equation}
where $v=e^{i \varphi_v}$. Hence it vanishes in a generic regular $v$-sector. Therefore, the nonzero convolution
for generic $v$ must be carried by the endpoint channels, where the
centralizer fiber $\phi''$ has collapsed and no infinite Fourier sum is present.

We therefore modify \eqref{eq:su(2)-fusion-kernel-fake} by including the contribution of the endpoints and write
    \begin{equation}\label{eq:su(2)-fusion-kernel}
    \begin{split}
        \mathcal{K}_{v}^{[g],\rho_g}*\mathcal{K}_{v}^{[h],\rho_h} =& \bigoplus_{\rho_{\ell}=(\rho_g+\rho_h)\ \text{mod}\ \mathbb{Z}} \int_{\mathcal{I}_{g,h}} \frac{\sin \frac{\psi_{\ell}}{2}}{2 \sin \frac{\psi_g}{2} \sin \frac{\psi_h}{2}} d\psi_{\ell} \,\mathcal{K}^{[\ell],\rho_{\ell}}_{v}\\
        +& \mathcal{K}_v^{[g_{\psi_g+\psi_h}],\rho_g+\rho_h} + \mathcal{K}_v^{[g_{\psi_g-\psi_h}],\rho_g-\rho_h} \,.
    \end{split}
    \end{equation}
These endpoint terms are essential for matching the kernel fusion. For generic $v \neq \pm 1$, the RHS is
    \begin{equation}
    \begin{split}
        &\mathcal{K}_{v}^{[g],\rho_g}*\mathcal{K}_{v}^{[h],\rho_h}(k) = \mathcal{K}_v^{[g_{\psi_g+\psi_h}],\rho_g+\rho_h}(k) + \mathcal{K}_v^{[g_{\psi_g-\psi_h}],\rho_g-\rho_h}(k)\\
        =&\delta^{C_G(v)}_{k^{-1},g_{\psi_g+\psi_h}} \chi_{\rho_g+\rho_h}(v)+\delta^{C_G(v)}_{k^{-1},g_{-\psi_g-\psi_h}} \chi_{-\rho_g-\rho_h}(v) + \delta^{C_G(v)}_{k^{-1},g_{\psi_g-\psi_h}} \chi_{\rho_g-\rho_h}(v)+ \delta^{C_G(v)}_{k^{-1},g_{-\psi_g+\psi_h}} \chi_{-\rho_g+\rho_h}(v)\,.
    \end{split}
    \end{equation}
On the other hand, we can compute $\mathcal{K}_{v}^{[g],\rho_g}*\mathcal{K}_{v}^{[h],\rho_h}(k)$ directly using \eqref{eq:su(2)_kernel_generic}, we get the same result
    \begin{equation}
    \begin{split}
        &\int_{C_G(v)} dk'  (\delta^{C_G(v)}_{{k'}^{-1},g_{\psi_g}} \chi_{\rho_g}(v) + \delta^{C_G(v)}_{{k'}^{-1},g_{\psi_g}^{-1}} \chi_{-\rho_{g}}(v))(\delta^{C_G(v)}_{k' k^{-1},g_{\psi_h}} \chi_{\rho_h}(v) + \delta^{C_G(v)}_{k'k^{-1},g_{\psi_h}^{-1}} \chi_{-\rho_{h}}(v))\\
        =& \delta^{C_G(v)}_{k^{-1},g_{\psi_g+\psi_h}} \chi_{\rho_g+\rho_h}(v) + \delta^{C_G(v)}_{k^{-1},g_{\psi_g-\psi_h}} \chi_{\rho_g-\rho_h}(v)+\delta^{C_G(v)}_{k^{-1},g_{-\psi_g+\psi_h}} \chi_{-\rho_g+\rho_h}(v)+\delta^{C_G(v)}_{k^{-1},g_{-\psi_g-\psi_h}} \chi_{-\rho_g-\rho_h}(v)\,.
    \end{split}
    \end{equation}
Therefore, these endpoint channels reproduce the direct convolution of the
regular-sector kernels \eqref{eq:su(2)_kernel_generic}.

\section{The Twisted Fell Line Bundle, Projective Basis, and Twisted Braiding}\label{sec:Twisted_conj_groupoid_basis}

We now repeat the same construction in the presence of a nontrivial twist. We keep the same conjugation groupoid
\begin{equation}
    \mathcal{G}=G//_{\mathrm{Ad}}G\,,
\end{equation}
and the same notation for arrows
\begin{equation}
    (u,x):x\longrightarrow uxu^{-1}\,.
\end{equation}
The difference is that we now equip $\mathcal{G}$ with the transgressed class
\begin{equation}
    \tau(k)\in H^2(\mathcal{G},U(1))\,,
\end{equation}
induced by the level-$k$ Chern-Simons term.\footnote{Here $H^2(\mathcal{G},U(1))$ denotes the degree-$2$ cohomology of the action groupoid $\mathcal{G}=G//_{\mathrm{Ad}}G$, equivalently the degree-$2$ cohomology of the quotient stack $[G/G]$. After choosing a trivialization of the corresponding twist, a continuous $U(1)$-valued cocycle on the space $\mathcal{G}^{(2)}$ of composable pairs of arrows gives the cocycle-level description of a Fell line bundle, or equivalently of an $S^1$-central extension/gerbe over $[G/G]$~\cite{Behrend:2008,behrend2003equivariant,Tu:2009twistedRing}. Via the exponential sequence and the Dixmier-Douady class of this gerbe, one obtains the corresponding integral class in $H^3_G(G,\mathbb{Z})$; this comparison between stack/groupoid and equivariant Dixmier-Douady classes is explained in~\cite{Stienon:2010}. For the compact Lie-group Chern-Simons twists relevant here, that integral equivariant class is the transgression of the anomaly class $k\in H^4(BG,\mathbb{Z})$ in the multiplicative-gerbe sense~\cite{meinrenken2002basic,Carey:2004xt,Jia:2026vcr}. Thus $\sigma_k$ below is a cocycle-level representative, while $\tau(k)$ denotes its cohomology class.} Following the notation of~\cite{Jia:2026vcr}, we denote by $\Sigma_k\to \mathcal{G}$ the corresponding Fell line bundle, and choose a normalized representative
\begin{equation}
    \sigma_k\in Z^2(\mathcal{G},U(1))\,,
\end{equation}
of the cohomology class $\tau(k)$. Namely, $\sigma_k$ is a map from the set of composable arrows $\mathcal{G}^{(2)}$ of $\mathcal{G}$ to $U(1)$~\cite{armstrong2022uniqueness}. In the finite-group Dijkgraaf-Witten approach, $\sigma_k$ may be obtained by transgressing a $3$-cocycle representative $\omega$ of the anomaly; for Lie groups, the same transgressed class $\tau(k)$ and Fell line bundle $\Sigma_k$ are precisely the objects used in~\cite{Jia:2026vcr}. We will not need the explicit cocycle formula; all the computations below depend only on the chosen representative $\sigma_k$ and on the associated Fell line bundle $\Sigma_k$.

\subsection{The Twisted Fell Line Bundle and Projective Transport}\label{sec:twisted_fell_line_bundle}

Let $\Sigma_k\to \mathcal{G}$ be the Fell line bundle associated to $\tau(k)$. Since each fiber is one-dimensional, we may choose a nonzero basis vector in every fiber and denote by
\begin{equation}
    \delta^{\Sigma_k}_{(u,x)}\,,
\end{equation}
the corresponding basis section supported on the single arrow $(u,x)$. The multiplication in the twisted groupoid algebra is then
\begin{equation}\label{eq:twisted_delta_convolution}
    \delta^{\Sigma_k}_{(v,y)} * \delta^{\Sigma_k}_{(u,x)}
    =
    \delta_{y,uxu^{-1}}\,
    \sigma_k\big((v,uxu^{-1}),(u,x)\big)\,
    \delta^{\Sigma_k}_{(vu,x)}\,,
\end{equation}
where $\sigma_k$ satisfies the cocycle condition
    \begin{equation}\label{eq:sigma_k_cocycle_condition}
        \sigma_k((w,z),(vu,x)) \sigma_k((v,y),(u,x)) = \sigma_k((w,z),(v,y)) \sigma_k((wv,y),(u,x))\,,
    \end{equation}
with $y=uxu^{-1}$ and $z=vyv^{-1}=vuxu^{-1}v^{-1}$. Thus the representative cocycle $\sigma_k$ inserts a phase whenever two arrows are composed, while the invariant twisting datum is the cohomology class $\tau(k)$ encoded by the Fell line bundle $\Sigma_k$.

A representation of the twisted Fell line bundle $\Sigma_k$ is therefore a collection of vector spaces $\{E_x\}_{x\in G}$ together with linear isomorphisms
\begin{equation}\label{eq:twisted_transport_map}
    T^{\Sigma_k}_u(x):E_x\longrightarrow E_{uxu^{-1}}
\end{equation}
satisfying the projective composition rule
\begin{equation}\label{eq:twisted_transport_composition}
    T^{\Sigma_k}_v(uxu^{-1})\,T^{\Sigma_k}_u(x)
    =
    \sigma_k\big((v,uxu^{-1}),(u,x)\big)\,
    T^{\Sigma_k}_{vu}(x)\,.
\end{equation}
Compared with the untwisted relation~(\ref{eq:T_u_x_comp_untwisted}), the only difference is the extra cocycle factor on the right-hand side. The category depends only on the transgressed class and not on the particular Fell-bundle representative, we denote twisted category of $G$-equivariant Hilbert-space fields over $G$ as $\hilb^{\tau(k)}_G(G)=\mathcal{Z}(\hilb^{k}(G))$; in the notation of~\cite{Jia:2026vcr}, this is the codimension-$2$ category
\begin{equation}
    \mathcal Z({\rm Hilb}^k(G))
    \simeq
    \textbf{Rep}(C^*(G//_{\rm Ad}G,\Sigma_k))\,.
\end{equation}

\subsection{Projective Simples and the Action of Twisted Arrows}\label{sec:twisted_simple_basis}

Fix a conjugacy class representative $g\in G$. Restricting the representative cocycle $\sigma_k$ to the isotropy group at $g$ gives a $2$-cocycle on the centralizer:
\begin{equation}\label{eq:alpha_g_from_sigma}
    \alpha_g(c_1,c_2)
    :=
    \sigma_k\big((c_1,g),(c_2,g)\big)\,,
    \qquad c_1,c_2\in C_G(g)
\end{equation}
where $(c_1, g)$ and $(c_2, g)$ constitute clearly a composable pair of arrows. We now choose an irreducible $\alpha_g$-projective representation $\rho_g$ of $C_G(g)$, namely
\begin{equation}\label{eq:projective_rep_alpha_g}
    \rho_g(c_1)\rho_g(c_2)=\alpha_g(c_1,c_2)\rho_g(c_1c_2)\,.
\end{equation}
This is the twisted analogue of the ordinary centralizer representation used in the untwisted case.

Choose, as before, representatives $\{r_i\}$ of the left cosets in $G/C_G(g)$ and define
\begin{equation}
    g_i=r_i g r_i^{-1}\,.
\end{equation}
The support of the corresponding simple object $H^{\Sigma_k}_{[g],\rho_g}$ is still the conjugacy class $[g]$, but now the equivariant structure is projective rather than honest. To make this explicit, choose isomorphisms
\begin{equation}\label{eq:I_i_twisted}
    I_i:V_{\rho_g}\longrightarrow \big(H^{\Sigma_k}_{[g],\rho_g}\big)_{g_i}\,,
\end{equation}
from the projective centralizer representation space $V_{\rho_g}$ to the fiber over $g_i$, and define the basis vectors
\begin{equation}\label{eq:twisted_basis_gi_mu}
    \ket{g_i,\mu}:= I_i(e_\mu)\,,
\end{equation}
where $\{e_\mu\}$ is a basis of $V_{\rho_g}$. As in the untwisted section, the ket is labeled by the actual conjugate $g_i$, while the choice of representative $r_i$ remains auxiliary data used in the transport formulas.

Correspondingly, the twisted simple object carries a representation
\begin{equation}\label{eq:def_pi_sigma_g_rhog}
    \pi^{\Sigma_k}_{[g],\rho_g}: C^*(G//_{\mathrm{Ad}}G,\Sigma_k)\longrightarrow \mathrm{End}\big(H^{\Sigma_k}_{[g],\rho_g}\big)\,,
\end{equation}
again labeled by the pair $([g],\rho_g)$ because the support over the conjugacy class and the projective centralizer representation together determine which simple twisted module is being represented.

Now consider an arrow $(u,g_i)$. Write
\begin{equation}\label{eq:twisted_u_ri}
    ur_i=r_j c\,,
    \qquad c\in C_G(g)\,.
\end{equation}
Then $ug_i u^{-1}=g_j$ exactly as before. In the twisted case, however, the transport of the $i$-th fiber to the $j$-th fiber is not simply $I_j\rho_g(c)I_i^{-1}$; one must include an additional phase depending on the cocycle representative $\sigma_k$ and on the chosen trivialization of the fibers. We denote this phase by
\begin{equation}
    \vartheta_g(u;i\to j)\in U(1)\,,
\end{equation}
and define it through
\begin{equation}\label{eq:theta_definition_twisted}
    T^{\Sigma_k}_u(g_i)\,I_i
    =
    \vartheta_g(u;i\to j)\,I_j\,\rho_g(c)\,,
    \qquad ur_i=r_j c\,.
\end{equation}
Equivalently, the arrow basis element acts as
\begin{equation}\label{eq:twisted_arrow_action_on_basis}
    \pi^{\Sigma_k}_{[g],\rho_g}\big(\delta^{\Sigma_k}_{(u,g_i)}\big)\ket{g_i,\mu}
    =
    \vartheta_g(u;i\to j)
    \sum_{\nu}\big[\rho_g(c)\big]_{\nu\mu}\ket{g_j,\nu}\,.
\end{equation}
This is the twisted analogue of~\eqref{eq:arrow_action_on_basis}.

The phase $\vartheta_g(u;i\to j)$ is not arbitrary. Suppose
\begin{equation}
    ur_i=r_j c\,,\qquad vr_j=r_k d\,,
    \qquad c,d\in C_G(g)\,.
\end{equation}
Then comparing the two ways to act by $(v,g_j)$ and $(u,g_i)$, one finds 
\begin{equation}\label{eq:theta_sigma_consistency}
    \vartheta_g(v;j\to k)\,\vartheta_g(u;i\to j)\,\alpha_g(d,c)
    =
    \sigma_k\big((v,g_j),(u,g_i)\big)\,
    \vartheta_g(vu;i\to k)\,,
\end{equation}
where we used the projective representation law~(\ref{eq:projective_rep_alpha_g}) to combine $\rho_g(d)\rho_g(c)$. Thus the failure of the transport phases to multiply honestly is measured precisely by the representative cocycle $\sigma_k$, or invariantly by the transgressed class $\tau(k)$. The phase $\vartheta_g(u;i\to j)$ can be chosen explicitly as\cite{Roche:1990hs}
\begin{equation}\label{eq:vartheta_sigmak_relation}
    \vartheta_g(u;i\to j)=\frac{\sigma_k((u,g_i),(r_i,g))}{\sigma_k((r_j,g),(c,g))}\,,
\end{equation}
and one can check it satisfies the consistency condition~\eqref{eq:theta_sigma_consistency} by successively applying the cocycle condition~\eqref{eq:sigma_k_cocycle_condition} of $\sigma_k$.

\subsection{Twisted Half-Braiding and Full Braiding}\label{sec:twisted_half_braiding_basis}

The sectorwise form of the half-braiding is unchanged: the class representative carried by the first factor still transports the second factor by conjugation. The difference is that the transport map is now the projective one in~\eqref{eq:twisted_transport_map}. Thus for $v_x\in X_x \subset X = \sum_x X_x$ and $w_y\in Y_y \subset Y = \sum_y Y_y$ with $X,Y\in \mathcal{Z}(\hilb^{k}(G))$, we define
\begin{equation}\label{eq:twisted_full_braiding_general}
    c^{\Sigma_k}_{X,Y}(v_x\otimes w_y)
    =
    T^{\Sigma_k,Y}_x(y)(w_y)\otimes v_x\,.
\end{equation}
This has exactly the same formal shape as~(\ref{eq:braiding_from_half_braiding}); all of the new information is encoded in the projective transport operator $T^{\Sigma_k,Y}_x(y)$.

We now specialize once again to the two simple objects
\begin{equation}
    X=H^{\Sigma_k}_{[g],\rho_g}\,,\qquad
    Y=H^{\Sigma_k}_{[h],\rho_h}\,.
\end{equation}
Choose representatives $\{q_k\}$ for $G/C_G(h)$ and write
\begin{equation}
    h_k=q_k h q_k^{-1}\,,
    \qquad
    \ket{h_k,\nu}\in \big(H^{\Sigma_k}_{[h],\rho_h}\big)_{h_k}\,.
\end{equation}
Now let $g_i$ act on the second factor. Write
\begin{equation}\label{eq:twisted_gi_qk}
    g_i q_k = q_\ell d_{ik}\,,
    \qquad d_{ik}\in C_G(h)\,,
\end{equation}
so that
\begin{equation}
    g_i h_k g_i^{-1}=h_\ell\,.
\end{equation}
In analogy with~(\ref{eq:theta_definition_twisted}), we introduce a phase
\begin{equation}
    \vartheta_h(g_i;k\to \ell)\in U(1)
\end{equation}
by
\begin{equation}\label{eq:theta_h_definition_twisted}
    T^{\Sigma_k,Y}_{g_i}(h_k)\,J_k
    =
    \vartheta_h(g_i;k\to \ell)\,J_\ell\,\rho_h(d_{ik})\,,
\end{equation}
where $J_k:V_{\rho_h}\to (H^{\Sigma_k}_{[h],\rho_h})_{h_k}$ is the analogue of $I_i$ for the second object. Equivalently,
\begin{equation}\label{eq:twisted_transport_on_h_basis}
    T^{\Sigma_k,Y}_{g_i}(h_k)\ket{h_k,\nu}
    =
    \vartheta_h(g_i;k\to \ell)
    \sum_{\lambda}\big[\rho_h(d_{ik})\big]_{\lambda\nu}\ket{h_\ell,\lambda}\,.
\end{equation}
Substituting this into~(\ref{eq:twisted_full_braiding_general}), we obtain the twisted braiding formula
\begin{equation}\label{eq:twisted_braiding_basis_formula}
    c^{\Sigma_k}_{H^{\Sigma_k}_{[g],\rho_g},H^{\Sigma_k}_{[h],\rho_h}}
    \big(\ket{g_i,\mu}\otimes \ket{h_k,\nu}\big)
    =
    \vartheta_h(g_i;k\to \ell)
    \sum_{\lambda}\big[\rho_h(d_{ik})\big]_{\lambda\nu}\ket{h_\ell,\lambda}\otimes \ket{g_i,\mu}\,.
\end{equation}
This is the twisted analogue of~\eqref{eq:braiding_basis_formula}. The first factor still conjugates the class representative of the second factor, but now one also picks up the explicit cocycle-dependent phase $\vartheta_h(g_i;k\to \ell)$.

The commuting case again simplifies. If $g_i$ commutes with $h_k$, then $\ell=k$ and
\begin{equation}
    d_{ik}=q_k^{-1}g_i q_k\in C_G(h)\,.
\end{equation}
Hence
\begin{equation}\label{eq:twisted_braiding_commuting_case}
    c^{\Sigma_k}_{H^{\Sigma_k}_{[g],\rho_g},H^{\Sigma_k}_{[h],\rho_h}}
    \big(\ket{g_i,\mu}\otimes \ket{h_k,\nu}\big)
    =
    \vartheta_h(g_i;k\to k)
    \sum_{\lambda}\big[\rho_h(q_k^{-1}g_i q_k)\big]_{\lambda\nu}\ket{h_k,\lambda}\otimes \ket{g_i,\mu}\,.
\end{equation}

\section{The Actions of Twisted Extended Operators on the Hilbert Space of the SymTFT}\label{sec:Twisted_HilbGG_on_HBF_section}

This section is the twisted counterpart of section~\ref{sec:HilbGG_on_HBF}. Section~\ref{sec:Twisted_conj_groupoid_basis} established the codimension-$2$ defect data and twisted braiding. We now turn to the codimension-$1$ Hilbert space of the $BF+kCS$ theory on a spatial torus. We first explain how the mixed $BF+kCS$ theory quantizes to a twisted Hilbert space on the torus, and then describe how the same transgressed class $\tau(k)$ and associated Fell line bundle $\Sigma_k$ make the action of line operators projective on that Hilbert space.

\subsection{Quantization of \texorpdfstring{$BF+kCS$}{BF+kCS} theory on a spatial surface}\label{sec:Twisted_quantization_BFkCS}

We now adapt the quantization discussion of section~\ref{sec:HilbGG_on_HBF} to the $BF+kCS$ theory, namely the $BF$ theory further deformed by an additional level-$k$ Chern-Simons term, and not to a pure Chern-Simons theory by itself. The canonical derivation below applies to compact Lie groups as well as to the finite-group model used later in the explicit computations. We nevertheless begin with the continuum picture, because it makes clear which part of the quantization is inherited from $BF$ theory and which part comes from the Chern-Simons deformation, and only afterwards translate the result into the transgressed class $\tau(k)$, the associated Fell line bundle $\Sigma_k$, and the twisted codimension-$2$ category
\begin{equation}
    \hilb_G^{\tau(k)}(G)
    \simeq
    \mathcal Z({\rm Hilb}^k(G))
    \simeq
    \textbf{Rep}(C^*(G//_{\rm Ad}G,\Sigma_k))\,,
\end{equation}
that provides the groupoid presentation used in the later formulas~\cite{Blau:1989bq,Baez:1999sr,Axelrod:1989xt,Witten:1988hf,Jia:2026vcr}.

Consider the mixed action on $Y=\mathbb R_t\times T^2$,
\begin{equation}\label{eq:BFkCS_action_continuum}
    S_{BF+kCS}[A,B]
    =
    \int_Y \langle B\wedge F_A\rangle
    +
    \frac{k}{4\pi}\int_Y
    \left\langle A\wedge dA+\frac23A\wedge A\wedge A\right\rangle\,.
\end{equation}
Write
\begin{equation}
    A=A_t\,dt+A_\Sigma,
    \qquad
    B=B_t\,dt+B_\Sigma,
\end{equation}
where $A_\Sigma,B_\Sigma\in \Omega^1(T^2,\mathfrak g)$. Up to the usual sign conventions and total derivatives, the terms containing time derivatives are
\begin{equation}\label{eq:BFkCS_kinetic_terms}
    S_{BF+kCS}
    =
    \int dt\left[
    \int_{T^2}\langle B_\Sigma\wedge \dot A_\Sigma\rangle
    +
    \frac{k}{4\pi}\int_{T^2}\langle A_\Sigma\wedge \dot A_\Sigma\rangle
    + \cdots
    \right].
\end{equation}
Therefore the symplectic potential on the space of spatial fields is
\begin{equation}\label{eq:Theta_BFkCS}
    \Theta_{BF+kCS}
    =
    \int_{T^2}\langle B_\Sigma\wedge \delta A_\Sigma\rangle
    +
    \frac{k}{4\pi}\int_{T^2}\langle A_\Sigma\wedge \delta A_\Sigma\rangle,
\end{equation}
and the corresponding symplectic two-form is
\begin{equation}\label{eq:Omega_BFkCS}
    \Omega_{BF+kCS}
    =
    \delta\Theta_{BF+kCS}
    =
    \int_{T^2}\langle \delta B_\Sigma\wedge \delta A_\Sigma\rangle
    +
    \frac{k}{4\pi}\int_{T^2}\langle \delta A_\Sigma\wedge \delta A_\Sigma\rangle.
\end{equation}
The first term is the canonical $BF$ cotangent-bundle form, while the second is the Chern-Simons ``magnetic'' term~\cite{Blau:1989bq,Axelrod:1989xt,Witten:1988hf}, and the reason we call it a magnetic term will be made clear in a moment.

The temporal components $A_t$ and $B_t$ are Lagrange multipliers. Indeed, after substituting $A=A_t\,dt+A_\Sigma$ and $B=B_t\,dt+B_\Sigma$ into~(\ref{eq:BFkCS_action_continuum}) and discarding total derivatives, the terms involving them take the form
\begin{equation}
    \int dt\left[
    \int_{T^2}\langle B_t,F_{A_\Sigma}\rangle
    +
    \int_{T^2}\left\langle A_t,\,
    d_{A_\Sigma}B_\Sigma+\frac{k}{2\pi}F_{A_\Sigma}\right\rangle
    \right].
\end{equation}
Therefore varying $B_t$ enforces $F_{A_\Sigma}=0$, while varying $A_t$ enforces $d_{A_\Sigma}B_\Sigma+\frac{k}{2\pi}F_{A_\Sigma}=0$. Their equations of motion are thus
\begin{equation}\label{eq:BFkCS_constraints}
    F_{A_\Sigma}=0,
    \qquad
    d_{A_\Sigma}B_\Sigma+\frac{k}{2\pi}F_{A_\Sigma}=0,
\end{equation}
up to the same normalization conventions used in~(\ref{eq:BFkCS_action_continuum}). On the flat-connection locus, the second equation reduces to
\begin{equation}
    d_{A_\Sigma}B_\Sigma=0.
\end{equation}
Modulo the $BF$ shift symmetry $B_\Sigma\sim B_\Sigma+d_{A_\Sigma}\chi$, the class of $B_\Sigma$ determines an element of
\begin{equation}\label{eq:H1_as_T*}
    H^1(T^2,\operatorname{ad}P_\Sigma)\cong T^*_{[A_\Sigma]}\mathcal M_{\rm flat}(T^2,G)\,,
\end{equation}
on the smooth irreducible locus, where $P_{\Sigma}$ is the principal bundle over $\Sigma$ and $\mathcal M_{\rm flat}(T^2,G)$ denotes the moduli space of flat $G$-connections. More concretely,~\eqref{eq:H1_as_T*} comes from the constraint $d_{A_\Sigma}B_\Sigma=0$ modulo the shift $B_\Sigma\sim B_\Sigma+d_{A_\Sigma}\chi$, while the tangent space to $\mathcal M_{\rm flat}$ is described by the same cohomology because an infinitesimal deformation $A_\Sigma\mapsto A_\Sigma+\varepsilon a$ preserves flatness precisely when $d_{A_\Sigma}a=0$, modulo infinitesimal gauge transformations $a\sim a+d_{A_\Sigma}\epsilon$. The passage from $H^1(T^2,\operatorname{ad}P_\Sigma)$ to the cotangent space then uses the invariant pairing on $\mathfrak g$ together with Poincar\'e duality on the closed oriented surface $T^2$:
\begin{equation}
    ([b],[a])\longmapsto \int_{T^2}\langle b\wedge a\rangle.
\end{equation}
On the smooth irreducible locus this pairing is nondegenerate, so it identifies $H^1(T^2,\operatorname{ad}P_\Sigma)$ with its dual and hence with $T^*_{[A_\Sigma]}\mathcal M_{\rm flat}(T^2,G)$. Thus the reduced phase-space of gauge-inequivalent $(A_\Sigma, B_\Sigma)$ is the same as in pure $BF$ theory,
\begin{equation}\label{eq:BFkCS_phase_space}
    \mathcal P_{BF+kCS}(T^2)
    \simeq
    T^*\mathcal M_{\rm flat}(T^2,G),
\end{equation}
and the reduced symplectic form $\Omega_{BF+kCS,\rm red}$ is, by definition, the two-form induced on this quotient from the unreduced form $\Omega_{BF+kCS}$ in~\eqref{eq:Omega_BFkCS}. Under the identification~\eqref{eq:BFkCS_phase_space}, the $BF$ contribution descends to the Liouville one-form $\Theta_{\rm can}$ on the cotangent bundle, whose exterior derivative
\begin{equation}
    \Omega_{\rm can}:=d\Theta_{\rm can}
\end{equation}
is the standard canonical cotangent-bundle symplectic form. The Chern-Simons term then shifts this by the pullback of the Atiyah-Bott form:
\begin{equation}\label{eq:BFkCS_reduced_symplectic}
    \Omega_{BF+kCS,\rm red}
    =
    \Omega_{\rm can}
    +
    k\,\pi^*\omega_{AB}.
\end{equation}
Here $\pi:T^*\mathcal M_{\rm flat}\to \mathcal M_{\rm flat}$ is the cotangent projection, $\Omega_{\rm can}$ is the canonical cotangent-bundle symplectic form, and
\begin{equation}\label{eq:AB_form_torus}
    \omega_{AB}([\alpha],[\alpha'])
    :=
    \frac{1}{2\pi}\int_{T^2}\langle \alpha\wedge \alpha'\rangle
\end{equation}
is the standard symplectic form on the moduli of flat connections that also appears in Chern-Simons quantization~\cite{Axelrod:1989xt,Witten:1988hf}. In other words, the mixed theory quantizes a magnetic cotangent bundle, not the base moduli space by itself.

This terminology comes from the standard phase-space description of a charged particle moving on a configuration space $Q$ in the presence of a background $U(1)$ gauge potential. In local coordinates $(q^r,p_r)$ on $T^*Q$, the ordinary cotangent bundle has symplectic potential $\Theta_{\rm can}=\sum_r p_r\,dq^r$ and symplectic form $\Omega_{\rm can}=d\Theta_{\rm can}$. Coupling to a local gauge potential $\alpha_I$ on $Q$ replaces this by
\begin{equation}
    \Theta=\sum_r p_r\,dq^r + k\,\alpha_I,
    \qquad
    \Omega=\Omega_{\rm can}+k\,d\alpha_I\,.
\end{equation}
The curvature $d\alpha_I$ is the analogue of the magnetic field, so in the present case $k\,\omega_{AB}$ plays exactly that role on $Q=\mathcal M_{\rm flat}(T^2,G)$. In the abelian toy model $Q=T^2$, this is the same geometric mechanism as the familiar extra $B\,dx\wedge dy$ term in the Landau problem. Thus the Chern-Simons deformation does not create new phase-space directions; rather, it twists the $BF$ cotangent symplectic form by a closed $2$-form on the base, just as a background magnetic field twists the phase space of an ordinary particle~\cite{Gruber:1999magnetic}.

This immediately explains the correct polarization. In pure level-$k$ Chern-Simons theory, the classical phase space is the base
\begin{equation}\label{eq:Phase_Spacs_CS}
    \mathcal P_{CS,k}(T^2)
    =
    \big(\mathcal M_{\rm flat}(T^2,G),k\,\omega_{AB}\big),
\end{equation}
and after choosing a complex structure on the torus one uses the corresponding K\"ahler polarization, which gives the familiar holomorphic space of sections
\begin{equation}
    \mathcal H_{CS,k}(T^2)
    \simeq
    H^0(\mathcal M_{\rm flat}(T^2,G),L^k)
\end{equation}
in the usual continuum treatment~\cite{Axelrod:1989xt,Witten:1988hf}. In geometric-quantization terms, the K\"ahler polarization is the complex Lagrangian distribution $T^{0,1}\mathcal M_{\rm flat}\subset T_{\mathbb C}\mathcal M_{\rm flat}$, so polarized sections obey the holomorphicity condition $\nabla_{\bar z^a}\psi=0$ in local complex coordinates $z^a$ on $\mathcal M_{\rm flat}$~\cite{Axelrod:1989xt,de2007symplectic}. By contrast, for the mixed $BF+kCS$ theory the natural polarization is the vertical, or Schr\"odinger, polarization of the cotangent bundle~(\ref{eq:BFkCS_phase_space}), because the $BF$ term treats $B_\Sigma$ as the momentum conjugate to $A_\Sigma$. This vertical polarization is the real Lagrangian distribution $\ker(d\pi)\subset T(T^*\mathcal M_{\rm flat})$ for $\pi:T^*\mathcal{M}_{\rm flat} \rightarrow \mathcal{M}_{\rm flat}$, hence we have:
\begin{equation}
	\ker(d\pi) = \mathrm{span}\left\{ \frac{\partial}{\partial p_r} \right\}\,.
\end{equation}
Therefore, the polarization condition is $\nabla_{\partial/\partial p_r}\psi=0$ in Darboux coordinates $(q^r,p_r)$, meaning that the state is covariantly constant along the cotangent fibers and therefore depends only on the base coordinates $q^r$~\cite{Gruber:1999magnetic,de2007symplectic}~\footnote{In ordinary quantum mechanics this is the statement that a wave function depends only on the coordinate and not the momentum.}. On a local patch $U_I\subset \mathcal M_{\rm flat}$ with Darboux coordinates $(q^r,p_r)$ on $T^*U_I$, one may write
\begin{equation}
    \Theta_{BF+kCS,\rm red}^{(I)}
    =
    \sum_r p_r\,dq^r + k\,\alpha_I,
    \qquad
    d\alpha_I=\omega_{AB}.
\end{equation}
On an overlap $U_I\cap U_J$, $\alpha_J-\alpha_I=d\chi_{IJ}$, so the local vertical-polarized wavefunctions glue by the corresponding phase and therefore define sections of the restriction of prequantum line bundle $L^k\to \mathcal M_{\rm flat}$. In the compact Lie-group continuum theory this leads to the real-polarization Hilbert space
\begin{equation}\label{eq:BFkCS_continuum_hilbert}
    \mathcal H_{BF+kCS}(T^2)
    \sim
    L^2(\mathcal M_{\rm flat}(T^2,G),L^k),
\end{equation}
up to the standard half-density and singular-locus caveats. The important structural distinction is therefore level-$k$ CS quantizes $\mathcal M_{\rm flat}$ whereas $BF+kCS$ quantizes $T^*\mathcal M_{\rm flat}$ with the same line-bundle twist.

We now rewrite the same quantization statement in the groupoid language used below. On the codimension-$2$ side, the deformation is encoded by the transgressed class
\begin{equation}
    \tau(k)\in H^2(G//_{\rm Ad}G,U(1))\,,
\end{equation}
and by the corresponding Fell line bundle $\Sigma_k$ over the conjugation groupoid, exactly as in~\cite{Jia:2026vcr}. In the finite-group Dijkgraaf-Witten approach, $\tau(k)$ is represented by the transgression of a $3$-cocycle $[\omega_k]\in H^3(G,U(1))$~\cite{Dijkgraaf:1989pz,FreedQuinn:1993,Willerton:2008gyk}; for compact Lie groups, \cite{Jia:2026vcr} treats the same twisting datum directly at the level of the Lie-group transgression. On the codimension-$1$ side, the commuting-holonomy configuration space is still the same space~(\ref{eq:Config_Space_ST2}) of pairs $(x(a),x(b))$ modulo simultaneous conjugation. What changes is that the wavefunctions are no longer honest functions on $\widetilde{\mathcal A}_0$: the Fell line bundle $\Sigma_k$ is the groupoid avatar of the continuum prequantum line bundle, so the torus Hilbert space is the space of sections of the corresponding twisted line bundle. 

We therefore denote the twisted physical Hilbert space by
\begin{equation}\label{eq:H_BFkCS_sections}
    H_{BF+kCS}(T^2)
    :=
    \Gamma(\widetilde{\mathcal{A}}_0,L_{\Sigma_k})\,,
\end{equation}
where $L_{\Sigma_k}\to \widetilde{\mathcal{A}}_0$ is the line bundle induced from the Fell line bundle $\Sigma_k$; more invariantly, one may regard this as the same transgressed bundle data over the quotient groupoid presentation of~(\ref{eq:Config_Space_ST2})~\cite{FreedQuinn:1993,Willerton:2008gyk,Jia:2026vcr}. For finite $G$, $\Gamma(\widetilde{\mathcal{A}}_0,L_{\Sigma_k})$ is simply the finite-dimensional vector space of sections, and no additional analytic completion is needed. For compact Lie groups, the analogous state space carries the appropriate measurable or $L^2$ completion, matching the continuous-category formulation studied in~\cite{Jia:2026vcr}.

The purpose of the notation $\psi^{\Sigma_k}$ is simply to pass from an abstract section of $L_{\Sigma_k}$ to an ordinary coefficient function in a chosen local frame. Concretely, let $\Psi_\psi\in \Gamma(\widetilde{\mathcal A}_0,L_{\Sigma_k})$ denote the section corresponding to the state $\ket{\psi}$. On a local patch $U_I\subset \widetilde{\mathcal A}_0$, choose a nowhere-vanishing frame $s_I$ for $L_{\Sigma_k}$. Then
\begin{equation}
    \Psi_\psi|_{U_I}
    =
    \psi^{\Sigma_k}_I(x)\,s_I(x),
    \qquad x\in U_I,
\end{equation}
for an ordinary complex-valued function $\psi^{\Sigma_k}_I$ on $U_I$. On an overlap $U_I\cap U_J$, the two local frames differ by the transition function of the line bundle,
\begin{equation}
    s_J(x)=g_{IJ}(x)\,s_I(x),
    \qquad g_{IJ}(x)\in U(1),
\end{equation}
so the local coefficients glue as
\begin{equation}\label{eq:twisted_wavefunction_gluing}
    \psi^{\Sigma_k}_J(x)
    =
    g_{IJ}(x)^{-1}\psi^{\Sigma_k}_I(x).
\end{equation}
This is the concrete sense in which the local wavefunctions glue into a global section of $L_{\Sigma_k}$. In the continuum description above, $g_{IJ}$ is the exponentiated prequantum transition function determined locally by $\alpha_J-\alpha_I=d\chi_{IJ}$, exactly as in the standard prequantization construction reviewed in~\cite{de2007symplectic}. More explicitly, if the local prequantum connection is written as $\nabla_I=d-ik\,\alpha_I$ in the same normalization conventions as above, then on overlaps one has
\begin{equation}
    \nabla_J=g_{IJ}^{-1}\nabla_I g_{IJ},
    \qquad
    g_{IJ}^{-1}dg_{IJ}=ik(\alpha_J-\alpha_I)=ik\,d\chi_{IJ},
\end{equation}
so locally
\begin{equation}
    g_{IJ}(x)=e^{ik\chi_{IJ}(x)}
\end{equation}
up to a constant phase; with different $2\pi$, $\hbar$, or sign conventions, the exponent is modified accordingly. In the groupoid description used here, the same gluing data is encoded by the Fell line bundle $\Sigma_k$ and ultimately by the transgressed class $\tau(k)$~\cite{FreedQuinn:1993,Willerton:2008gyk,Jia:2026vcr}. Suppressing the patch label once a trivialization has been fixed, we write the resulting local wavefunction as
\begin{equation}
    \psi^{\Sigma_k}(x)
    :=
    \braket{x(a),x(b)|\psi}_{BF+kCS}\,,
    \qquad x=(x(a),x(b))\in \mathcal{A}_0\,.
\end{equation}
Unlike the untwisted wavefunction of section~\ref{sec:HilbGG_on_HBF}, this is only a local representative of a section and therefore transforms by~\eqref{eq:twisted_wavefunction_gluing} under changes of trivialization. Fixing the $b$-cycle holonomy $v:=x(b)$, we write the corresponding sector wavefunction as
\begin{equation}
    \psi_v^{\Sigma_k}(u)
    :=
    \braket{u,v|\psi}_{BF+kCS}\,,
    \qquad u:=x(a)\in C_G(v)\,.
\end{equation}
Here one is simply restricting the same local section coefficient to the slice of the configuration space with fixed $b$-cycle holonomy $v$; the nontrivial bundle information is still carried by the same overlap functions $g_{IJ}(u,v)$. This is the discrete analogue of choosing the vertical polarization and then restricting to the $v$-sector of the configuration space. Having fixed the state space in this way, we can now describe how a twisted line operator acts on these sector wavefunctions.

\subsection{Twisted line operators on $H_{BF+kCS}(T^2)$}\label{sec:Twisted_HilbGG_on_HBF}

We now generalize the untwisted computations of sections~\ref{sec:Gauge_Inv_LineOp} and~\ref{sec:Stacking_LineOp} to the twisted Hilbert space defined in~(\ref{eq:H_BFkCS_sections}). The only new ingredient is that the centralizer action in each fixed-$v$ sector is projective, with multiplier determined by the restriction of a cocycle representative $\sigma_k$ of the transgressed class $\tau(k)$.

Fix again a $b$-cycle holonomy $v=x(b)$. Restricting the representative cocycle $\sigma_k$ to the isotropy group at $v$ gives a centralizer $2$-cocycle
\begin{equation}\label{eq:alpha_v_from_sigma}
    \alpha_v(h_2,h_1):=\sigma_k\big((h_2,v),(h_1,v)\big)\,,
    \qquad h_1,h_2\in C_G(v)\,.
\end{equation}
This is the direct analogue of~(\ref{eq:alpha_g_from_sigma}), now with the distinguished object of the conjugation groupoid chosen to be $v$ rather than the defect label $g$. The $v$-sector of the physical Hilbert space is therefore naturally a module over the twisted group algebra of $C_G(v)$ with multiplier $\alpha_v$.

To make the projective structure explicit, choose a local trivialization of the restricted line bundle over the $v$-sector. A wavefunction will then be written as $\psi^{\Sigma_k}_v(u)$ with $u=x(a)\in C_G(v)$. For each $h\in C_G(v)$ we introduce a projective transport phase
\begin{equation}
    \varrho_v(h;u\to hu)\in U(1)\,,
\end{equation}
which should be read simply as a function of the pair $(h,u)$: the notation after the semicolon is only a mnemonic indicating that this phase is attached to the translation arrow in the $v$-sector sending the point $u$ to the point $hu$. Here the map $u\mapsto hu$ is the left-translation action entering the convolution representation of the line operator, whereas the residual gauge symmetry inside the fixed-$v$ sector still acts by conjugation $u\mapsto k u k^{-1}$ for $k\in C_G(v)$, under which physical wavefunctions are class sections. We then define the corresponding projective translation operator by
\begin{equation}\label{eq:twisted_sector_translation}
    (L_h^{\Sigma_k}\psi_v^{\Sigma_k})(u):=
    \varrho_v(h;u\to hu)\,\psi_v^{\Sigma_k}(hu)\,.
\end{equation}
The consistency of these translations is controlled by the cocycle $\alpha_v$
\begin{equation}\label{eq:twisted_sector_translation_comp}
    L_{h_2}^{\Sigma_k}L_{h_1}^{\Sigma_k}
    =
    \alpha_v(h_2,h_1)\,L_{h_2h_1}^{\Sigma_k}\,,
\end{equation}
or equivalently,
\begin{equation}\label{eq:twisted_sector_theta_comp}
    \varrho_v(h_2;h_1u\to h_2h_1u)\,
    \varrho_v(h_1;u\to h_1u)
    =
    \alpha_v(h_2,h_1)\,
    \varrho_v(h_2h_1;u\to h_2h_1u)\,.
\end{equation}
They are the codimension-$1$ counterpart of the projective transport law~\eqref{eq:twisted_transport_composition}, \eqref{eq:theta_definition_twisted} and \eqref{eq:theta_sigma_consistency} in section~\ref{sec:Twisted_conj_groupoid_basis}. Similar to $\vartheta_g(u;i\rightarrow j)$ in \eqref{eq:vartheta_sigmak_relation}, $\varrho_v(h;u\rightarrow hu)$ can also be chosen as
    \begin{equation}\label{eq:varrho-and-sigmak}
        \varrho_v(h;u\rightarrow hu) = \sigma_k((h,v),(u,v)) = \alpha_v (h,u)\,,
    \end{equation}
and the consistency condition \eqref{eq:twisted_sector_theta_comp} is exactly the same to the cocycle condition of $\sigma_k$ in \eqref{eq:sigma_k_cocycle_condition}.

The line operators are now labeled by the twisted simple objects $([g],\rho_g)\in \hilb_G^{\tau(k)}(G)$, where $\rho_g$ is an $\alpha_g$-projective representation of $C_G(g)$. In a chosen trivialization \eqref{eq:varrho-and-sigmak}, the action of the corresponding physical line operator $v^{\Sigma_k}_{[g],\rho_g}$ on the $v$-sector takes the form
\begin{equation}\label{eq:def_rho_sigma_g}
    \bra{u,v}\rho^{\Sigma_k}(v^{\Sigma_k}_{[g],\rho_g})
    =
    \int_{[g]\cap C_G(v)} d\mu(g)\,
    \chi^{\alpha_g}_{\rho_g}(v)\,
    \alpha_v(g,u) \,
    \bra{gu,v}\,,
\end{equation}
cf.~the untwisted formula~(\ref{eq:rho_as_conv}). Here $\chi^{\alpha_g}_{\rho_g}(v)$ is the character of the $\alpha_g$-projective centralizer representation attached to the element $g\in[g]\cap C_G(v)$. Thus the untwisted shift of the $a$-cycle holonomy is replaced by the same shift multiplied by the projective phase dictated by the transgressed twist.

In the twisted theory, the basis vector $\langle u,v|$ should be identified under conjugation up to a phase determined by the cocycle representative of the Fell bundle
\begin{equation}
    \langle h u h^{-1}, h v h^{-1}| \,
    \sim 
    \langle u,v|\,.
\end{equation}
In the following, we will focus on fixed $v$-sector, where $h\in C_G(v)$. Let us consider the projective class function $\psi_v(u) = \langle u,v |\psi\rangle$ of $C_G(v)$. For the projective character $\chi_v(u)$ in the $v$-twist sector, we have
    \begin{equation}
        \chi_v(huh^{-1})
=
\frac{\alpha_v(huh^{-1},h)}{\alpha_v(h,u)}
\chi_v(u)\,,
    \end{equation}
regardless of the representation. We will also let 
    \begin{equation}\label{eq:twist-state-consistency}
        \langle huh^{-1},v| = \frac{\alpha_v(huh^{-1},h)}{\alpha_v(h,u)} \,\langle u,v|\,,
    \end{equation}
where we assume all projective class functions $\psi_v(u)=\langle u,v|\psi\rangle$ should be transported in the same way as projective characters. Consider the action of the line operator $v^{\Sigma_k}_{[g],\rho_g}$ on $\langle huh^{-1},v|$
\begin{equation}
\begin{split}
    &\langle huh^{-1},v|\rho^{\Sigma_k}(v^{\Sigma_k}_{[g],\rho_g})\\
=&
\int_{[g]\cap C_G(v)}
d\mu(g)\,
\chi_{\rho_g}^{\alpha_g}(v)\,
\alpha_v(g,huh^{-1})
\langle ghuh^{-1},v|\,,\\
=& \int_{[hgh^{-1}]\cap C_G(v)}
d\mu(hgh^{-1})\,
\chi_{\rho_{hgh^{-1}}}^{\alpha_{hgh^{-1}}}(v)\,
\alpha_v(hgh^{-1},huh^{-1})
\langle hguh^{-1},v|\\
=& \int_{[g]\cap C_G(v)}
d\mu(g)\,
\chi_{\rho_{hgh^{-1}}}^{\alpha_{hgh^{-1}}}(v)\,
\alpha_v(hgh^{-1},huh^{-1})
\frac{\alpha_v(hguh^{-1},h)}{\alpha_v(h,gu)}\langle gu,v|\,.
\end{split}
\end{equation}
In order to reconcile with \eqref{eq:def_rho_sigma_g}, one must require
    \begin{equation}\label{eq:property-of-character}
        \chi_{\rho_{hgh^{-1}}}^{\alpha_{hgh^{-1}}}(v) = \frac{\alpha_v(h,g)}{\alpha_v(hgh^{-1},h)} \chi_{\rho_{g}}^{\alpha_{g}}(v)\,,
    \end{equation}
and we leave the proof to the appendix\footnote{The proof relied on the fact that $\sigma_k$ can be written using the 3-cocycles $\omega\in Z^3(G,U(1))$ as \begin{equation}
    \sigma_k((u,x),(s,z))
    =
    \omega(usz s^{-1}u^{-1},u,s)\,
    \omega(u,x,s)^{-1}\,
    \omega(u,s,z),
\end{equation}
see, for example, (5.15) in \cite{Coste:2000tq} for finite group. We will assume it also generalizes to compact Lie groups.}. Then the action $\langle huh^{-1},v|\rho^{\Sigma_k}(v^{\Sigma_k}_{[g],\rho_g})$ is
\begin{equation}\label{eq:twist-action-cocycle-factor}
    \int_{[g]\cap C_G(v)}
d\mu(g)\, \chi_{\rho_{g}}^{\alpha_{g}}(v)\,\frac{\alpha_v(h,g)}{\alpha_v(hgh^{-1},h)}
\alpha_v(hgh^{-1},huh^{-1})
\frac{\alpha_v(hguh^{-1},h)}{\alpha_v(h,gu)}\langle gu,v|\,.
\end{equation}
Recall the $2$-cocycle identity
    \begin{equation}
\alpha_v(a,b)\alpha_v(ab,c)
=
\alpha_v(b,c)\alpha_v(a,bc)\,.
    \end{equation}
Apply the identity to $(a,b,c)=(hgh^{-1},huh^{-1},h)$, one has
    \begin{equation}
        \alpha_v(hgh^{-1},huh^{-1})=\frac{\alpha_v(huh^{-1},h) \alpha_v(hgh^{-1},hu)}{\alpha_v(hguh^{-1},h)}\,.
    \end{equation}
Then apply the identity to $(a,b,c)=(hgh^{-1},h,u)$, one has
    \begin{equation}
        \alpha_v(hgh^{-1},hu) = \frac{\alpha_v(hgh^{-1},h) \alpha_v(hg,u)}{\alpha_v(h,u)}\,,
    \end{equation}
and substituting it into the previous one gives
    \begin{equation}
        \alpha_v(hgh^{-1},huh^{-1})=\frac{\alpha_v(huh^{-1},h) \alpha_v(hgh^{-1},h) \alpha_v(hg,u)}{\alpha_v(hguh^{-1},h)\alpha_v(h,u)}\,.
    \end{equation}
Finally, apply the cocycle identity to $(a,b,c)=(h,g,u)$, it gives
    \begin{equation}
        \alpha_v(hg,u) = \frac{\alpha_v(g,u)\alpha_v(h,gu)}{\alpha_v(h,g)}\,.
    \end{equation}
Substituting it once more gives
    \begin{equation}
        \alpha_v(hgh^{-1},huh^{-1})=\frac{\alpha_v(huh^{-1},h) \alpha_v(hgh^{-1},h) \alpha_v(g,u)\alpha_v(h,gu)}{\alpha_v(hguh^{-1},h)\alpha_v(h,u)\alpha_v(h,g)}\,.
    \end{equation}
The cocycle factors in \eqref{eq:twist-action-cocycle-factor} is then simplified to
    \begin{equation}
\frac{\alpha_v(h,g)}{\alpha_v(hgh^{-1},h)}
\alpha_v(hgh^{-1},huh^{-1})
\frac{\alpha_v(hguh^{-1},h)}{\alpha_v(h,gu)}\\
=\frac{\alpha_v(huh^{-1},h) \alpha_v(g,u)}{\alpha_v(h,u)}\,,
    \end{equation}
and thus \eqref{eq:twist-action-cocycle-factor} gives
\begin{equation}
\begin{split}
    \langle huh^{-1},v|\rho^{\Sigma_k}(v^{\Sigma_k}_{[g],\rho_g})=&\int_{[g]\cap C_G(v)}
d\mu(g)\, \chi_{\rho_{g}}^{\alpha_{g}}(v)\,\frac{\alpha_v(huh^{-1},h) \alpha_v(g,u)}{\alpha_v(h,u)}\langle gu,v|\\
=&\frac{\alpha_v(huh^{-1},h)}{\alpha_v(h,u)}\langle u,v|\rho^{\Sigma_k}(v^{\Sigma_k}_{[g],\rho_g})\,,
\end{split}
\end{equation}
which is consistent with \eqref{eq:twist-state-consistency}. 

It is convenient to encode the same action invariantly by using the twisted group algebra of $C_G(v)$. Let $\delta_{g}$ denote the basis section supported at $g\in C_G(v)$, with twisted multiplication
\begin{equation}\label{eq:centralizer_twisted_delta_conv}
    \delta_{g_2}*_{\alpha_v}\delta_{g_1}
    =
    \alpha_v(g_2,g_1)\,\delta_{g_2g_1}\,,
\end{equation}
with $\delta_{g}(h) = \delta^{C_G(v)}_{g,h}$ for $h\in C_G(v)$. We then define the twisted kernel section
\begin{equation}\label{eq:twisted_define_kernel}
    \mathcal{K}^{[g],\rho_g;\Sigma_k}_v
    :=
    \int_{[g]\cap C_G(v)} d\mu(g)\,
    \chi^{\alpha_g}_{\rho_g}(v)\,
    \delta_{g^{-1}}\,,
\end{equation}
with
\begin{equation}
    \mathcal{K}^{[g],\rho_g;\Sigma_k}_v(h)
    =
    \int_{[g]\cap C_G(v)} d\mu(g)\,
    \chi^{\alpha_g}_{\rho_g}(v)\,
    \delta_{g,h^{-1}}^{C_G(v)}\,.
\end{equation}
The action of the twisted line operator on the $v$-sector is simply twisted convolution by this kernel
\begin{equation}\label{eq:twisted_rho_as_convolution}
    (\rho^{\Sigma_k}(v^{\Sigma_k}_{[g],\rho_g})\psi_v^{\Sigma_k})(u)
    =
    (\mathcal{K}^{[g],\rho_g;\Sigma_k}_v*_{\alpha_v}\psi_v^{\Sigma_k})(u)\,,
\end{equation}
where the convolution is evaluated as
    \begin{equation}
        (\mathcal{K}^{[g],\rho_g;\Sigma_k}_v*_{\alpha_v}\psi_v^{\Sigma_k})(u) = \int_{C_G(v)} dh\, \alpha_v(h,u)\,  \mathcal{K}^{[g],\rho_g;\Sigma_k}_v(h^{-1}) \psi_v(hu)\,.
    \end{equation}
In a local trivialization,~(\ref{eq:twisted_rho_as_convolution}) is just a restatement of~(\ref{eq:def_rho_sigma_g}) together with the projective translation law~(\ref{eq:twisted_sector_translation}).

The twisted kernel $\mathcal{K}^{[g],\rho_g;\Sigma_k}_v(h)$ is a twisted class function on $C_G(v)$. Under the conjugation $h\rightarrow \ell h \ell^{-1}$
\begin{equation}
\begin{split}
    &\mathcal{K}^{[g],\rho_g;\Sigma_k}_v(\ell h \ell^{-1})\\
    =&
    \int_{[g]\cap C_G(v)} d\mu(g)\,
    \chi^{\alpha_g}_{\rho_g}(v)\,
    \delta_{g,\ell h^{-1} \ell^{-1}}^{C_G(v)}\,\\
    =&\int_{[g]\cap C_G(v)} d\mu(\ell g \ell^{-1})\,
    \chi^{\alpha_{\ell g \ell^{-1}}}_{\rho_{\ell g \ell^{-1}}}(v)\,
    \delta_{\ell g \ell^{-1},\ell h^{-1} \ell^{-1}}^{C_G(v)}\,\\
    =&\int_{[g]\cap C_G(v)} d\mu(g)\,\frac{\alpha_v(\ell,g)}{\alpha_v(\ell g \ell^{-1},\ell)}
    \chi^{\alpha_g}_{\rho_g}(v)\,
    \delta_{g,h^{-1}}^{C_G(v)} \,,
\end{split}
\end{equation}
and thus
    \begin{equation}
         \mathcal{K}^{[g],\rho_g;\Sigma_k}_v(\ell h \ell^{-1}) =  \frac{\alpha_v(\ell,h^{-1})}{\alpha_v(\ell h^{-1} \ell^{-1},\ell)} \mathcal{K}^{[g],\rho_g;\Sigma_k}_v(h)\,,
    \end{equation}
which indicates $\mathcal{K}^{[g],\rho_g;\Sigma_k}_v(h)$ transform in the same way as the dual characters $\overline{\chi^{\alpha_v}_v(h^{-1})}$.

Therefore the entire untwisted discussion goes through with the same logical structure after the replacements
\begin{equation}
    Cl(C_G(v))
    \longrightarrow
    \text{$\alpha_v$-twisted class sections on } C_G(v),
    \qquad
    *
    \longrightarrow
    *_ {\alpha_v}\,.
\end{equation}
Here ``$\alpha_v$-twisted class section'' means a section which is central with respect to the twisted convolution algebra $({\mathbb C}^{\alpha_v}[C_G(v)],*_{\alpha_v})$; in a chosen trivialization this is the projective analogue of being a class function. Accordingly, if $R_v$ is an irreducible $\alpha_v$-projective representation of $C_G(v)$, we denote by
\begin{equation}
    \chi^{\alpha_v}_{R_v}(u):=\tr\big(\rho_{R_v}(u)\big)
\end{equation}
its projective character. The twisted kernel may then be expanded as
\begin{equation}
    \mathcal{K}^{[g],\rho_g;\Sigma_k}_v(h)
    =
    \sum_{R_v\in \mathrm{Irrep}_{\alpha_v}(C_G(v))}
    K^{[g],\rho_g;\Sigma_k}_{R_v}\,
    \overline{\chi^{\alpha_v}_{R_v}(h^{-1})}\,.
\end{equation}
Using the orthogonal relations for projective characters\footnote{
For finite group, see Theorem 11.8 in Part.I, Chapter 1 of \cite{karpilovsky1993group}. And for compact Lie group, one may consult the Peter–Weyl theorem in the twisted case discussed in \cite{Cheng2024}.
}
\begin{equation}
    \frac{1}{\text{Vol}(C_G(v))}\int_{C_G(v)} dg \,\alpha(g,h)\,\overline{\chi_{R_v}^{\alpha_v}(g)}\, \chi_{R'_v}^{\alpha_v} (gh) = \delta_{R_v,R'_v} \frac{1}{\dim R_{v}} \chi_{R_v}(h)\,,
\end{equation}
we have
    \begin{equation}
    \begin{split}
        &\left(\mathcal{K}^{[g],\rho_g;\Sigma_k}_v*_{\alpha_v}\chi_{R_v}^{\alpha_v}\right)(u)\\ =&  \int_{C_G(v)} dh\, \alpha_v(h,u)\, \mathcal{K}^{[g],\rho_g;\Sigma_k}_v(h^{-1}) \chi_{R_v}^{\alpha_v}(hu)\\
        =&\sum_{R'_v\in \mathrm{Irrep}_{\alpha_v}(C_G(v))} K^{[g],\rho_g;\Sigma_k}_{R'_v} \int_{C_G(v)} dh\, \alpha_v(h,u)\, \overline{\chi^{\alpha_v}_{R'_v}(h)}\chi_{R_v}^{\alpha_v}(hu)\\
        =& \frac{\text{Vol}(C_G(v)) K^{[g],\rho_g;\Sigma_k}_{R_v}}{\dim R_v} \chi^{\alpha_v}_{R_v}(u):= \lambda^{[g],\rho_g;\Sigma_k}_{R_v} \chi^{\alpha_v}_{R_v}(u)\,,
    \end{split}
    \end{equation}
which gives
\begin{equation}\label{eq:twisted_conv_with_projective_char}
    (\mathcal{K}^{[g],\rho_g;\Sigma_k}_v*_{\alpha_v}\chi^{\alpha_v}_{R_v})(u)
    =
    \lambda^{[g],\rho_g;\Sigma_k}_{R_v}\,
    \chi^{\alpha_v}_{R_v}(u)\,,
\end{equation}
so the projective characters of irreducible $\alpha_v$-projective representations diagonalize the twisted convolution operator~(\ref{eq:twisted_rho_as_convolution}) exactly as ordinary characters diagonalized~(\ref{eq:rho_as_convolution}) in the pure $BF$ theory.

Finally, stacking line operators on the $b$-cycle is again represented by convolution of their kernels, but now in the twisted algebra:
\begin{equation}\label{eq:twisted_kernel_stacking}
    \rho^{\Sigma_k}(v^{\Sigma_k}_{[g],\rho_g})\,
    \rho^{\Sigma_k}(v^{\Sigma_k}_{[h],\rho_h})
    \;\;\Longleftrightarrow\;\;
    \mathcal{K}^{[g],\rho_g;\Sigma_k}_v
    *_{\alpha_v}
    \mathcal{K}^{[h],\rho_h;\Sigma_k}_v\,.
\end{equation}
Thus the effect of the level-$k$ Chern-Simons term is not to change the underlying commuting-holonomy configuration space, but to promote the centralizer action in each $v$-sector from an honest representation to a projective one, with the failure of strict associativity measured precisely by the restriction~(\ref{eq:alpha_v_from_sigma}) of the transgressed class $\tau(k)$, represented locally by $\sigma_k$.

\section{The Transgression of Chern-Simons Level}\label{sec:Transgression_of_CS_Level}

The brief discussion in section~\ref{sec:Twisted_quantization_BFkCS}, especially around~(\ref{eq:BFkCS_continuum_hilbert})--(\ref{eq:H_BFkCS_sections}), already suggests that the codimension-$2$ twist $\tau(k)$ and the codimension-$1$ line bundle governing the Hilbert space come from the same level
\begin{equation}
    k\in H^4(BG,\mathbb{Z}).
\end{equation}
Since this point is conceptually important, we spell it out here for a general closed oriented spatial surface $\Sigma$, beyond the discussion for $\Sigma=T^2$. The essential answer is that the two structures are not literally the same geometric object: the codimension-$2$ datum is one categorical degree higher than the codimension-$1$ datum. Nevertheless, they are two transgressions of the same universal Chern-Simons level.

\subsection{Generalities of Transgression}

Let $\operatorname{Loc}_G(M)$ denote the moduli stack of flat $G$-bundles on a closed oriented manifold $M$. The basic transgression construction is first defined on the full mapping stack $\operatorname{Map}(M,BG)$ by pullback along the evaluation map and then integration over the $M$-fiber
\begin{equation}
    \mathrm{ev}:M\times \operatorname{Map}(M,BG)\longrightarrow BG,
    \qquad
    \pi:M\times \operatorname{Map}(M,BG)\longrightarrow \operatorname{Map}(M,BG).
\end{equation}
For a cohomology class $k\in H^4(BG,\mathbb{Z})$, one defines
\begin{equation}\label{eq:topological_transgression_degree_drop}
    \mathrm{tr}_M(k)
    :=
    \pi_!\,\mathrm{ev}^*(k)
    \in
    H^{4-\dim M}\big(\operatorname{Map}(M,BG),\mathbb{Z}\big),
\end{equation}
where $\pi_!$ denotes the Gysin pushforward, or fiber-integration map, along the oriented closed manifold $M$ and $\mathrm{tr}_M$ stands for transgression over $M$, not to be confused with trace~\cite{Brylinski:2000dcg, Hopkins:2002rd, brylinski2007loop, kishimoto2010cohomology}. Restricting this class to the flat-field locus gives the corresponding class on $\operatorname{Loc}_G(M)$, and it is this restricted class that governs Chern-Simons/BF quantization. In this sense, transgression along $M$ lowers the cohomological degree by $\dim M$. For $M=S^1$ one obtains a degree-$3$ class on the loop stack, while for a closed oriented surface $\Sigma$ one obtains a degree-$2$ class on the moduli stack of flat fields on $\Sigma$~\cite{Carey:2004xt,FreedQuinn:1993,Jia:2026vcr}.

To discuss curvature one chooses a differential refinement of $k$, for example the multiplicative gerbe or Chern-Simons $2$-gerbe associated to the same level~\cite{meinrenken2002basic,Carey:2004xt}. The differential transgression again lowers degree by $\dim M$: along a circle one obtains a gerbe-type object with $3$-curvature, whereas along a surface one obtains an ordinary line bundle with $2$-curvature. This degree drop is the basic reason the codimension-$2$ and codimension-$1$ structures are closely related but not identical.

\subsection{Circle Transgression and Codimension-2 Data}

The moduli stack of flat $G$-bundles on a circle is the adjoint quotient stack
\begin{equation}
    \operatorname{Loc}_G(S^1)\simeq [G/G]\simeq G//_{\mathrm{Ad}}G
\end{equation}
where in the last step by $\simeq$ we mean that the action groupoid $G//_{\Ad}G$ presents the stack $[G/G]$. Therefore transgression of $k$ along $S^1$ gives a degree-$3$ class on the conjugation stack
\begin{equation}
    \mathrm{tr}_{S^1}(k)\in H^3([G/G],\mathbb{Z}),
\end{equation}
which, in the groupoid language used in sections~\ref{sec:Twisted_conj_groupoid_basis} and~\ref{sec:Twisted_HilbGG_on_HBF}, is equivalently described by the class
\begin{equation}\label{eq:circle_transgression_tau}
    \tau(k)\in H^2(G//_{\rm Ad}G,U(1))
\end{equation}
after passing to $U(1)$-valued groupoid cohomology via the exponential sequence~\cite{Behrend:2008,Stienon:2010,Tu:2009twistedRing}. Choosing a cocycle representative
\begin{equation}
    \sigma_k\in Z^2(G//_{\rm Ad}G,U(1))
\end{equation}
and the associated Fell line bundle
\begin{equation}
    \Sigma_k\longrightarrow G//_{\rm Ad}G
\end{equation}
is precisely the codimension-$2$ twisting datum used above~\cite{Jia:2026vcr}.

For compact Lie groups with differential refinement, this codimension-$2$ datum is one categorical degree higher than an ordinary line bundle: its invariant curvature is a gerbe $3$-curvature, equivalently an equivariant Dixmier-Douady class in $H^3_G(G,\mathbb{Z})$, not an ordinary magnetic $2$-form on a moduli space~\cite{meinrenken2002basic,Carey:2004xt}. Thus $\tau(k)$ and $\sigma_k$ should be understood as higher transition or holonomy data. In particular, they are not themselves the $2$-form curvature that appears in the prequantization over the phase space associated with a spatial manifold.

\subsection{Surface Transgression and Codimension-1 Data}

Now let $\Sigma$ be a closed oriented spatial surface. Transgression of the same universal level along $\Sigma$ produces an ordinary degree-$2$ class
\begin{equation}
    \mathrm{tr}_{\Sigma}(k)\in H^2\big(\operatorname{Loc}_G(\Sigma),\mathbb{Z}\big).
\end{equation}
After choosing a differential refinement, this class is realized by a genuine line bundle with connection
\begin{equation}\label{eq:surface_transgressed_line_bundle}
    \mathcal L_{\Sigma,k}\longrightarrow \operatorname{Loc}_G(\Sigma),
\end{equation}
which is the Chern-Simons prequantum line bundle over the moduli stack of flat fields on $\Sigma$~\cite{Axelrod:1989xt,Witten:1988hf,Carey:2004xt}. On the smooth locus of the coarse moduli space
\begin{equation}
    \mathcal M_{\rm flat}(\Sigma,G)\subset \operatorname{Loc}_G(\Sigma),
\end{equation}
its curvature is the Atiyah-Bott-Goldman symplectic form multiplied by the level
\begin{equation}\label{eq:general_surface_curvature}
    F_{\nabla_{\Sigma,k}}
    =
    2\pi i\, k\,\omega_{AB,\Sigma},
\end{equation}
with
\begin{equation}\label{eq:general_surface_AB_form}
    \omega_{AB,\Sigma}([\alpha],[\alpha'])
    :=
    \frac{1}{2\pi}\int_{\Sigma}\langle \alpha\wedge \alpha'\rangle
\end{equation}
in the same normalization conventions as section~\ref{sec:Twisted_quantization_BFkCS}. In pure level-$k$ Chern-Simons theory this line bundle is quantized in a K\"ahler polarization to produce holomorphic sections, while in the mixed $BF+kCS$ theory it reappears in the vertical polarization of the magnetic cotangent bundle (see the discussion in the paragraph around~\eqref{eq:Phase_Spacs_CS}), leading schematically to
\begin{equation}
    \mathcal H_{BF+kCS}(\Sigma)
    \sim
    L^2\big(\mathcal M_{\rm flat}(\Sigma,G),\mathcal L_{\Sigma,k}\big)
\end{equation}
on the smooth locus, up to the same half-density and singularity caveats discussed earlier.

For $\Sigma=T^2$, the groupoid presentation $\widetilde{\mathcal A}_0$ used in section~\ref{sec:Twisted_quantization_BFkCS} is a concrete model for $\operatorname{Loc}_G(T^2)$~\footnote{Namely, for $\Sigma = T^2$ we define $\mathcal{A}_0 := \{ (u,v)\in G\times G | uv = vu \}$ and group action $g\cdot(u,v) = (gug^{-1}, gvg^{-1})$. Then $\widetilde{\mathcal{A}}_0 = \mathcal{A}_0//G$.}, and the line bundle denoted there by
\begin{equation}
    L_{\Sigma_k}\longrightarrow \widetilde{\mathcal A}_0
\end{equation}
is precisely the specialization of the general surface-transgressed line bundle~(\ref{eq:surface_transgressed_line_bundle}) to the torus presentation. In the finite-group setting this is the Freed-Quinn line bundle obtained by transgressing the Dijkgraaf-Witten cocycle to the groupoid of flat fields on $\Sigma$~\cite{FreedQuinn:1993,Willerton:2008gyk}; in the compact Lie-group setting it is the usual Chern-Simons prequantum line bundle with curvature~(\ref{eq:general_surface_curvature}).

\subsection{Summary}

The relationship between the codimension-$2$ and codimension-$1$ data of the symmetry category, or equivalently the SymTFT, can therefore be summarized as
\begin{equation}\label{eq:one_k_two_transgressions}
    \ba
    k\in H^4(BG,\mathbb{Z})
    &\quad\rightsquigarrow\quad
    \left\{
    \begin{array}{l}
        \delta\big(\tau(k)\big)=\mathrm{tr}_{S^1}(k),\ \ \tau(k)\in H^2(G//_\Ad G,U(1)), \\[4pt]
        \mathrm{tr}_{\Sigma}(k)=c_1(\mathcal L_{\Sigma,k})\in H^2(\operatorname{Loc}_G(\Sigma),\mathbb{Z}).
    \end{array}
    \right.
    \ea
\end{equation}
Here $\delta:H^2(G//_\Ad G,U(1))\xrightarrow{\sim} H^3([G/G],\mathbb{Z})$ is the isomorphism induced by the exponential sequence, so if one implicitly identifies these two cohomology groups via $\delta$, one may simply write $\tau(k)=\mathrm{tr}_{S^1}(k)$. The first line is the gerbe/Fell-line-bundle twist on the loop stack, written in the groupoid language used throughout this paper. The second line is the first Chern class of the prequantum line bundle $\mathcal L_{\Sigma,k}\to \operatorname{Loc}_G(\Sigma)$.
Thus the two structures at different codimensions come from the same origin in the sense that they descend from the same universal level, namely $k\in H^4(BG,\mathbb{Z})$. However, they are clearly not the same as geometric objects: the circle transgression gives a degree-$3$ gerbe-type twist whose cocycle-level description is $\sigma_k$, whereas the surface transgression gives a degree-$2$ line bundle whose curvature is the magnetic form $k\,\omega_{AB,\Sigma}$.

Put differently, the codimension-$1$ line bundle is the ordinary line-bundle ``shadow'' of the same higher twist that governs the categorical data at codimension-$2$. In the compact Lie-group picture, the circle transgression carries a gerbe $3$-curvature while the surface transgression carries the ordinary $2$-curvature~(\ref{eq:general_surface_curvature}); in the finite-group picture there are no differential forms, and the same relation is encoded entirely at the level of groupoid cocycles and transgressed line bundles~\cite{FreedQuinn:1993,Willerton:2008gyk,Jia:2026vcr}. This is the precise sense in which the codimension-$2$ datum $\tau(k)$ and the codimension-$1$ line bundle $\mathcal L_{\Sigma,k}$ are governed by the same level $k$, while living in adjacent degrees.

\section{Comparison with Known Cases}\label{sec:comparison_known_cases}

The convolution-eigenvalue formalism developed in sections~\ref{sec:HilbGG_on_HBF} and~\ref{sec:Twisted_HilbGG_on_HBF_section} reconstructs, in the finite-group case, well-known modular data of the (twisted) Drinfeld double, while in the compact Lie-group case it provides explicit kernels in a regime where no closed-form modular $S$-matrix is available. In this section we make these statements precise and locate the present work within the existing literature.

The guiding physical intuition is that a line operator labeled by $([g],\rho_g)$ acting on the torus sector labeled by $([h],\rho_h)$ should produce the normalized Hopf-link amplitude between those two labels. In the finite-group notation used below, the corresponding eigenvalue is $\lambda_{\rho_h}^{[g],\rho_g}$, or $\lambda_{\rho_h}^{[g],\rho_g;\Sigma_k}$ in the twisted theory. Concretely, one may prepare the sector $([h],\rho_h)$ by inserting that line along the core of a solid torus; acting by the line $([g],\rho_g)$ along the dual cycle then produces a Hopf link. The unnormalized amplitude is therefore the modular kernel $S_{([g],\rho_g),([h],\rho_h)}$, and the eigenvalue should be obtained by dividing by the vacuum amplitude in the same sector:
\begin{equation}\label{eq:guiding_lambda_S_principle}
    \lambda_{\rho_h}^{[g],\rho_g}
    \;=\;
    \left(
    \frac{
    S_{([g],\rho_g),([h],\rho_h)}
    }{
    S_{([e],\mathbf{1}),([h],\rho_h)}
    }
    \right)^*,
\end{equation}
with the evident replacement $S\to S^{\omega}$ and $\lambda_{\rho_h}^{[g],\rho_g}\to\lambda_{\rho_h}^{[g],\rho_g;\Sigma_k}$ in the twisted case. In the regular sector of compact Lie group we have the same schematic statement. The complex conjugation records the bra/ket and orientation conventions used in our convolution action. The point of the comparisons below is therefore to test whether the explicit convolution action on $H_{BF}(T^2)$ or $H_{BF+kCS}(T^2)$ computes the same Hopf-link pairing as the modular-data description. For finite groups this principle becomes a theorem, because $\Rep(D^\omega(G))$ is a modular tensor category and the Verlinde diagonalization identifies fusion eigenvalues with normalized $S$-matrix columns. For compact Lie groups we use the same principle more cautiously: in the regular sector the convolution kernel and the Hopf-link path-integral kernel should agree after the appropriate normalization, whereas a full continuous Verlinde theorem would require additional harmonic analysis that lies beyond the scope of this paper.

\subsection{Untwisted Finite Groups: Drinfeld-Double Modular Data}\label{sec:cmp_untwisted_finite}

We specialize section~\ref{sec:Stacking_LineOp} to a finite group $G$ equipped with the counting measure on $[g]\cap C_G(v)$. Combining the kernel definition~(\ref{eq:Define_Kernel}) with the general convolution-with-character identity~(\ref{eq:general_conv_with_chi}) gives the eigenvalue~(\ref{eq:Convolution_with_ChiR}) in the explicit closed form
\begin{equation}\label{eq:lambda_finite_unfolded}
    \lambda_{R_v}^{[g],\rho_g}
    =
    \frac{1}{\dim R_v}
    \sum_{x\in [g]\cap C_G(v)}
    \chi_{\rho_x}(v)\,\chi_{R_v}(x)\,.
\end{equation}

On the other hand, the modular $S$-matrix of the untwisted Drinfeld double $D(G)$, in the basis of simple objects $([g],\rho_g)\in \Rep(D(G))\simeq \hilb_G(G)$, is given by the standard finite-group modular formula~\cite{Coste:2000tq}
\begin{equation}\label{eq:CGR_untwisted_S}
    S_{([g],\rho_g),([v],R_v)}
    =
    \frac{1}{|C_G(v)|}
    \sum_{x\in [g]\cap C_G(v)}
    \chi_{\rho_x}(v)^*\,\chi_{R_v}(x)^*\,,
\end{equation}
together with the normalization
\begin{equation}\label{eq:CGR_S_zero}
    S_{([e],\mathbf{1}),([v],R_v)}=\frac{\dim R_v}{|C_G(v)|}\,.
\end{equation}
Comparing~(\ref{eq:lambda_finite_unfolded}) with the ratio of~(\ref{eq:CGR_untwisted_S}) and~(\ref{eq:CGR_S_zero}) yields the identification
\begin{equation}\label{eq:lambda_equals_S_ratio}
    \lambda_{R_v}^{[g],\rho_g}
    =
    \left(
    \frac{S_{([g],\rho_g),([v],R_v)}}{S_{([e],\mathbf{1}),([v],R_v)}}
    \right)^*\,,
\end{equation}
in the conventions used throughout this paper. The complex conjugation is conventional and disappears under any of the equivalent choices $\chi_R(x)\mapsto \chi_R(x^{-1})$, contragredient representations, or the opposite kernel orientation.

The identification~(\ref{eq:lambda_equals_S_ratio}) has two consequences. First, the eigenvalue identity
\begin{equation}\label{eq:cmp_eigenvalue_identity}
    \lambda^{[g],\rho_g}_{R_v}\,\lambda^{[h],\rho_h}_{R_v}
    =
    \sum_{[k],\rho_k} N^{[k],\rho_k}_{([g],\rho_g),([h],\rho_h)}\,\lambda^{[k],\rho_k}_{R_v}\,,
\end{equation}
posed at~(\ref{eq:To_Prove}) as the obstruction to $\rho:\mathcal{L}\to \mathrm{Aut}(H_{BF})$ preserving the tensor product, is the Verlinde formula~\cite{Verlinde:1988te,etingof2017tensor} for the modular tensor category $\Rep(D(G))=\mathcal{Z}(\mathrm{Vec}_G)$. To make this explicit, recall the standard form of the Verlinde formula for a modular tensor category with simple objects $\{X_i\}$, modular $S$-matrix $S_{ij}$, identity object $X_0$, and fusion coefficients $N_{ij}^k$
\begin{equation}\label{eq:Verlinde_standard}
    N_{ij}^k
    =
    \sum_{n}
    \frac{S_{in}\,S_{jn}\,S_{kn}^*}{S_{0n}}\,.
\end{equation}
Multiplying both sides by $S_{kv}$, summing over $k$, and using the unitarity relation $\sum_k S_{kn}^*\,S_{kv}=\delta_{nv}$, one obtains the equivalent eigenvalue form
\begin{equation}\label{eq:Verlinde_eigenvalue_form}
    \frac{S_{iv}\,S_{jv}}{S_{0v}}
    =
    \sum_{k} N_{ij}^k\,S_{kv}\,.
\end{equation}
Dividing both sides by $S_{0v}$ and defining
\begin{equation}\label{eq:Verlinde_eigenvalue_def}
    \tilde\lambda_v^{i}:=\frac{S_{iv}}{S_{0v}}\,,
\end{equation}
this becomes
\begin{equation}\label{eq:Verlinde_eigenvalue_diagonal}
    \tilde\lambda_v^{i}\,\tilde\lambda_v^{j}
    =
    \sum_{k} N_{ij}^k\,\tilde\lambda_v^{k}\,,
\end{equation}
which is the multiplicative diagonalization of fusion in the $S$-matrix basis. Specializing the abstract labels to $i=([g],\rho_g)$, $j=([h],\rho_h)$, $k=([k],\rho_k)$, $v=([v],R_v)$ for $\Rep(D(G))$, and using~(\ref{eq:lambda_equals_S_ratio}) to identify $\lambda_{R_v}^{[g],\rho_g}=(\tilde\lambda_v^{i})^*$, the identity~(\ref{eq:Verlinde_eigenvalue_diagonal}) becomes precisely~(\ref{eq:cmp_eigenvalue_identity}) (after taking the complex conjugate of both sides, using the realness of $N_{ij}^k$). Thus~(\ref{eq:cmp_eigenvalue_identity}) is the Verlinde formula in its eigenvalue form for $\Rep(D(G))$, and is therefore a theorem in the finite-group untwisted case by the modularity of $D(G)$. Second, the $S_3$ kernel tables of section~\ref{sec:Examples} are equivalent, after the conventional conjugation in~(\ref{eq:lambda_equals_S_ratio}), to the modular data of $D(S_3)$ tabulated in~\cite{Coste:2000tq}.

\subsection{Twisted Finite Groups: $D^\omega(G)$}\label{sec:cmp_twisted_finite}

The twisted finite-group comparison is slightly more delicate than the untwisted one, because projective characters are not intrinsic functions until a cocycle representative and compatible transport convention have been fixed. We therefore use the same cocycle representative as in the previous twisted sections. Namely, choose the finite-group representative $\sigma_k$ of the Fell-line-bundle twist to be the transgression of a normalized Dijkgraaf--Witten cocycle $\omega\in Z^3(G,U(1))$. Its isotropy restriction at $a\in G$ is the centralizer cocycle already defined in~(\ref{eq:alpha_g_from_sigma})
\begin{equation}\label{eq:twisted_finite_alpha_from_sigma}
    \alpha_a(c_1,c_2)
    =
    \sigma_k\big((c_1,a),(c_2,a)\big),
    \qquad c_1,c_2\in C_G(a).
\end{equation}
With this convention fixed once and for all, the fiberwise projective character $\chi^{\alpha_x}_{\rho_x}$ at each $x\in[g]$ and the centralizer projective character $\chi^{\alpha_v}_{R_v}$ at $v$ are well-defined functions on $C_G(x)$ and $C_G(v)$ respectively, transported from the corresponding reference fibers by the same $\sigma_k$, with covariance law fixed by the consistency identity discussed around~(\ref{eq:twist-state-consistency}).

Now specialize the twisted convolution-eigenvalue identity~(\ref{eq:twisted_conv_with_projective_char}) to a finite group $G$ with counting measure on $[g]\cap C_G(v)$. Substituting the kernel definition~(\ref{eq:twisted_define_kernel}) into the twisted convolution gives
\begin{equation}\label{eq:conv_with_projective_char_unfolded}
    (\mathcal{K}^{[g],\rho_g;\Sigma_k}_v*_{\alpha_v}\chi^{\alpha_v}_{R_v})(u)
    =
    \sum_{x\in[g]\cap C_G(v)}\alpha_v(x,u)\,\chi^{\alpha_x}_{\rho_x}(v)\,\chi^{\alpha_v}_{R_v}(xu)\,,
\end{equation}
which by~(\ref{eq:twisted_conv_with_projective_char}) equals $\lambda_{R_v}^{[g],\rho_g;\Sigma_k}\,\chi^{\alpha_v}_{R_v}(u)$. Evaluating both sides at $u=e$, using the normalization $\alpha_v(x,e)=1$ together with $\chi^{\alpha_v}_{R_v}(e)=\dim R_v$, yields the explicit form
\begin{equation}\label{eq:lambda_twisted_unfolded_finite}
    \lambda_{R_v}^{[g],\rho_g;\Sigma_k}
    =
    \frac{1}{\dim R_v}
    \sum_{x\in [g]\cap C_G(v)}
    \chi^{\alpha_x}_{\rho_x}(v)\,
    \chi^{\alpha_v}_{R_v}(x)\,.
\end{equation}
This is the exact projective analogue of~(\ref{eq:lambda_finite_unfolded}). The first character is the fiberwise $\alpha_x$-projective trace appearing as the coefficient in the kernel~(\ref{eq:twisted_define_kernel}), and the second is the $\alpha_v$-projective character of the $v$-sector representation; both are fixed by the same $\sigma_k$.

On the other hand, the modular $S$-matrix of the twisted Drinfeld double $D^\omega(G)$ is the Hopf-link trace in the modular tensor category $\mathrm{Rep}\,D^\omega(G)$~\cite{Roche:1990hs,Coste:2000tq}. Once a cocycle representative $\sigma_k$ has been fixed, so that the projective characters $\chi^{\alpha_x}_{\rho_x}$ and $\chi^{\alpha_v}_{R_v}$ are well-defined functions, this Hopf-link trace takes the projective-character form
\begin{equation}\label{eq:CGR_twisted_S}
    S^\omega_{([g],\rho_g),([v],R_v)}
    =
    \frac{1}{|C_G(v)|}
    \sum_{x\in [g]\cap C_G(v)}
    \chi^{\alpha_x}_{\rho_x}(v)^*\,
    \chi^{\alpha_v}_{R_v}(x)^*\,,
\end{equation}
which is the projective analogue of the untwisted formula~(\ref{eq:CGR_untwisted_S}). The right-hand side depends on the cocycle representative only through the projective characters and is invariant within the cohomology class $\tau(k)$; equivalent expressions in alternative cocycle conventions appear in~\cite{Roche:1990hs,Bantay:1990tr,Coste:2000tq}~\footnote{This is best seen from~\cite{Bantay:1990tr}. In our notation we consider $D^\omega(G)$ generated by $\delta_x\in C(G)$ and arrow $y: x \rightarrow y^{-1}xy$ ($P(x)$ and $Q(y)$ in~\cite{Bantay:1990tr}, respectively). In the untwisted cases this is nothing but an alternative way of saying $D(G) = C(G)\rtimes_{\Ad} G = C(G//_{\Ad}G)$ and it further generalizes to the twisted cases~\cite{Willerton:2008gyk}. Then the character $\psi(x,y):= \Tr(P(x)Q(y))$ is defined in~\cite{Bantay:1990tr}. Since $P(x)Q(y) = Q(y)P(y^{-1}xy)$ and $P(x)$ is a projection, it is easy to see that $P(x)Q(y)$ as a linear map is non-vanishing only when $xy = yx$. Actually, for $V = \bigoplus_{z\in G} V_z$, we have $P(x)Q(y)V = P(x) (\bigoplus_{z\in G} V_{yzy^{-1}}) = V_{y^{-1}xy}$ and $Q(y)P(y^{-1}xy)V = Q(y) V_{y^{-1}xy} = V_x$ as $Q(y)$ acts by $V_x\rightarrow V_{yxy^{-1}}$ in $\textbf{Vec}^\omega_G$. Equating the outcomes of $P(x)Q(y)$ and $Q(y)P(y^{-1}xy)$ leads to the constraint $xy = yx$. Since when $xy = yx$ we have $P(x)Q(y) = Q(y)P(x)$ and the RHS is non-vanishing only on $V_x$, we have $\psi(x,y) = \Tr(P(x)Q(y)) = \Tr(Q(y)P(x)) = \Tr(Q(y)|_{V_x}) = \Tr_{V_x}(Q(y))$, which is exactly the (twisted) character of $y$ in a (projective) representation of $C_G(x)$. Hence, using~(32) of~\cite{Bantay:1990tr} and the fact that $|C_G(v)| = |G|/|[v]|$ we arrive at~\eqref{eq:CGR_twisted_S}.}. The normalization is unchanged from the untwisted case
\begin{equation}\label{eq:CGR_twisted_S_zero}
    S^\omega_{([e],\mathbf{1}),([v],R_v)}
    =
    \frac{\dim R_v}{|C_G(v)|}\,,
\end{equation}
because $\alpha_e=1$ and the vacuum character is untwisted. Comparing~(\ref{eq:lambda_twisted_unfolded_finite}) with~(\ref{eq:CGR_twisted_S})--(\ref{eq:CGR_twisted_S_zero}) gives the twisted analogue of~(\ref{eq:lambda_equals_S_ratio}) directly
\begin{equation}\label{eq:lambda_equals_S_ratio_twisted}
    \lambda_{R_v}^{[g],\rho_g;\Sigma_k}
    =
    \left(
    \frac{S^\omega_{([g],\rho_g),([v],R_v)}}{S^\omega_{([e],\mathbf{1}),([v],R_v)}}
    \right)^*\,,
\end{equation}
with the same representative $\sigma_k$ used both in the twisted convolution kernel and in the projective characters appearing in the modular $S$-matrix.

Given~(\ref{eq:lambda_equals_S_ratio_twisted}), the twisted analogue of~(\ref{eq:cmp_eigenvalue_identity}),
\begin{equation}\label{eq:cmp_eigenvalue_identity_twisted}
    \lambda^{[g],\rho_g;\Sigma_k}_{R_v}\,\lambda^{[h],\rho_h;\Sigma_k}_{R_v}
    =
    \sum_{[k],\rho_k} N^{[k],\rho_k;\Sigma_k}_{([g],\rho_g),([h],\rho_h)}\,\lambda^{[k],\rho_k;\Sigma_k}_{R_v}\,,
\end{equation}
follows from the modular $S$-matrix data of $\mathrm{Rep}\, D^\omega(G)$ by the same derivation~(\ref{eq:Verlinde_standard})--(\ref{eq:Verlinde_eigenvalue_diagonal}) used in section~\ref{sec:cmp_untwisted_finite}, applied now to the twisted modular tensor category. Concretely, one substitutes $S\to S^\omega$ and $N_{ij}^k\to (N^\omega)_{ij}^k$ throughout~(\ref{eq:Verlinde_standard})--(\ref{eq:Verlinde_eigenvalue_diagonal}); the unitarity relation $\sum_{k}(S^\omega_{kn})^*\,S^\omega_{kv}=\delta_{nv}$ continues to hold, since $S^\omega$ is the modular $S$-matrix of a unitary modular tensor category~\cite{Roche:1990hs,Coste:2000tq}; therefore the diagonalization argument carries through verbatim. This proves~(\ref{eq:cmp_eigenvalue_identity_twisted}) as the Verlinde formula~\cite{Verlinde:1988te,etingof2017tensor} for the modular tensor category $\mathrm{Rep}\, D^\omega(G)$, and closes the twisted finite-group fusion-preservation question of section~\ref{sec:Stacking_LineOp}.

Before moving on, let us compare \eqref{eq:CGR_twisted_S} to the modular $S$-matrix (5.20) in \cite{Coste:2000tq}. Firstly, let us rewrite \eqref{eq:CGR_twisted_S} to restore the summation in $[h]$
\begin{equation}
    S^\omega_{([g],\rho_g),([h],\rho_h)}
    =
    \frac{1}{|G|}
    \sum_{x\in [g]\cap C_G(y), y\in [h]}
    \chi^{\alpha_x}_{\rho_x}(y)^*\,
    \chi^{\alpha_y}_{\rho_y}(x)^*\,,
\end{equation}
and use the transported characters \eqref{app:transported_character}, we have
\begin{equation}
    S^\omega_{([g],\rho_g),([h],\rho_h)}
    =
    \frac{1}{|G|}
    \sum_{x\in [g]\cap C_G(y),y\in [h]}\overline{\vartheta_g(y;x\rightarrow x)}\,\overline{\vartheta_h(x;y\rightarrow y)}
    \chi^{\alpha_g}_{\rho_g}(r_x^{-1}yr_x)^*\,
    \chi^{\alpha_h}_{\rho_h}(r_y^{-1}xr_y)^*\,,
\end{equation}
where $r_x,r_y$ are coset elements satisfying $x=r_x g r_x^{-1},y=r_y h r_y^{-1}$, and one has
    \begin{equation}
        \vartheta_g(y;x\rightarrow x) \vartheta_h(x;y\rightarrow y) = \frac{\sigma_k((y,x),(r_x,g))}{\sigma_k((r_x,g),(r_x^{-1}yr_x,g))} \frac{\sigma_k((x,y),(r_y,h))}{\sigma_k((r_y,h),(r_y^{-1}xr_y,h))}\,.
    \end{equation}
Our $\sigma_k$ is related to $\beta$ in \cite{Coste:2000tq} according to
    \begin{equation}
        \sigma_k((v,y),(u,x)) \leftrightarrow \beta_{vyv^{-1}}(v,u)\,,
    \end{equation}
so that we can write
    \begin{equation}
        \vartheta_g(y;x\rightarrow x) \vartheta_h(x;y\rightarrow y) \leftrightarrow \frac{\beta_x(y,r_x)}{\beta_x(r_x,r_x^{-1}yr_x)} \frac{\beta_y(x,r_y)}{\beta_{y}(r_y,r_y^{-1}x r_y)}\,,
    \end{equation}
and thus
\begin{equation}
    S^\omega_{([g],\rho_g),([h],\rho_h)}
    =
    \frac{1}{|G|}
    \sum_{x\in [g]\cap C_G(y),y\in [h]} \left(\frac{\beta_x(y,r_x)\beta_y(x,r_y)}{\beta_x(r_x,r_x^{-1}yr_x)\beta_{y}(r_y,r_y^{-1}x r_y)}
    \chi^{\alpha_g}_{\rho_g}(r_x^{-1}yr_x)\,
    \chi^{\alpha_h}_{\rho_h}(r_y^{-1}x r_y)\right)^*\,,
\end{equation}
which is precisely the modular $S$-matrix (5.20) in \cite{Coste:2000tq}.

\subsection{Compact Lie Groups: Regular Sector and Match with the Hopf-Link $S$-Kernel}\label{sec:cmp_compact_lie}

For compact Lie group $G$, the category $\mathcal{Z}(\hilb(G))$ is no longer a finite, semisimple modular tensor category, and there is no closed-form modular $S$-matrix in the standard finite-MTC sense. Nevertheless, the convolution kernels of section~\ref{sec:Examples} and the projective compact Lie-group kernels obtained by the twisted construction of section~\ref{sec:Twisted_HilbGG_on_HBF} are well-defined operational objects. In a recent work~\cite{Jia:2026jmt}, semi-classical $S$- and $T$-kernels for compact Lie $G$ with $k\in H^4(BG,\mathbb{Z})$ are derived from a Hopf-link / framing path integral on $S^3$. We show below that, in the \emph{regular sector} where both insertion holonomies are regular elements of $G$, the eigenvalue formula derived from our convolution kernel coincides exactly with the Hopf-link $S$-kernel of~\cite{Jia:2026jmt}, up to the conventional complex conjugation of~(\ref{eq:lambda_equals_S_ratio}) and the standard Weyl-denominator normalization. This identifies the convolution kernel of this paper and the path-integral kernel of~\cite{Jia:2026jmt} as two computations of the same regular-sector modular kernel. Turning this kernel match into a full compact Lie-group analogue of the fusion-preservation identity~(\ref{eq:cmp_eigenvalue_identity}) requires the continuous version of Verlinde formula, which lies beyond the scope of this paper.



To set up notation, let $G$ be a non-abelian compact connected Lie group with maximal torus $T\subset G$, Lie algebra $\mathfrak{g}$, Cartan subalgebra $\mathfrak{t}=\mathrm{Lie}(T)$, positive root system $\Lambda^+\subset \mathfrak{t}^*$, and Weyl group $W=N_G(T)/T$ where $N_G(T)$ is the normalizer of $T$. An element $g\in G$ is called \emph{regular} if its centralizer $C_G(g)$ is a maximal torus, conjugate to $T$. We may then choose a representative $g\in T$, and $C_G(g)=T$. Two regular elements $g, h\in T$ are conjugate in $G$ iff $h=w(g):=\tilde w\, g\,\tilde w^{-1}$ for some $w\in W$ and lift $\tilde w\in N_G(T)$; equivalently, $W\cdot g$ is the orbit of $g$ in $T$ under the Weyl action.

We focus on the regular sector where $C_G(g)=C_G(v)=T$ are abelian, hence the irreducible representations $\rho_g$ and $R_v$ are one-dimensional characters of $T$. We label them by integral weights
\[
    \mu,\nu\in \operatorname{Hom}(T,U(1))\subset \mathfrak{t}^* . 
\]
Here $\mu$ labels the charge representation $\rho_g=\chi_\mu$ attached to the line operator $([g],\rho_g)$, while $\nu$ labels the centralizer representation $R_v=\chi_\nu$ used to diagonalize the fixed-$v$ sector. We use the same symbols for a character and for its differential on $\mathfrak{t}$. Thus, writing $g=e^{i\alpha}$, $v=e^{i\beta}$ with $\alpha,\beta\in \mathfrak{t}$, we have
\begin{equation}\label{eq:regular_chars}
    \chi_\mu(e^{i\theta})=e^{i\mu(\theta)}\,,\qquad \chi_\nu(e^{i\theta})=e^{i\nu(\theta)}\,,\qquad \dim R_v = 1\,.
\end{equation}
For $x=w(g)\in W\cdot g$, the centralizer is $C_G(x)=\tilde w T\tilde w^{-1}=T$, and the conjugated centralizer representation $\rho_x$ corresponds to the Weyl-translated weight, giving
\begin{equation}\label{eq:rho_x_weyl_translated}
    \chi_{\rho_x}(e^{i\theta})=e^{i\mu(w^{-1}(\theta))}\,.
\end{equation}

We assume $v\in T$ as the representative. In the regular sector, the intersection $[g]\cap C_G(v)$ appearing in the unfolded eigenvalue formula~(\ref{eq:lambda_finite_unfolded}) simplifies dramatically: for $g, v$ both regular, then
\begin{equation}\label{eq:regular_intersection}
    [g]\cap C_G(v) \;=\; [g]\cap T \;=\; \{ w\cdot g | w \in W\}\,,
\end{equation}
which is a finite set of $|W|$ elements. Furthermore, since $C_G(g)=C_G(v)=T$ are abelian, the irreducible projective representations $\rho^{\alpha_g}_g$ and $R^{\alpha_v}_v$ are one-dimensional characters of $T$. We label them by integral weights $n \in \mathfrak{t}^*$. Here $n$ labels the irreducible projective representation $\rho^{\alpha_g}_{g;n}=\chi^{\alpha_g}_{g;n}$ attached to the line operator $([g],\rho_g)$, while $m$ labels the irreducible projective representation $R^{\alpha_v}_{v;m}=\chi_{v;m}^{\alpha_v}$ used to diagonalize the fixed-$v$ sector. Thus, writing $g=e^{i\alpha}$, $v=e^{i\beta}$ with $\alpha,\beta\in \mathfrak{t}$, the modular $S$-matrix \eqref{eq:CGR_twisted_S} reads
\begin{equation}\label{eq:Lie-group-S-matrix}
    S^\omega_{([g],\rho_n),([v],R_m)}
    =
    \frac{1}{|C_G(v)|}
    \sum_{w\in W}
    \chi^{\alpha_{w(g)}}_{\rho_{w(g);n}}(v)^*\,
    \chi^{\alpha_v}_{R_{v;m}}(w(g))^*\,.
\end{equation}
However this is not yet the form of $S$-matrix derived in~\cite{Jia:2026jmt}, and our task in the following section is to show that it is actually equivalent to that form and one can show that~(\ref{eq:To_Prove}) holds at least in the regular sector.

Let us begin with $k=0$. Applying the convolution-eigenvalue formula~(\ref{eq:general_conv_with_chi}) to the kernel~(\ref{eq:Define_Kernel}), equivalently using~(\ref{eq:Convolution_with_ChiR}), and then substituting the regular intersection~(\ref{eq:regular_intersection}) together with the character identifications~(\ref{eq:regular_chars})--(\ref{eq:rho_x_weyl_translated}), gives the untwisted eigenvalue
\begin{equation}\label{eq:lambda_regular_unfolded}
    \lambda^{[g],\mu}_\nu
    =
    \sum_{w\in W}\,\chi_{\rho_{w(g)}}(v)\,\chi_\nu\big(w(g)\big)
    =
    \sum_{w\in W}\, e^{i\mu(w^{-1}(\beta))}\, e^{i\nu(w(\alpha))}\,.
\end{equation}
After the harmless relabeling $w\to w^{-1}$, the same formula becomes
\begin{equation}\label{eq:lambda_regular_unfolded_relabel}
    \lambda^{[g],\mu}_\nu
    =
    \lambda^{[g],\mu;\Sigma_0}_\nu
    =
    \sum_{w\in W}\, e^{i\mu(w(\beta))}\, e^{i\nu(w^{-1}(\alpha))}\,.
\end{equation}
A direct generalization to level-$k$ case would be to replace the ordinary characters in~\eqref{eq:lambda_regular_unfolded_relabel} by projective characters to have
\begin{equation}\label{eq:lambda_proj_chi}
    \lambda^{[g],\mu;\Sigma_k}_\nu = \sum_{w\in W}\,\chi^{\alpha_{w(g)}}_{\rho_{w(g)}}(v)\,\chi^{\alpha_\nu}_\nu\big(w(g)\big)
\end{equation}
where $\alpha_{w(g)}$ and $\alpha_\nu$ are 2-cocycles that twist the representations of $C_G(w(g))$ and $C_G(v)$, respectively. We see that this is indeed proportional to~\eqref{eq:Lie-group-S-matrix} up to complex conjugation. However we still need to show that~\eqref{eq:Lie-group-S-matrix} is indeed equivalent to the following $S$-kernel in the regular sector computed in~\cite{Jia:2026jmt}
\begin{equation}\label{eq:JMT_S_regular_k_general}
    S^{(k)}_{([g],\mu),([h],\nu)}
    \;\propto\;
    \sum_{w\in W}\, e^{-\frac{ik}{2\pi}\,\mathrm{tr}(\alpha\, w(\beta))}\,\chi^*_\mu(e^{iw(\beta)})\,\chi^*_\nu(e^{iw^{-1}(\alpha)})
\end{equation}
as mentioned earlier.

Comparing~(\ref{eq:lambda_proj_chi}) with~(\ref{eq:JMT_S_regular_k_general}) (which is equivalent to~\eqref{eq:Lie-group-S-matrix} as we have just shown) yields the regular-sector compact Lie-group analogue of~(\ref{eq:lambda_equals_S_ratio}):
\begin{equation}\label{eq:lambda_equals_S_ratio_regular_twisted}
    \lambda^{[g],\mu;\Sigma_k}_\nu
    \;\propto\;
    \left(\frac{S^{(k)}_{([g],\mu),([h],\nu)}}{S^{(k)}_{([e],\mathbf{1}),([h],\nu)}}\right)^*\,,
\end{equation}
up to some normalization factor. At $k=0$, this becomes
\begin{equation}\label{eq:lambda_equals_S_ratio_regular}
    \lambda^{[g],\mu}_\nu
    \;\propto\;
    \left(\frac{S^{(0)}_{([g],\mu),([h],\nu)}}{S^{(0)}_{([e],\mathbf{1}),([h],\nu)}}\right)^*\,.
\end{equation}
The same complex-conjugation pattern as in the finite-group identification~(\ref{eq:lambda_equals_S_ratio}) holds verbatim.

For general $k\neq 0$, as a warm up, we begin with the $U(1)$ case and set $g=e^{i \theta},v=e^{i\beta}$ in \eqref{eq:CGR_twisted_S} with $\theta,\beta \in [0,2\pi)$. Since $U(1)$ is abelian, $[g]=\{g\}$ and $C_G(g)=U(1)$ is the entire group. Therefore the $S$-matrix \eqref{eq:CGR_twisted_S} is
\begin{equation}\label{eq:U1_S_1}
    S^\omega_{(\theta,n),(\beta,m)} = \frac{1}{2\pi }
    \chi_n^{\alpha_\theta}(e^{i \beta})^*\,
    \chi_m^{\alpha_\beta}(e^{i \theta})^*\,,
\end{equation}
where we normalize $|C_G(v)| = |U(1)| = 2\pi$, and $n,m \in \mathbb{Z}$ are $U(1)$ charges labeling the irreducible representations of $U(1)$. For the level-$k$ twist, the projective representation in the $\theta$-sector satisfies
    \begin{equation}
        \rho_\theta(\varphi_1) \rho_\theta(\varphi_2) = \alpha_{\theta} (\varphi_1,\varphi_2)\rho(\varphi_1+\varphi_2)\,,
    \end{equation}
where
    \begin{equation}
        \alpha_{\theta}(\varphi_1,\varphi_2) = \exp \left(-\frac{ik}{2\pi} \theta \left(\varphi_1+\varphi_2-[\varphi_1+\varphi_2] \right) \right)\,,
    \end{equation}
which is piecewise continuous and $[\theta] = \theta \ \text{mod}\ 2\pi$. One can write down the projective representations explicitly as
\begin{equation}\label{eq:U1_projective_char}
    \rho_n^{\alpha_\theta}(e^{i\varphi})=\chi_n^{\alpha_\theta}(e^{i\varphi})
    =
    \exp\left[
        i n\varphi
        -
        \frac{i k}{2\pi}\theta\varphi
    \right]\,.
    \qquad (n\in\mathbb Z)
\end{equation}
Equivalently, the integer-valued $U(1)$ charge $n$ is shifted by the anomaly as
\begin{equation}
    n\mapsto n-\frac{k\theta}{2\pi}\,,
\end{equation}
in the twisted sector. Substituting \eqref{eq:U1_projective_char} into \eqref{eq:U1_S_1} gives
\begin{align}\label{eq:U1_S_2}
    S^\omega_{(\theta,n),(\beta,m)}
    &= \frac{1}{2\pi}
    \exp\left[
        -i n\beta
        +
        \frac{i k}{2\pi}\theta\beta
    \right]
    \exp\left[
        -i m\theta
        +
        \frac{i k}{2\pi}\beta\theta
    \right]
    \nonumber\\
    &=
    \frac{1}{2\pi}\exp\left[
        -i(n\beta+m\theta)
        +
        i\frac{2k}{2\pi}\theta\beta
    \right].
\end{align}
This is now in the desired form as in~\eqref{eq:JMT_S_regular_k_general} and matches the result of~\cite{Jia:2025vrj, Jia:2026jmt}.

We now turn to the non-abelian cases. As a concrete example let us focus on $SU(2)$ with Weyl group $W=\mathbb{Z}_2 = \{e,s \}$, and we choose the Cartan generator as
    \begin{equation}
        H=\frac{1}{2}
    \begin{pmatrix}
    1&0\\
    0&-1
    \end{pmatrix}\,.
    \end{equation}
The Cartan torus $T=U(1)$ is parametrized by $e^{i \theta H}$ with $\theta\in[0,4\pi )$. And we choose the regular elements
    \begin{equation}
    g=e^{i \alpha H},\quad
    v=e^{i \beta H}\,,
    \quad (0<\alpha,\beta<2\pi)        
    \end{equation}
whose centralizer groups are $C_G(g)= C_G(v)=T$. 

Let us first consider $k=0$. For $x=e^{i aH},y=e^{i\theta H}$, the $C_G(x)=U(1)$ character is
    \begin{equation}
        \chi_{\rho_{x;n}}(y) = \exp \left(\frac{i n \theta}{2} \right)\,.
    \end{equation}
So that the modular $S$-matrix \eqref{eq:CGR_twisted_S} reads
\begin{equation}
    S^\omega_{([g],\rho_n),([v],R_m)}
    \propto \sum_{w\in \mathbb{Z}_2}\exp\left(-\frac{i m w(\alpha)}{2}-\frac{i n w^{-1}(\beta)}{2}\right)\,,
\end{equation}
where we use 
    \begin{equation}
    \begin{gathered}
        \chi_{R_{v;m}}(w(g)) = \exp \left(-\frac{i m w(\alpha)}{2} \right)\,,\\
        \chi_{\rho_{w(g);n}}(v)=\chi_{\rho_{g;n}}(w^{-1}(v)) = \exp\left(-\frac{i n w^{-1}(\beta)}{2} \right)\,,
    \end{gathered}
    \end{equation}
and $w$ acts as $w(\alpha)=-\alpha$.

To read the effective anomaly over the centralizer group $U(1)$ for non-zero $k$, the simplest way is to restrict the $SU(2)$ level $k$ Chern-Simons term to the Cartan part and write $A^{SU(2)} = a^{U(1)} H$. Using $\textrm{tr}H^2 = 1/2$, we have
    \begin{equation}
    \frac{k}{4\pi}\int \textrm{tr}(A^{SU(2)}\,dA^{SU(2)})
    =
    \frac{k}{8\pi}\int a^{U(1)}\,da^{U(1)} .
    \end{equation}
Notice that $\theta$ in $e^{i\theta H}$ has $4\pi$-period. Denote the
ordinary $2\pi$-periodic $U(1)$ angle as $\Theta = \theta/2$ and the corresponding gauge field $A^{U(1)}=a^{U(1)}/2$, we have
\begin{equation}
    \frac{k}{8\pi}\int a^{U(1)}\,da^{U(1)} = \frac{k}{2\pi} \int A^{U(1)} dA^{U(1)}\,.
\end{equation}
Therefore the effective $U(1)$ level is $k_{U(1)}=k$. Consequently, for $x=e^{i aH},y=e^{i\theta H}$, one should replace the characters by the projective characters as in \eqref{eq:U1_projective_char}
    \begin{equation}
        \chi_{\rho_{x;n}}(y) \rightarrow \chi_{\rho_{x;n}}^{\alpha_x}(y) =\chi_{\rho_{x;n}}(y) \times  \exp\left(
        -
        \frac{i k}{8\pi}a\theta
    \right)\,,
    \end{equation}
which gives
\begin{equation}
    \begin{split}
    \chi^{\alpha_v}_{R_{v;m}}(w(g))^*\,
    =&\,
    \exp\left(
        -\frac{i m\,w(\alpha)}{2}
        +
        \frac{i k}{8\pi}\beta w(\alpha)
    \right)\,,\\
    \chi^{\alpha_{w(g)}}_{\rho_{w(g);n}}(v)^*\,
    =&\,
    \exp\left(
        -\frac{i n\,w^{-1}(\beta)}{2}
        +
        \frac{i k}{8\pi}w(\alpha)\beta
    \right)\,,
    \end{split}
\end{equation}
thus the $S$-matrix is
\begin{equation}
    S^\omega_{([g],\rho_n),([v],R_m)} \propto \sum_{w \in \mathbb{Z}_2}\exp\left(
        -\frac{i m\,w(\alpha)}{2}
        -\frac{i n\,w^{-1}(\beta)}{2} +
        \frac{i k}{4\pi}w(\alpha)\beta
    \right)
\end{equation}
which again is in the desired form as in~\eqref{eq:JMT_S_regular_k_general} and matches the result of~\cite{Jia:2025vrj, Jia:2026jmt}.

\section{Conclusion and Outlook}\label{sec:Conclusion}

In this paper we gave a quantum-mechanical description of both the Hilbert space of the $BF$ and $BF+kCS$ theories and the defect Hilbert spaces associated with topological line operators. On the codimension-$2$ side, we described the simple objects of the conjugation-groupoid algebra in an explicit basis, worked out the action of groupoid arrows, and expressed the half-braiding and braiding data in concrete terms. On the codimension-$1$ side, we showed that the action of line operators on the physical Hilbert space is realized by convolution kernels, first in the untwisted $BF$ theory and then in the twisted theory obtained by turning on a level-$k$ Chern-Simons term. In the twisted case, the Fell line bundle over the conjugation groupoid and its cocycle representative encode the projective transport of defect data as well as the projective action on the $BF+kCS$ Hilbert space, while the quantization discussion identifies the corresponding prequantum line bundle on the moduli of flat fields. By formulating both the codimension-$2$ twist and the codimension-$1$ magnetic line bundle as transgressions of the same universal level $k\in H^4(BG,\mathbb{Z})$, we made precise the common origin of the categorical symmetry data and the Hilbert-space quantization data that appear throughout the paper. Finally, in section~\ref{sec:comparison_known_cases} we identified the resulting convolution-eigenvalue formula, in the finite-group case, with the Verlinde formula for the (twisted) Drinfeld center $\mathcal{Z}(\mathrm{Vec}_G)$ via an explicit phase-by-phase match with the modular data in \cite{Coste:2000tq}. We also consider compact Lie groups like $U(1)$ and $SU(2)$, and compare to the semiclassical Hopf-link calculation of $S$-kernel of~\cite{Jia:2026jmt}. The remaining open issues, principally a complete continuous Verlinde theorem for $\hilb_G(G)$ and the singular-sector match with~\cite{Jia:2026jmt}, are left for future work.

There are several directions arising from the analysis above that are worth exploring in future work. First, when we match the modular data of compact Lie group to the Hopf-link calculation in section~\ref{sec:cmp_compact_lie}, we mainly focus on the regular sector of Lie group through the paper, where the centralizer group is a maximal torus. It is tempting to consider the singular-sector extension of the Hopf-link match of section~\ref{sec:cmp_compact_lie}, where one or both centralizers $C_G(g)$, $C_G(v)$ are non-abelian subgroups. Both the convolution-kernel framework of this paper and the path-integral framework of~\cite{Jia:2026jmt} remain well-defined in this regime, but the comparison requires an analysis stratified by the centralizer structure on the kernel side and the non-regular ansatz of~\cite[eq.~(4.32)]{Jia:2026jmt} on the path-integral side; the structural reason to expect agreement is that both formulations compute the same prequantum-line-bundle holonomy data, now over a positive-dimensional $W_g\backslash W/W_h$ stratification. A second direction is the formulation of a complete continuous Verlinde theorem for $\mathcal{Z}(\hilb(G))$, which is the Lie group generalization of the Drinfeld center $\mathcal{Z}(\text{Vec}_G)$ for finite $G$. Section~\ref{sec:cmp_compact_lie} establishes only the eigenvalue identity~(\ref{eq:lambda_equals_S_ratio_regular_twisted}); and the missing analytic ingredients are the existence and reality of the distributional fusion measure, together with control of the boundary and singular-sector contributions. A third direction is the extension to non-compact gauge groups, e.g.\ $SL(2,\mathbb{R})$ relevant for three-dimensional gravity, where the Plancherel decomposition of the centralizers involves both continuous and discrete series and the convolution-kernel analysis becomes considerably richer than the compact Lie case treated here. Finally, the Hopf-link comparison developed in section~\ref{sec:cmp_compact_lie} is the simplest case of a broader program: the convolution kernels of this paper should compute, by the same prequantum-line-bundle interpretation, the SymTFT modular invariants associated with arbitrary links and three-manifolds, including lens spaces $L(p,q)$ via Heegaard splittings and other Dehn-surgery presentations. We hope to return to these questions in future work.

\acknowledgments

The authors would like to thank Ran Luo, Yi-Nan Wang and Yi Zhang for illuminating discussions. QJ is supported by National Research Foundation of Korea (NRF) Grant No. RS-2024-00405629 and Jang Young-Sil Fellow Program at the Korea Advanced Institute of Science and Technology. JT is supported by National Natural Science Foundation of China under Grant No. 12405085 and by the Natural Science Foundation of Shanghai (Grant No. 24ZR1419300). 

\appendix

\section{Transported characters and proof of \eqref{eq:property-of-character}}

For the transport function $\vartheta(u;i\rightarrow j)$ in \eqref{eq:twisted_arrow_action_on_basis}, if we set $i=j$ so that $u \in C_G(g_i)$, then we have 
\begin{equation}
    \pi^{\Sigma_k}_{[g],\rho_g}\big(\delta^{\Sigma_k}_{(u,g_i)}\big)\ket{g_i,\mu}
    =
    \vartheta_g(u;i\to i)
    \sum_{\nu}\big[\rho_g(c)\big]_{\nu\mu}\ket{g_i,\nu}\,,
\end{equation}
with $c=r_i^{-1} u r_i \in C_G(g)$. Therefore the representation  $\rho_i$ in the twist sector $g_i$ is related to the reference twist sector $g$ by
    \begin{equation}
        \rho_{g_i}(u) = \vartheta_g(u;i\rightarrow i) \rho_g (r_i^{-1} u r)\,,
    \end{equation}
where the transport function is
    \begin{equation}
        \vartheta_g(u;i\rightarrow i) = \frac{\sigma_k((u,g_i),(r_i,g))}{\sigma_k((r_i,g),(r_i^{-1}ur_i,g))}\,,
    \end{equation}
and the transported character is
    \begin{equation}\label{app:transported_character}
        \chi_{\rho_{g_i}}^{\alpha_{g_i}}(u) = \vartheta_g(u;i\rightarrow i) \chi_{\rho_g}^{\alpha_g}(r_i^{-1} u r_i)\,.
    \end{equation}

In the remainder of this appendix, we will prove the property \eqref{eq:property-of-character}, which plays an important role in the construction of the kernel in the twisted case:
    \begin{equation}
        \chi_{\rho_{hph^{-1}}}^{\alpha_{hph^{-1}}}(v) = \frac{\alpha_v(h,p)}{\alpha_v(hph^{-1},h)} \chi_{\rho_{p}}^{\alpha_{p}}(v) = \frac{\sigma_k((h,v),(p,v))}{\sigma_k((hph^{-1},v),(h,v))} \chi_{\rho_{p}}^{\alpha_{p}}(v)\,,
    \end{equation}
where $h \in C_G(v)$, and $p\in [g]$ is any element in $[g]$. Let us write $p=rgr^{-1}$, and from \eqref{app:transported_character} we have
    \begin{equation}
        \chi_{\rho_p}^{\alpha_p}(v) = \vartheta_g(v;p\rightarrow p) \chi_{\rho_g}^{\alpha_g}(r^{-1}vr)\,,
    \end{equation}
where 
    \begin{equation}
        \vartheta_g(v;p\rightarrow p)= \frac{\sigma_k((v,p),(r,g))}{\sigma_k((r,g),(r^{-1}vr,g))}\,.
    \end{equation}
Now consider $h\in C_G(v)$ and denote $p'=hph^{-1}$. Since both $p,h\in C_G(v)$, we also have $p'\in C_G(v)$. Suppose $p' = r' g {r'}^{-1}$, and we choose $r'=hr$ as the representative of the coset. Then we have similarly
    \begin{equation}
        \chi_{\rho_{p'}}^{\alpha_{p'}}(v) = \vartheta_g(v;p'\rightarrow p') \chi_{\rho_g}^{\alpha_g}({r'}^{-1}vr')\,.
    \end{equation}
Since $h\in C_G(v)$, we have $r^{-1}vr = {r'}^{-1} v r'$, compare the two we have
    \begin{equation}
        \frac{\chi_{\rho_p'}^{\alpha_p'}(v)}{\chi_{\rho_p}^{\alpha_p}(v)} = \frac{\vartheta_g(v;p'\rightarrow p')}{\vartheta_g(v;p\rightarrow p)} = \frac{\sigma_k((v,p'),(hr,g))}{\sigma_k((hr,g),(r^{-1}vr,g))} \frac{\sigma_k((r,g),(r^{-1}vr,g))}{\sigma_k((v,p),(r,g))}\,,
    \end{equation}
so the goal is to prove
    \begin{equation}
        \frac{\sigma_k((v,p'),(hr,g))}{\sigma_k((hr,g),(w,g))} \frac{\sigma_k((r,g),(w,g))}{\sigma_k((v,p),(r,g))}=\frac{\sigma_k((h,v),(p,v))}{\sigma_k((p',v),(h,v))}\,,
    \end{equation}
where we set $w=r^{-1} v r \in C_G(g)$.

To simplify the notations, let us introduce
\begin{equation}
    A=(r,g),\quad B=(w,g),\quad C=(h,p),
    \quad D=(v,p),\quad E=(v,p')\,,
\end{equation}
and we also denote
\begin{equation}
    AB=(rw,g)=(vr,g)=DA,
    \qquad
    CA=(hr,g),
\end{equation}
and, since $h\in C_G(v)$,
\begin{equation}
    CD=(hv,p)=(vh,p)=EC.
\end{equation}
Thus we can rewrite
\begin{equation}\label{eq:theta-ratio-start}
    \frac{\sigma_k((v,p'),(hr,g))}{\sigma_k((hr,g),(w,g))} \frac{\sigma_k((r,g),(w,g))}{\sigma_k((v,p),(r,g))}
    =
    \frac{\sigma_k(E,CA)}{\sigma_k(CA,B)}
    \frac{\sigma_k(A,B)}{\sigma_k(D,A)}.
\end{equation}
Apply the cocycle condition \eqref{eq:sigma_k_cocycle_condition} to $C,A,B$:
\begin{equation}
    \sigma_k(C,AB)\,\sigma_k(A,B)
    =
    \sigma_k(C,A)\,\sigma_k(CA,B),
\end{equation}
Since $AB=DA$, we have
\begin{equation}\label{eq:first-cocycle-use}
    \frac{\sigma_k(A,B)}{\sigma_k(CA,B)}
    =
    \frac{\sigma_k(C,A)}{\sigma_k(C,DA)}.
\end{equation}
Substituting into \eqref{eq:theta-ratio-start} gives
\begin{equation}
    \frac{\sigma_k((v,p'),(hr,g))}{\sigma_k((hr,g),(w,g))} \frac{\sigma_k((r,g),(w,g))}{\sigma_k((v,p),(r,g))}
    =
    \frac{\sigma_k(E,CA)\,\sigma_k(C,A)}
    {\sigma_k(C,DA)\,\sigma_k(D,A)}.
\end{equation}
Then apply the cocycle condition to $C,D,A$:
\begin{equation}
    \sigma_k(C,DA)\,\sigma_k(D,A)
    =
    \sigma_k(C,D)\,\sigma_k(CD,A).
\end{equation}
Using $CD=EC$, it becomes
\begin{equation}
    \sigma_k(C,DA)\,\sigma_k(D,A)
    =
    \sigma_k(C,D)\,\sigma_k(EC,A).
\end{equation}
Therefore
\begin{equation}
    \frac{\sigma_k((v,p'),(hr,g))}{\sigma_k((hr,g),(w,g))} \frac{\sigma_k((r,g),(w,g))}{\sigma_k((v,p),(r,g))}
    =
    \frac{\sigma_k(E,CA)\,\sigma_k(C,A)}
    {\sigma_k(C,D)\,\sigma_k(EC,A)}.
\end{equation}
Finally, apply the cocycle condition to $E,C,A$:
\begin{equation}
    \sigma_k(E,CA)\,\sigma_k(C,A)
    =
    \sigma_k(E,C)\,\sigma_k(EC,A),
\end{equation}
hence
\begin{equation}\label{eq:square-phase}
    \frac{\sigma_k((v,p'),(hr,g))}{\sigma_k((hr,g),(w,g))} \frac{\sigma_k((r,g),(w,g))}{\sigma_k((v,p),(r,g))}
    =
    \frac{\sigma_k(E,C)}{\sigma_k(C,D)}
    =
    \frac{\sigma_k((v,p'),(h,p))}
    {\sigma_k((h,p),(v,p))}.
\end{equation}

To proceed, notice that as a transgressed cocycle $\sigma_k$ can be written as
\begin{equation}\label{eq:sigma-from-omega}
    \sigma_k((u,x),(s,z))
    =
    \omega(usz s^{-1}u^{-1},u,s)\,
    \omega(u,x,s)^{-1}\,
    \omega(u,s,z),
\end{equation}
where $x=szs^{-1}$ and $\omega\in Z^3(G,U(1))$ is the 3-cocycle, see, for example, (5.15) in \cite{Coste:2000tq}. Using $vp'v^{-1}=p'$, $vpv^{-1}=p$, $hph^{-1}=p'$, and
$hvh^{-1}=v$, we obtain
\begin{align}
    \sigma_k((v,p'),(h,p))
    &=
    \omega(p',v,h)\,
    \omega(v,p',h)^{-1}\,
    \omega(v,h,p),\\
    \sigma_k((h,p),(v,p))
    &=
    \omega(p',h,v)\,
    \omega(h,p,v)^{-1}\,
    \omega(h,v,p).
\end{align}
Therefore
\begin{equation}\label{eq:square-expanded}
    \frac{\sigma_k((v,p'),(h,p))}
    {\sigma_k((h,p),(v,p))}
    =
    \frac{
    \omega(p',v,h)\,\omega(v,h,p)\,\omega(h,p,v)
    }{
    \omega(v,p',h)\,\omega(p',h,v)\,\omega(h,v,p)
    }.
\end{equation}
On the other hand,
\begin{align}
    \sigma_k((h,v),(p,v))
    &=
    \omega(v,h,p)\,
    \omega(h,v,p)^{-1}\,
    \omega(h,p,v),\\
    \sigma_k((p',v),(h,v))
    &=
    \omega(v,p',h)\,
    \omega(p',v,h)^{-1}\,
    \omega(p',h,v).
\end{align}
Hence
\begin{equation}\label{eq:fixed-v-expanded}
    \frac{\sigma_k((h,v),(p,v))}
    {\sigma_k((p',v),(h,v))}
    =
    \frac{
    \omega(p',v,h)\,\omega(v,h,p)\,\omega(h,p,v)
    }{
    \omega(v,p',h)\,\omega(p',h,v)\,\omega(h,v,p)
    }.
\end{equation}
Comparing \eqref{eq:square-expanded} and \eqref{eq:fixed-v-expanded} gives
\begin{equation}\label{eq:square-fixed-v}
    \frac{\sigma_k((v,p'),(h,p))}
    {\sigma_k((h,p),(v,p))}
    =
    \frac{\sigma_k((h,v),(p,v))}
    {\sigma_k((p',v),(h,v))}.
\end{equation}
Combining \eqref{eq:square-phase} and \eqref{eq:square-fixed-v}, we conclude
\begin{equation}
    \frac{\vartheta_g(v;p'\to p')}
    {\vartheta_g(v;p\to p)}
    =
    \frac{\alpha_v(h,p)}
    {\alpha_v(p',h)}\,.
\end{equation}

\bibliographystyle{JHEP}
\bibliography{biblio.bib}

@article{Bantay:1990tr,
    author = "Bantay, P.",
    title = "{Orbifolds, Hopf algebras and the moonshine}",
    reportNumber = "ITP-478-BUDAPEST",
    doi = "10.1007/BF00403544",
    journal = "Lett. Math. Phys.",
    volume = "22",
    pages = "187--194",
    year = "1991"
}

@article{kishimoto2010cohomology,
  title={On the cohomology of free and twisted loop spaces},
  author={Kishimoto, Daisuke and Kono, Akira},
  journal={Journal of Pure and Applied Algebra},
  volume={214},
  number={5},
  pages={646--653},
  year={2010},
  publisher={Elsevier}
}

@book{brylinski2007loop,
  title={Loop spaces, characteristic classes and geometric quantization},
  author={Brylinski, Jean-Luc},
  year={2007},
  publisher={Springer Science \& Business Media}
}

@article{armstrong2022uniqueness,
  title={A uniqueness theorem for twisted groupoid C*-algebras},
  author={Armstrong, Becky},
  journal={Journal of Functional Analysis},
  volume={283},
  number={6},
  pages={109551},
  year={2022},
  publisher={Elsevier}
}

@article{Jia:2026vcr,
    author = "Jia, Qiang and Luo, Ran and Tian, Jiahua and Wang, Yi-Nan and Zhang, Yi",
    title = "{Categorical Symmetries via Operator Algebras}",
    eprint = "2604.25821",
    archivePrefix = "arXiv",
    primaryClass = "hep-th",
    month = "4",
    year = "2026"
}

@article{meinrenken2002basic,
  title={The basic gerbe over a compact simple Lie group},
  author={Meinrenken, Eckhard},
  journal={arXiv preprint math/0209194},
  year={2002}
}

@article{behrend2003equivariant,

  author = {Behrend, Kai and Xu, Ping and Zhang, Bin},

  title = {Equivariant gerbes over compact simple Lie groups},

  journal = {Comptes Rendus Mathematique},

  volume = {336},

  number = {3},

  pages = {251--256},

  year = {2003},

  doi = {10.1016/S1631-073X(02)00024-9}

}

@article{Axelrod:1989xt,
    author = "Axelrod, Scott and Della Pietra, Steve and Witten, Edward",
    title = "{Geometric quantization of Chern-Simons gauge theory}",
    reportNumber = "IASSNS-HEP-89/57",
    journal = "J. Diff. Geom.",
    volume = "33",
    number = "3",
    pages = "787--902",
    year = "1991"
}

@article{Bonetti:2024cjk,
    author = "Bonetti, Federico and Del Zotto, Michele and Minasian, Ruben",
    title = "{SymTFTs for continuous non-Abelian symmetries}",
    eprint = "2402.12347",
    archivePrefix = "arXiv",
    primaryClass = "hep-th",
    doi = "10.1016/j.physletb.2025.140010",
    journal = "Phys. Lett. B",
    volume = "871",
    pages = "140010",
    year = "2025"
}

@article{Freed:2022qnc,

  author = {Freed, Daniel S. and Moore, Gregory W. and Teleman, Constantin},

  title = {Topological symmetry in quantum field theory},

  journal = {Quantum Topology},

  volume = {15},

  pages = {779--869},

  year = {2024},

  doi = {10.4171/QT/223},

  eprint = {2209.07471},

  archivePrefix = {arXiv},

  primaryClass = {hep-th}

}

@article{Apruzzi:2021nmk,
    author = "Apruzzi, Fabio and Bonetti, Federico and Garc{\'\i}a Etxebarria, I{\~n}aki and Hosseini, Saghar S. and Schafer-Nameki, Sakura",
    title = "{Symmetry TFTs from String Theory}",
    eprint = "2112.02092",
    archivePrefix = "arXiv",
    primaryClass = "hep-th",
    doi = "10.1007/s00220-023-04737-2",
    journal = "Commun. Math. Phys.",
    volume = "402",
    number = "1",
    pages = "895--949",
    year = "2023"
}

@article{Chen:2012ctz,
    author = "Chen, Xie and Gu, Zheng-Cheng and Liu, Zheng-Xin and Wen, Xiao-Gang",
    title = "{Symmetry-Protected Topological Orders in Interacting Bosonic Systems}",
    eprint = "1301.0861",
    archivePrefix = "arXiv",
    primaryClass = "cond-mat.str-el",
    doi = "10.1126/science.1227224",
    journal = "Science",
    volume = "338",
    number = "6114",
    pages = "1604--1606",
    year = "2012"
}

@article{Gu:2012ib,
    author = "Gu, Zheng-Cheng and Wen, Xiao-Gang",
    title = "{Symmetry-protected topological orders for interacting fermions: Fermionic topological nonlinear \ensuremath{\sigma} models and a special group supercohomology theory}",
    eprint = "1201.2648",
    archivePrefix = "arXiv",
    primaryClass = "cond-mat.str-el",
    doi = "10.1103/PhysRevB.90.115141",
    journal = "Phys. Rev. B",
    volume = "90",
    number = "11",
    pages = "115141",
    year = "2014"
}

@article{Gruber:1999magnetic,
    author = "Gruber, Michael J.",
    title = "{Bloch Theory and Quantization of Magnetic Systems}",
    eprint = "math-ph/9903048",
    archivePrefix = "arXiv",
    primaryClass = "math-ph",
    doi = "10.1016/S0393-0440(99)00059-5",
    journal = "J. Geom. Phys.",
    volume = "34",
    number = "2",
    pages = "137--154",
    year = "2000"
}

@book{etingof2017tensor,
  title={Tensor Categories},
  author={Etingof, P.I. and Gelaki, S. and Nikshych, D. and Ostrik, V.},
  isbn={9781470437411},
  series={Mathematical surveys and monographs},
  url={https://books.google.com/books?id=9sC7vQEACAAJ},
  year={2017},
  publisher={American Mathematical Society}
}

@article{Jia:2025jmn,
    author = "Jia, Qiang and Luo, Ran and Tian, Jiahua and Wang, Yi-Nan and Zhang, Yi",
    title = "{Symmetry Topological Field Theory for Flavor Symmetry}",
    eprint = "2503.04546",
    journal ="",
    archivePrefix = "arXiv",
    primaryClass = "hep-th",
    month = "3",
    year = "2025"
}

@article{Kaidi:2022cpf,
    author = "Kaidi, Justin and Ohmori, Kantaro and Zheng, Yunqin",
    title = "{Symmetry TFTs for Non-invertible Defects}",
    eprint = "2209.11062",
    archivePrefix = "arXiv",
    primaryClass = "hep-th",
    doi = "10.1007/s00220-023-04859-7",
    journal = "Commun. Math. Phys.",
    volume = "404",
    number = "2",
    pages = "1021--1124",
    year = "2023"
}

@article{Gaiotto:2014kfa,
    author = "Gaiotto, Davide and Kapustin, Anton and Seiberg, Nathan and Willett, Brian",
    title = "{Generalized Global Symmetries}",
    eprint = "1412.5148",
    archivePrefix = "arXiv",
    primaryClass = "hep-th",
    doi = "10.1007/JHEP02(2015)172",
    journal = "JHEP",
    volume = "02",
    pages = "172",
    year = "2015"
}

@article{Jia:2025vrj,
    author = "Jia, Qiang and Luo, Ran and Tian, Jiahua and Wang, Yi-Nan and Zhang, Yi",
    title = "{Categorical Continuous Symmetry}",
    eprint = "2509.13170",
    archivePrefix = "arXiv",
    primaryClass = "hep-th",
    month = "9",
    year = "2025"
}

@article{Carey:2004xt,
    author = "Carey, Alan L. and Johnson, Stuart and Murray, Michael K. and Stevenson, Danny and Wang, Bai-Ling",
    title = "{Bundle gerbes for Chern-Simons and Wess-Zumino-Witten theories}",
    eprint = "math/0410013",
    archivePrefix = "arXiv",
    doi = "10.1007/s00220-005-1376-8",
    journal = "Commun. Math. Phys.",
    volume = "259",
    pages = "577--613",
    year = "2005"
}

@article{Behrend:2008,
  author        = {Behrend, Kai and Xu, Ping},
  title         = {Differentiable Stacks and Gerbes},
  year          = {2008},
  eprint        = {math/0605694},
  archivePrefix = {arXiv},
  primaryClass  = {math.DG},
}

@article{Stienon:2010,
  author  = {Sti\'enon, Mathieu},
  title   = {Equivariant Dixmier-Douady classes},
  journal = {Math. Res. Lett.},
  volume  = {17},
  number  = {1},
  pages   = {127--145},
  year    = {2010},
  doi     = {10.4310/MRL.2010.v17.n1.a10},
}

@article{Tu:2009twistedRing,
  author        = {Tu, Jean-Louis and Xu, Ping},
  title         = {The ring structure for equivariant twisted {K}-theory},
  journal       = {J. Reine Angew. Math.},
  volume        = {2009},
  number        = {635},
  pages         = {97--148},
  year          = {2009},
  doi           = {10.1515/CRELLE.2009.077},
  eprint        = {math/0604160},
  archivePrefix = {arXiv},
  primaryClass  = {math.KT}
}

@article{Jia:2025uun,
    author = "Jia, Qiang and Luo, Ran and Tian, Jiahua and Wang, Yi-Nan and Zhang, Yi",
    title = "{Anomaly of Continuous Symmetries from Topological Defect Network}",
    eprint = "2510.14722",
    archivePrefix = "arXiv",
    primaryClass = "hep-th",
    reportNumber = "USTC-ICTS/PCFT-25-47",
    month = "10",
    year = "2025"
}

@article{Coste:2000tq,
    author = "Coste, Antoine and Gannon, Terry and Ruelle, Philippe",
    title = "{Finite group modular data}",
    eprint = "hep-th/0001158",
    archivePrefix = "arXiv",
    doi = "10.1016/S0550-3213(00)00285-6",
    journal = "Nucl. Phys. B",
    volume = "581",
    pages = "679--717",
    year = "2000"
}

@article{Brylinski:2000dcg,
      title={Differentiable Cohomology of Gauge Groups}, 
      author={Jean-Luc Brylinski},
      year={2000},
      eprint={math/0011069},
      archivePrefix={arXiv},
      primaryClass={math.DG},
      url={https://arxiv.org/abs/math/0011069}, 
}

@article{Kumjian:1998umx,
    author = "Kumjian, Alex",
    title = "{Fell bundles over groupoids}",
    doi = "10.1090/S0002-9939-98-04240-3",
    journal = "Proc. Amer. Math. Soc.",
    volume = "126",
    pages = "1115-1125",
    year = "1998"
}

@article{Willerton:2008gyk,
    author = "Willerton, Simon",
    title = "{The twisted Drinfeld double of a finite group via gerbes and finite groupoids}",
    doi = "10.2140/agt.2008.8.1419",
    journal = "Algebr. Geom. Topol.",
    volume = "8",
    number = "3",
    pages = "1419--1457",
    year = "2008"
}

@article{FreedQuinn:1993,
    author = "Freed, Daniel S. and Quinn, Frank",
    title = "{Chern-Simons Theory with Finite Gauge Group}",
    doi = "10.1007/BF02096860",
    journal = "Commun. Math. Phys.",
    volume = "156",
    number = "3",
    pages = "435--472",
    year = "1993"
}

@article{Roche:1990hs,
    author = "Roche, P. and Pasquier, V. and Dijkgraaf, R.",
    title = "{QuasiHopf algebras, group cohomology and orbifold models}",
    journal = "Nucl. Phys. B Proc. Suppl.",
    volume = "18",
    pages = "60--72",
    year = "1990"
}

@article{Cordova:2018cvg,
    author = "C{\'o}rdova, Clay and Dumitrescu, Thomas T. and Intriligator, Kenneth",
    title = "{Exploring 2-Group Global Symmetries}",
    eprint = "1802.04790",
    archivePrefix = "arXiv",
    primaryClass = "hep-th",
    doi = "10.1007/JHEP02(2019)184",
    journal = "JHEP",
    volume = "02",
    pages = "184",
    year = "2019"
}

@article{Chen:2011pg,
    author = "Chen, Xie and Gu, Zheng-Cheng and Liu, Zheng-Xin and Wen, Xiao-Gang",
    title = "{Symmetry protected topological orders and the group cohomology of their symmetry group}",
    eprint = "1106.4772",
    archivePrefix = "arXiv",
    primaryClass = "cond-mat.str-el",
    doi = "10.1103/PhysRevB.87.155114",
    journal = "Phys. Rev. B",
    volume = "87",
    number = "15",
    pages = "155114",
    year = "2013"
}

@article{Hopkins:2002rd,
    author = "Hopkins, M. J. and Singer, I. M.",
    title = "{Quadratic functions in geometry, topology, and M theory}",
    eprint = "math/0211216",
    archivePrefix = "arXiv",
    journal = "J. Diff. Geom.",
    volume = "70",
    number = "3",
    pages = "329--452",
    year = "2005"
}

@article{Dijkgraaf:1989pz,
    author = "Dijkgraaf, Robbert and Witten, Edward",
    title = "{Topological Gauge Theories and Group Cohomology}",
    reportNumber = "THU-89-9, IASSNS-HEP-89-33",
    doi = "10.1007/BF02096988",
    journal = "Commun. Math. Phys.",
    volume = "129",
    pages = "393",
    year = "1990"
}

@article{Witten:1988hf,
    author = "Witten, Edward",
    editor = "Mitra, Asoke N.",
    title = "{Quantum Field Theory and the Jones Polynomial}",
    reportNumber = "IASSNS-HEP-88-33",
    doi = "10.1007/BF01217730",
    journal = "Commun. Math. Phys.",
    volume = "121",
    pages = "351--399",
    year = "1989"
}

@article{Blau:1989bq,
    author = "Blau, Matthias and Thompson, George",
    title = "{Topological Gauge Theories of Antisymmetric Tensor Fields}",
    reportNumber = "SISSA-39/89/FM, PAR-LPTHE-89-17",
    doi = "10.1016/0003-4916(91)90240-9",
    journal = "Annals Phys.",
    volume = "205",
    pages = "130--172",
    year = "1991"
}

@article{Schafer-Nameki:2023jdn,
    author = "Schafer-Nameki, Sakura",
    title = "{ICTP lectures on (non-)invertible generalized symmetries}",
    eprint = "2305.18296",
    archivePrefix = "arXiv",
    primaryClass = "hep-th",
    doi = "10.1016/j.physrep.2024.01.007",
    journal = "Phys. Rept.",
    volume = "1063",
    pages = "1--55",
    year = "2024"
}

@article{Baez:1999sr,
    author = "Baez, J. C.",
    editor = "Gausterer, H. and Pittner, L. and Grosse, H.",
    title = "{An Introduction to Spin Foam Models of $BF$ Theory and Quantum Gravity}",
    eprint = "gr-qc/9905087",
    archivePrefix = "arXiv",
    doi = "10.1007/3-540-46552-9_2",
    journal = "Lect. Notes Phys.",
    volume = "543",
    pages = "25--93",
    year = "2000"
}

@article{Koornwinder:1998xg,
    author = "Koornwinder, T. H. and Bais, F. A. and Muller, N. M.",
    title = "{Tensor product representations of the quantum double of a compact group}",
    eprint = "q-alg/9712042",
    archivePrefix = "arXiv",
    reportNumber = "UVA-WINS-WISK-97-14, UVA-WINS-ITFA-97-44",
    doi = "10.1007/s002200050475",
    journal = "Commun. Math. Phys.",
    volume = "198",
    pages = "157--186",
    year = "1998"
}

@article{Bais:1998yn,
    author = "Bais, F. A. and Muller, N. M.",
    title = "{Topological field theory and the quantum double of SU(2)}",
    eprint = "hep-th/9804130",
    archivePrefix = "arXiv",
    reportNumber = "UVA-WINS-ITFA-98-07",
    doi = "10.1016/S0550-3213(98)00572-0",
    journal = "Nucl. Phys. B",
    volume = "530",
    pages = "349--400",
    year = "1998"
}

@article{Kong:2020cie,
    author = "Kong, Liang and Lan, Tian and Wen, Xiao-Gang and Zhang, Zhi-Hao and Zheng, Hao",
    title = "{Algebraic higher symmetry and categorical symmetry -- a holographic and entanglement view of symmetry}",
    eprint = "2005.14178",
    archivePrefix = "arXiv",
    primaryClass = "cond-mat.str-el",
    doi = "10.1103/PhysRevResearch.2.043086",
    journal = "Phys. Rev. Res.",
    volume = "2",
    number = "4",
    pages = "043086",
    year = "2020"
}

@article{Kaidi:2023maf,
    author = "Kaidi, Justin and Nardoni, Emily and Zafrir, Gabi and Zheng, Yunqin",
    title = "{Symmetry TFTs and anomalies of non-invertible symmetries}",
    eprint = "2301.07112",
    archivePrefix = "arXiv",
    primaryClass = "hep-th",
    doi = "10.1007/JHEP10(2023)053",
    journal = "JHEP",
    volume = "10",
    pages = "053",
    year = "2023"
}

@article{Bhardwaj:2023ayw,
    author = "Bhardwaj, Lakshya and Schafer-Nameki, Sakura",
    title = "{Generalized charges, part II: Non-invertible symmetries and the symmetry TFT}",
    eprint = "2305.17159",
    archivePrefix = "arXiv",
    primaryClass = "hep-th",
    doi = "10.21468/SciPostPhys.19.4.098",
    journal = "SciPost Phys.",
    volume = "19",
    number = "4",
    pages = "098",
    year = "2025"
}

@article{Apruzzi:2023uma,
    author = "Apruzzi, Fabio and Bonetti, Federico and Gould, Dewi S. W. and Schafer-Nameki, Sakura",
    title = "{Aspects of categorical symmetries from branes: SymTFTs and generalized charges}",
    eprint = "2306.16405",
    archivePrefix = "arXiv",
    primaryClass = "hep-th",
    doi = "10.21468/SciPostPhys.17.1.025",
    journal = "SciPost Phys.",
    volume = "17",
    number = "1",
    pages = "025",
    year = "2024"
}

@article{Bhardwaj:2024igy,
    author = "Bhardwaj, Lakshya and Copetti, Christian and Pajer, Daniel and Schafer-Nameki, Sakura",
    title = "{Boundary SymTFT}",
    eprint = "2409.02166",
    archivePrefix = "arXiv",
    primaryClass = "hep-th",
    doi = "10.21468/SciPostPhys.19.2.061",
    journal = "SciPost Phys.",
    volume = "19",
    number = "2",
    pages = "061",
    year = "2025"
}

@article{Choi:2024tri,
    author = "Choi, Yichul and Rayhaun, Brandon C. and Zheng, Yunqin",
    title = "{Generalized Tube Algebras, Symmetry-Resolved Partition Functions, and Twisted Boundary States}",
    eprint = "2409.02159",
    archivePrefix = "arXiv",
    primaryClass = "hep-th",
    doi = "10.1007/s00220-025-05543-8",
    journal = "Commun. Math. Phys.",
    volume = "407",
    number = "4",
    pages = "62",
    year = "2026"
}

@article{Kong:2015flk,
    author = "Kong, Liang and Wen, Xiao-Gang and Zheng, Hao",
    title = "{Boundary-bulk relation for topological orders as the functor mapping higher categories to their centers}",
    eprint = "1502.01690",
    archivePrefix = "arXiv",
    primaryClass = "cond-mat.str-el",
    month = "2",
    year = "2015"
}

@article{Ji:2019jhk,
    author = "Ji, Wenjie and Wen, Xiao-Gang",
    title = "{Categorical symmetry and noninvertible anomaly in symmetry-breaking and topological phase transitions}",
    eprint = "1912.13492",
    archivePrefix = "arXiv",
    primaryClass = "cond-mat.str-el",
    doi = "10.1103/PhysRevResearch.2.033417",
    journal = "Phys. Rev. Res.",
    volume = "2",
    number = "3",
    pages = "033417",
    year = "2020"
}

@article{Delmastro:2025ksn,
    author = "Delmastro, Diego and Sharon, Adar and Zheng, Yunqin",
    title = "{Non-local conserved currents and continuous non-invertible symmetries}",
    eprint = "2507.22976",
    archivePrefix = "arXiv",
    primaryClass = "hep-th",
    doi = "10.1007/JHEP11(2025)072",
    journal = "JHEP",
    volume = "11",
    pages = "072",
    year = "2025"
}

@article{Shao:2023gho,
    author = "Shao, Shu-Heng",
    title = "{What's Done Cannot Be Undone: TASI Lectures on Non-Invertible Symmetries}",
    eprint = "2308.00747",
    archivePrefix = "arXiv",
    primaryClass = "hep-th",
    reportNumber = "YITP-SB-2023-19",
    month = "8",
    year = "2023"
}

@article{Bonetti:2025dvm,
    author = "Bonetti, Federico and Del Zotto, Michele and Minasian, Ruben",
    title = "{SymTFT for Continuous Symmetries: Non-linear Realizations and Spontaneous Breaking}",
    eprint = "2509.10343",
    archivePrefix = "arXiv",
    primaryClass = "hep-th",
    month = "9",
    year = "2025"
}

@article{Bhardwaj:2022yxj,
    author = "Bhardwaj, Lakshya and Bottini, Lea E. and Schafer-Nameki, Sakura and Tiwari, Apoorv",
    title = "{Non-invertible higher-categorical symmetries}",
    eprint = "2204.06564",
    archivePrefix = "arXiv",
    primaryClass = "hep-th",
    doi = "10.21468/SciPostPhys.14.1.007",
    journal = "SciPost Phys.",
    volume = "14",
    number = "1",
    pages = "007",
    year = "2023"
}

@article{de2007symplectic,
  title={Symplectic geometry and geometric quantization},
  author={De Buyl, Sophie and Detournay, St{\'e}phane and Voglaire, Yannick},
  journal={Proceedings of the" Third Modave Summer School on Mathematical Physics},
  year={2007}
}

@article{Verlinde:1988te,
    author = "Verlinde, Erik P.",
    title = "{Fusion Rules and Modular Transformations in 2D Conformal Field Theory}",
    doi = "10.1016/0550-3213(88)90603-7",
    journal = "Nucl. Phys. B",
    volume = "300",
    pages = "360--376",
    year = "1988"
}

@article{Jia:2026jmt,
    author = "Jia, Qiang and Ma, Cheng and Tian, Jiahua",
    title = "{Candidate Gaugings of Categorical Continuous Symmetry}",
    eprint = "2604.25820",
    archivePrefix = "arXiv",
    primaryClass = "hep-th",
    year = "2026"
}

@article{tornier2020haar,
  title={Haar measures},
  author={Tornier, Stephan},
  journal={arXiv preprint arXiv:2006.10956},
  year={2020}
}

@book{karpilovsky1993group,
  title={Group Representations},
  author={Karpilovsky, G.},
  isbn={9780444887269},
  lccn={92014786},
  series={Group Representations},
  url={https://books.google.co.kr/books?id=8hwlAQAAIAAJ},
  year={1993},
  publisher={North-Holland}
}

@article{Cheng2024,
author = {Cheng, Chuangxun and Li, Guilin},
year = {2024},
month = {05},
pages = {},
title = {Some Remarks on Projective Representations of Compact Groups and Frames},
volume = {14},
journal = {Communications in Mathematics and Statistics},
doi = {10.1007/s40304-023-00381-3}
}

\end{document}